\documentclass[12pt]{article}
\usepackage{amssymb}

\usepackage{jeep}
\usepackage{citesort}

\input tcilatex

\begin{document}

\section{INTRODUCTION}

It is generally considered that the aim of Statistical Mechanics of
many-body systems away from equilibrium is to determine their thermodynamic
properties, and the evolution in time of their macroscopic observables, in
terms of the dynamical laws which govern the motion of their constitutive
elements. This implies, first, in the construction of an irreversible
thermodynamics and a thermo-hydrodynamics (the latter meaning the particle
and energy motion in fluids, rheological properties, etc., with the
transport coefficients depending on the macroscopic thermodynamic state of
the system). Second, we need to face the all-important derivation of a
generalized nonlinear quantum kinetic theory and a response function theory,
which are of fundamental relevance to connect theory with observation and
experiment, basic for the corroboration of any theory [1], that is, the
synthesis leg in the scientific method born in the seventeenth century.

Oliver Penrose [2] has noted that Statistical Mechanics is notorious for
conceptual problems to which is difficult to give a convincing answer,
mainly: what is the physical significance of a Gibbs' ensemble?; How can we
justify the standard ensembles used in equilibrium theory?; What are the
right ensembles for nonequilibrium \ problems?; How can we reconcile the
reversibility of microscopic mechanics with the irreversibility of
macroscopic behavior? Moreover, related to the case of many-body systems out
of equilibrium, the late Ryogo Kubo, in the oppening addres in the Oji
Seminar [3], told us that statistical mechanics of nonlinear nonequilibrium
phenomena is just in its infancy and further progress can only be hoped by
closed cooperation with experiment. Some progress has been achieved since
then, and we try in this paper to describe, in a simple manner, some
attempts in the direction to provide a path for one particular initial
programme to face the questions posited above.

This is the so-called \textbf{Nonequilibrium Statistical Operator Method} (%
\textbf{NE-SOM} for short), which, however initially built on intuitive and
heuristic arguments, apparently can be incorporated within an interesting
approach to the rationalization of statistical mechanics, as contained in
the Maximization of (informa-tional-statistical) entropy (\textbf{MaxEnt}
for short) and Bayesian methods. In that way, NESOM may be considered as
covered under the umbrella of the so-called Jaynes'\textit{\ }\textbf{%
Predictive Statistical Mechanics} [4]. In the present paper, we attempt a
description of the MaxEnt-based NESOM (from now on MaxEnt-NESOM for short)
however mainly from a, say, pragmatical point of view: In section \textbf{3}%
, we show its use as a variational principle which allows to codify the
different Gibbs' statistical operators for the traditional ensembles in
equilibrium. In section \textbf{4}, the application to nonequilibrium\
situations is described, mainly on intuitive basis through analogy with the
case of equilibrium, and the six fundamental steps required to have a
complete reliable theory are presented and discussed. Previously, in next
section \textbf{2}, some general considerations on the rationale of the
approach are presented. In section \textbf{5} are briefly described some
applications of the formalism, like a MaxEnt-NESOM-based Thermodynamics and
Thermo-Hydrodynamics of irreversible processes, and an all important
Response Function Theory. The latter allows to make contact with
experiments, and some successful applications are summarized (they refer to
the case of the highly excited photoinjected plasma in semiconductors and of
polymers). Finally, in section \textbf{6}, are presented some concluding
remarks together with a brief account of some associated epistemological
aspects of the formalism and criticisms that have been labelled on some of
its aspects.

\section{GENERAL CONSIDERATIONS}

In the study of the macroscopic state of nonequilibrium systems we face
greater difficulties than those present in the theory of equilibrium
systems. This is mainly due to the fact that a more detailed analysis is
necessary to determine the temporal dependence of measurable properties, and
to calculate transport coefficients which are time-dependent (that is,
depending on the evolution in time of the nonequilibrium macrostate of the
system where dissipative processes are unfolding), and which\ are also space
dependent. That dependence is nonlocal in space and non-instantaneous in
time, as it encompasses space and time correlations. Robert Zwanzig [5] has
summarized the basic goals of nonequilibrium statistical mechanics as
consisting of: (i) To derive transport equations and to grasp their
structure; (ii) To understand how the approach to equilibrium occurs in
natural isolated systems; (iii) To study the properties of steady states;
and (iv) To calculate the instantaneous values and the temporal evolution of
the physical quantities which specify the macroscopic state of the system.
Also according to Zwanzig, for the purpose to face these items, there exist
several approaches which can be classified as: (a) Intuitive techniques; (b)
Techniques based on the generalization of the theory of gases; (c)
Techniques \ based on the theory of stochastic processes; (d) Expansions
from an initial equilibrium ensemble; (e) Generalization of Gibbs' ensemble
formalism.

The last item (e) is connected with Penrose's question noticed in the
Introduction concerning if there are, and what are, right ensembles for
nonequilibrium problems. In the absence of a Gibbs-style ensemble approach,
for a long time different kinetic theories were used, with variable success,
to deal with the great variety of nonequilibrium phenomena occurring in
physical systems in nature. We describe here a proposition for a
nonequilibrium statistical ensemble formalism, namely, the already mentioned
Nonequilibrium Statistical Operator Method, or NESOM, which appears to
provide grounds for a general prescription to choose appropriate ensembles
for nonequilibrium systems. The formalim has an accompanying nonlinear
quantum transport theory of a large scope (which encompasses as particular
limiting cases Boltzmann's and Mori's approaches), a response function
theory for arbitrarily-away-from-equilibrium systems, a statistical
thermodynamics (the so-called Informational Statistical Thermodynamics), and
an accompanying thermo-hydrodynamic for quantum fluids.

NESOM appears as a very powerful, concise, based on sound principles, and
elegant formalism of a broad scope to deal with systems arbitrarily away
from equilibrium. Zwanzig stated that the formalism ``has by far the most
appealing structure, and may yet become the most effective method for
dealing with nonlinear transport processes'' [5]. Later developments have
confirmed Zwanzig's prediction. The present structure of the formalism
consists in a vast extension and generalization of earlier pioneering
approaches, among which we can pinpoint the works of Kirkwood [6], Green
[7], Mori-Oppenheim-Ross [8], Mori [9], and Zwanzig [10]. NESOM has been
approached from different points of view: some are based on heuristic
arguments [8,11-14], others on projection operator techniques [15-17] (the
former following Kirkwood and Green and the latter following Zwanzig and
Mori). The formalism has been particularly systematized and largely improved
by the Russian School of statistical physics, which can be considered to
have been initiated by the renowned Nicolai Nicolaievich Bogoliubov [18],
and we may also name Nicolai Sergievich Krylov [19], and more recently
mainly through the relevant contributions by Dimitrii Zubarev [20,21],
Sergei Peletminskii [12,13], and others.

These different approaches to NESOM can be brought together under a unique
variational principle. This has been originally done by Zubarev and
Kalashnikov [22], and later on reconsidered in Ref. [23] (see also Refs.
[24] and [25]). It consists on the maximization, in the context of
Information Theory, of Gibbs statistical entropy (to be called fine-grained
informational-statistical entropy), subjected to certain constraints, and
including non-locality in space, retro-effects, and irreversibility on the
macroscopic level. This is the foundation of the nonequilibrium statistical
ensemble formalism that we describe in general terms in following sections.
The topic has surfaced in the section ``Questions and Answers'' of the Am.
J. Phys. [26,27]. The question by Baierlein [26], ``A central organizing
principle for statistical and thermal physics?'', was followed by Semura's
answer [27] that ``the best central organizing principle for statistical and
thermal physics is that of maximum [informational] entropy [...]. The
principle states that the probability should be chosen to maximize the
average missing information of the system, subjected to the constraints
imposed by the [available] information. This assignment is consistent with
the least biased estimation of probabilities.''

As already noticed in Section \textbf{1}, the formalism may be considered as
covered under the umbrella provided by the scheme of Jaynes' Predictive
Statistical Mechanics [4,28]. This is a powerful approach based on the
Bayesian method in probability theory, together with the principle of
maximization of informational entropy (MaxEnt), and the resulting
statistical ensemble formalism here described is the already referred-to as 
\textbf{MaxEnt-NESOM. }Jaynes' scheme implies in a predictive statistics
that is built only on the access to the relevant information that there
existis of the system [4,29-32]. As pointed out by Jaynes [28]. ``How shall
we best think about Nature and most efficiently predict her behavior, given
only our incomplete knowledge [of the microscopic details of the system]?
[...]. We need to see it, not as an example of the N-body equations of
motion, but as an example of the logic of scientific inference, which
by-passes all details by \textit{going directly from our macroscopic
information to the best macroscopic predictions that can be made from that
information'' }(emphasis is ours) [...]. ``Predictive Statistical Mechanics
is not a physical theory, but a method of reasoning that accomplishes this
by finding, not the particular that the equations of motion say in any
particular case, but the general things that they say in `almost all' cases
consisting with our information; for those are the reproducible things''.

Again following Jaynes' reasoning, the constuction of a statistical approach
is based on ``a rather basic principle [...]: If any macrophenomenon is
found to be reproducible, then it follows that all microscopic details that
were not under the experimenters' control must be irrelevant for
understanding and predicting it''. Further, ``the difficulty of prediction
from microstates lies [..] in our own lack of the information needed to
apply them. We never know the microstate; only a few aspects of the
macrostate. Nevertheless, the aforementioned principle of [macroscopic]
reproductibility convinces us that this should be enough; \textit{the
relevant information is there, if only we can see how to recognize it and
use it}'' [emphasis is ours].

As noticed, Predictive Statistical Mechanics is founded on the Bayesian
approach in probability theory. According to Jaynes, the question of what
are theoretically valid, and pragmatically useful, ways of applying
probability theory in science has been approached by Sir Harold Jeffreys
]33,34], in the sense that he stated the general philosophy of what
scientific inference is and proceeded to carry both the mathematical theory
and its implementations. Together with\ Jaynes and others, the Nobelist
Philip W. Anderson [35] mantains that what seems to be the most appropriate
probability theory for the sciences is the Bayesian approach. The Bayesian
interpretation is that probability is the degree of belief which is
consistent to hold in considering a proposition as being true, once other
conditioning proposition are taken as true [36]. Or, also according to
Anderson: ``What Bayesian does is to focus one's attention on the question
one wants to ask of the data. It says in effect, how do these data affect my
previous knowlodge of the situation? It is sometimes called \textit{maximum
likelihood thinking, but the essence of it is to clearly identify the
possible answers, assign reasonable a priori probabilities to them and then
ask which answers have been done more likely by the data''} [emphasis is
ours].

The question that arises is, as stated by Jaynes, ``how shall we use
probability theory to help us do plausible reasoning in situations where,
because of incomplete information we cannot use deductive reasoning?'' In
other words, the main question is how to obtain the probability assignment
compatible with the available information, while avoiding unwarranted
assumptions. This is answered by Jaynes who formulated the criterion that:
the least biased probability assignment $\{p_{j}\}$, for a set of mutually
exclusive events $\{x_{l}\}$, is the one that maximizes the quantity $S_{I}$%
, sometimes referred to as the \textit{informational entropy}, given by 
\begin{equation}
S_{I}=-\sum\limits_{j}p_{j}\ln p_{j}\qquad ,  \label{eq1}
\end{equation}
conditioned by the constraints imposed by the available information. This is
based on Shannon's ideas in the mathematical theory of communications [37],
who first demonstrated that, for an exhaustive set of mutually exclusive
propositions, there exists a unique function measuring the uncertainty of
the probability assignment. This is the already mentioned principle of
maximization of the informational-statistical entropy, MaxEnt for short. It
provides the variational principle which results in a unifying theoretical
framework for the NESOM, thus introducing, as we have noticed, the
MaxEnt-NESOM as a nonequilibrium statistical ensemble formalism. It should
be stressed \textit{that the maximization of }$S_{I}$\textit{\ implies in
making maximum the uncertainty in the information available} (in
Shannon-Brillouin's sense [37,38]), to have in fact the least biased
probability assignment.

We proceed in the next two sections to describe the construction of the
MaxEnt-NESOM: First we briefly review the case corresponding to Gibbs'
equilibrium ensemble formalism, for, in continuation to consider the case of
systems arbitrarily away from equilibrium presenting strong dissipative
effects.

\section{MaxEnt-NESOM IN EQUILIBRIUM CONDITIONS}

Let us consider a many-body system in equilibrium with a given set of ideal
reservoirs, and, to be specific, let us take the case of thermal and
particle reservoirs, with the equilibrium between the system and the two
reservoirs implying that they have equal temperature and chemical potential
respectively. Moreover, because of Liouville theorem, in any case the
statistical operator, designated by $\varrho \left( t\right) $, satisfies
Liouville-Dirac equation, namely 
\begin{equation}
\frac{\partial }{\partial t}\varrho \left( t\right) +\frac{1}{i\hslash }%
\left[ \varrho \left( t\right) ,\hat{H}\right] =0\quad ,  \label{eq2}
\end{equation}
where $\hat{H}$ is the system Hamiltonian. In equilibrium $\varrho $ does
not depend on time, commutes with the Hamiltonian and, consequently, as well
known, it must be a superoperator depending on the dynamical operators
corresponding to constants of motion of the system.

The derivation in MaxEnt-NESOM of the equilibrium statistical operator [39]
(and also in the local equilibrium approximation [40]) is already a textbook
matter [41]. For that purpose, in Eq. (\ref{eq1}), index $j$ is to be
interpreted as the complete set of quantum numbers which characterizes the
eigenstates of the Hamiltonian $\hat{H},$ and $p_{j}$ is the diagonal matrix
element of $\varrho $ in such states.\ In this case Eq. (\ref{eq1}) can
alternatively be written in the form 
\begin{equation}
S_{G}=-Tr\left\{ \varrho \ln \varrho \right\} \quad ,  \label{eq3}
\end{equation}
that is, in terms of the trace operator, what thus makes the calculation
independent of the quantum representation. This $S_{G}$ is denominated
Gibbs' statistical entropy which multiplied by Boltzmann constant $k_{B}$
provides in equilibrium the proper thermodynamic entropy in Clausius-Carnot
sense [42]. According to MaxEnt, $S_{G}$ is maximized however subjected to
the appropriate normalization of $\varrho $, that is 
\begin{equation}
Tr\left\{ \varrho \right\} =1\qquad ,  \label{eq4a}
\end{equation}
and the other constraints consist into the choice of the relevant constants
of motion to be used. But the \textit{one and only information} we do have
is the imposed macroscopic condition in the given experiment, namely, the
preparation of the system in equilibrium with a thermal and a particle
reservoirs. In these conditions are fixed the temperature $T$, the chemical
potential $\mu $, and, of course, the volume $V$ is given. [We recall that
the thermodynamic state is then fully described by a thermodynamic
potential, which in this case is the so-called grand-canonical free energy $%
\mathcal{F}\left( T,V,\mu \right) $]. Therefore, on the basis of the
information we do have, the constraints in the maximization process are the
expected values for energy $E$ and particle number $N$, at the given $T$ and 
$\mu $ (for fixed $V$), namely 
\begin{mathletters}
\begin{eqnarray}
E &=&Tr\left\{ \hat{H}\varrho \right\} \qquad ,  \label{eq4b} \\
&&  \nonumber \\
N &=&Tr\left\{ \hat{N}\varrho \right\} \qquad ,  \label{eq4c}
\end{eqnarray}
where $\hat{N}$ is the particle number operator. Resorting to the Lagrange
variational method for the calculation of the $\varrho $ which makes maximum 
$S_{G}$ of Eq. (\ref{eq3}) together with the constraints imposed by Eqs. (%
\ref{eq4b}), one easily finds that (see Appendix \textbf{I}, first part) 
\end{mathletters}
\begin{equation}
\varrho =\exp \left\{ -\Phi -F_{1}\hat{H}-F_{2}\hat{N}\right\} \quad ,
\label{eq5}
\end{equation}
where $\Phi $, $F_{1}$ and $F_{2}$ are the \textit{Lagrange multipliers}
that the variational method introduces, and which are related to the three
constraints in Eqs. (\ref{eq4a}), (\ref{eq4b}) and (\ref{eq4c})
respectively. Moreover, it is usually written 
\begin{equation}
\Phi \left( T,V,\mu \right) =\ln Z\left( T,V,\mu \right) \quad ,  \label{eq6}
\end{equation}
introducing the grand-partition function $Z$. The other Lagrange multipliers
are related to partial derivatives of the entropy, that is 
\begin{equation}
F_{1}=\partial S_{G}/\partial E\qquad ;\qquad F_{2}=\partial S_{G}/\partial
N\quad ,  \label{eq7}
\end{equation}
and are functions of the basic extensive variables $E,V,N.$ Equations (\ref
{eq7}) are equations of state connecting extensive and intensive
thermodynamic variables, since these Lagrange multipliers $F_{1}$ and $F_{2}$
are related to the temperature $T$ and the chemical potencial $\mu $: In
fact, as known, building Gibbs statistical thermodynamics in terms of $%
\varrho $ and comparing it with the results of phenomenological
thermodynamics, it follows that 
\begin{equation}
F_{1}=\beta =1/k_{B}T\qquad ;\qquad F_{2}=-\mu /k_{B}T\quad .  \label{eq8}
\end{equation}
Moreover, the third Lagrange multiplier $\Phi $, which ensures the
normalization of $\varrho $, determines the grand-canonical free energy 
\begin{equation}
\mathcal{F}\left( T,V,\mu \right) =-k_{B}T\Phi \left( T,V,\mu \right)
=-k_{B}T\ln Z\left( T,V,\mu \right) .  \label{eq9}
\end{equation}
Hence, the equilibrium statistical operator reads as 
\begin{equation}
\varrho =Z^{-1}\left( T,V,\mu \right) \ \exp \left\{ -\beta \left( \hat{H}%
-\mu \hat{N}\right) \right\} \quad ,  \label{eq10}
\end{equation}
which is precisely Gibbs' grand-canonical statistical operator, with 
\begin{equation}
\Phi \left( T,V,\mu \right) =\ln Z\left( T,V,\mu \right) =\ln Tr\left\{ \exp 
\left[ -\beta \left( \hat{H}-\mu \hat{N}\right) \right] \right\} .
\label{eq11}
\end{equation}
Along a quite similar line of reasoning, we can derive any Gibbs' canonical
statistical operator, simply introducing the appropriate \textit{%
informational constraints }determined by the given macroscopic conditions of
preparation of the system, meaning \textit{the knowledge of the set of
reservoirs with which the system is in contact and in equilibrium} (an
equilibrium, we stress, defined by equal values of the corresponding
intensive thermodynamic variables - the Lagrange multipliers in the
formalism - in system and reservoir).

Without further considerations, we simply notice that the local equilibrium
situation is described by the statistical operator 
\begin{equation}
\varrho \left( t\right) =\exp \left\{ -\Phi \left( t\right) -\int d^{3}r\
\beta \left( \vec{r},t\right) \left[ \hat{h}\left( \vec{r}\right) -\mu
\left( \vec{r},t\right) \hat{n}\left( \vec{r}\right) \right] \right\} ,
\label{eq12}
\end{equation}
where $\hat{h}\left( \vec{r}\right) $ and $\hat{n}\left( \vec{r}\right) $
are the locally conserved densities of energy and particle number, $\beta
\left( \vec{r},t\right) $, the reciprocal of a field of local temperature
and $\mu \left( \vec{r},t\right) $ the field of a local-chemical potential
[41]. The corresponding Gibbs' statistical thermodynamics gives microscopic
foundations to Classical (sometimes called Linear or Onsagerian)
Irreversible Thermodynamics, once the constitutive Fick and Fourier laws for
the fluxes of matter and energy are introduced (see for example the
classical textbook of Ref. [43]).

The associated Response Function Theory and Transport Theories for weak
excitations near the equilibrium state, follow from perturbation theory
applied on the equilibrium state described by Eq. (\ref{eq10}) corresponding
to the equilibrium state of initial preparation of the system in the given
experiment [44-47].

\section{MaxEnt-NESOM FOR DISSIPATIVE PROCESSES}

We consider now a many-body system out of equilibrium, described by a
statistical operator $\varrho \left( t\right) $, to be derived in
MaxEnt-NESOM. It satisfies Liouville equation, Eq. (\ref{eq2}), but
differently to equilibrium it depends explicity on time, as it describes the
evolution of the macroscopic state of the system while dissipative processes
unfold in the isolated media. Hence, in this case, we do not have the
information available in the case of equilibrium that $\varrho $ is
dependent only on constants of motion. Consequently, the first fundamental
step in the present situation is to decide on the basic set of dynamical
variables appropriate for the description of the macroscopic state of the
nonequilibrium system. At this point enters the fundamental \textit{%
Bogoliubov's principle of correlation weakening,} and the accompanying
hierarchy of relaxation times [48,49]. According to this view, a series of
successively contracted descriptions is possible because the existence, in
many cases, of an array of relaxation times, say $\tau _{\mu }<\tau _{1}<...$%
, such that after each one has elapsed, correlations with lifetimes smaller
that each one of these time lengths are damped out (that is, the associated
dissipative processes have died down) and can be ignored. Then, an ever
shortened set of dynamical variables can be used for a proper description of
the macrostate of the system; at a sufficiently long time the equilibrium
condition is approached (with all correlations being wiped out), and the
most contracted descriptions is to be used (i.e., in terms of only $\hat{H}$
and $\hat{N}$ as shown in Section \textbf{2}). Uhlenbeck [50] has pointed
out that it seems likely that successive contractions of the description are
an essential feature of the theory of irreversile processes, and that such
contraction must be a property of the basic equations of the system
(illustrations in the case of a spin-lattice system and of the photoinjected
plasma in semiconductors, are given in Refs. [51] and [52], respectively).

This is fundamental in MaxEnt-NESOM, and has been in practice introduced
according to the proposal set forward by, among others, Mori [9,47], Zubarev
[14,21], and Peletminskii [12,13]. It consists in the \textbf{basic first
step} in the formalism consisting into introducing a separation of the total
Hamiltonian into two parts, namely 
\begin{equation}
\hat{H}=\hat{H}_{o}+\hat{H}^{\prime }\qquad ,  \label{eq13}
\end{equation}
where $\hat{H}_{o}$ is the so-called ``relevant'' (or secular, or
quasi-conserving) part, composed by energy operators involving the kinetic
energies and a part of the interactions, namely, those strong enough to be
responsible for fast dissipative processes with \textit{very short
relaxation times, }meaning those smaller than the characteristic time scale
to be used for the description of the system, essentially the resolution
time in any given experiment under consideration. Hence, as pointed out by
Grandy [40], the ground level in the formulation of a theory is to properly
describe the system and the kind of experiment(s) to be analized. It is
worth noticing the nowadays impressive development of experimental
techniques in ultrafast laser spectroscopy [53] in the pico- and
femto-second time scale, and soon, resorting to atomic instead of molecular
transitions, extended to the atto-second $\left[ 10^{-18}\text{ }\sec \right]
$ scale, and also the development of means of detection with spatial
resolution in the nanometer scale. The other contribution, $\hat{H}^{\prime
},$ in Eq. (\ref{eq13}) contains the remainig interaction potentials,
associated to interactions responsible for processes with \textit{long-time
relaxations times }(meaning larger than the experimental characteristic
time), in the dissipative processes that develop in the system.

This leads to the \textbf{second basic step} in the formalism, namely, to
introduce a basic set of dynamical variables, call them $\left\{ \hat{P}%
_{j}\right\} $,\ $j=1,2,...$ (the triangular hat indicating Hermitian
operator in Quantum Mechanics, or in Classical Mechanics a dynamical
function defined over the phase space). This is done, in analogy with
equilibrium, introducing those that are quasi-conserved (quasi-constants of
motion) under the dynamics generated by the secular part of the Hamiltonian,
namely, $\hat{H}_{o}$, that is, satisfying what we call \textit{%
Zubarev-Peletminskii selection rule,} consisting in that the evolution of
the basic variables under $\hat{H}_{o}$ produces linear combinations of the
type 
\begin{equation}
\frac{1}{i\hslash }\left[ \hat{P}_{j},\hat{H}_{o}\right] =\sum\limits_{k}%
\alpha _{jk}\hat{P}_{k}\qquad ,  \label{eq14}
\end{equation}
where $j,k=1,2,...$, and, in an appropriate quantum representation (the use
of reciprocal space) the $\alpha $'s are c-numbers. In other
representations, quantities $\hat{P}$ can be dependent on the space
variable, that is, when considering local densities of dynamical variables,
and then the $\alpha $'s can depend on the space variable or be differential
operators [54-56]. As already noticed, this introduces the quasi-conserved
quantities $\hat{P}_{j}$, in the sense of having associated -- at the
macroscopic level defined by their average values over the nonequilibrium
ensemble -- a near dissipationless character, meaning that their relaxation
times are much larger than the experimental resolution time (or the
corresponding characteristic time in the theoretical analysis being
performed). It also needs be stressed that Eq. (\ref{eq14}) encompasses the
case of quantities $\hat{P}$ for which all coefficients $\alpha $'s on the
right are null, i.e. they are full constants of motion under the dynamics
generated by $\hat{H}_{o}$, which itself falls under this condition, and, as
a rule, must always be taken as a basic variable. It is relevant to notice
that Zubarev-Peletminskii law provides a kind of a \textit{closure condition 
}in the accompanying nonlinear-nonlocal-memory dependent kinetic theory,
that is, a closed set of equations of evolution, as shown later on as we
proceed. As discussed elsewhere [57], this is the statistical counterpart
(in Informational Statistical Thermodynamics) of the principle of
equipresence in phenomenological Thermodynamics [58,59].

The practical use of the law of Eq. (\ref{eq14}) runs as follows: First, the
secular part of the Hamiltonian, viz. $\hat{H}_{o}$, is chosen in the
particular problem and conditions under consideration (as noted, it contains
the kinetic energies plus the interactions strong enough to produce damping
of correlations, responsible for a certain set of dissipative processes, in
times smaller than that of the characteristic time of the experiment one has
in mind; hence, here Bogoliubov's principle of correlation weakening is
fully at work). Second, one introduces a few dinamical variables $\hat{P}$
deemed relevant for the description of the physical problem in hands
(typically are chosen the densities of energy and of particle number, to
introduce - via the process described below - a kind of nonequilibrium
generalized grand-canonical ensemble). Next, the commutator of these firstly
chosen variables is performed and the new variables - which are different
from those already introduced - appearing in the linear combination
indicated by the right-hand side of Eq. (\ref{eq14}) are incorporated to the
basic set. This procedure is repeated until a closure is attained. For a
first choice consisting of the energy density and particle number density,
application of the law of Eq. (\ref{eq14}) requires to introduce as basic
variables the fluxes of all order of energy and mass [55], what constitutes
the basis for the construction of a MaxEnt-NESOM-based Thermo-Hydrodynamics
[54], that is, a nonclassical hydrodynamics founded in the nonequilibrium
thermodynamic state described by IST [60-62].

Briefly summarizing the points described above, we may say that in a first
step are separated fast-relaxating processes from the slow ones. Next, on
the basis of this, Zubarev-Peletminskii selection rule implies taking into
account all dynamical quantities (mechanical observables) that, under the
dynamics generated by $\hat{H}_{o}$, are kept in a subspace of the Hilbert
space, and then are referred to as quasi-conserved variables. They are the
relevant ones to be retained in the informational-based approach of
MaxEnt-NESOM. As already noticed, the procedure is the analog of the choice
of the basic variables in the case of systems in equilibrium when it is done
on the basis of taking the wholly conserved ones. The point shall be better
clarified as we proceed.

\subsection{The Variational Method}

Once the basic set of variables $\left\{ \hat{P}_{j}\right\} $ has been
chosen, in analogy with the case of equilibrium one should, in principle,
proceed to obtain the nonequilibrium statistical operator by - according to
MaxEnt - maximizing the informational-statistical entropy at time $t$, given
by 
\begin{equation}
\bar{S}\left( t\right) =-Tr\left\{ \bar{\varrho}\left( t,0\right) \ln \bar{%
\varrho}\left( t,0\right) \right\} \qquad ,  \label{eq15}
\end{equation}
subjected to the constraints consisting in the requirement of its
normalization 
\begin{equation}
Tr\left\{ \bar{\varrho}\left( t,0\right) \right\} =1\qquad ,  \label{eq16a}
\end{equation}
and the chosen set of average values 
\begin{equation}
Q_{j}\left( \mathbf{r},t\right) =Tr\left\{ \hat{P}_{j}\left( \mathbf{r}%
\right) \bar{\varrho}\left( t,0\right) \right\} \qquad ,  \label{eq16b}
\end{equation}
where we have explicitly introduced a possible dependence on the space
coordinate. Here, we are indicating by $\bar{\varrho}\left( t,0\right) $ the
statistical operator to follow from application of the variational method in
this instantaneous (at time $t$) procedure; the first $t$ in the argument of 
$\bar{\varrho}$ stands for the time dependence of the basic thermodynamic
variables, and\ the zero indicates that the dynamical operators are taken in
Schroedinger representation (this point will be better clarify below). The
set $\left\{ Q_{j}\left( \mathbf{r},t\right) \right\} $ constitutes the one
composed by the macrovariables that are the\ basic ones in the
nonequilibrium thermodynamic description that the MaxEnt-NESOM provides (the
so-called Informational Statistical Thermodynamics), or, in other words,
they define Gibbs - or nonequilibrium thermodynamic - state space. As
already noticed, the selection rule of Eq. (\ref{eq14}) provides a set of
macrovariables for a closed kinetic theory which accounts for the
thermodynamic description of the system and its evolution, the equivalent,
as indicated, of the principle of equipresence imposed in some
phenomenological thermodynamic theories.

Using the Lagrange multipliers method, it follows that 
\begin{equation}
\bar{\varrho}\left( t,0\right) =\exp \left\{ -\Phi \left( t\right)
-\sum\limits_{j}\int d^{3}r\ F_{j}\left( \mathbf{r},t\right) \hat{P}%
_{j}\left( \mathbf{r}\right) \right\} \ ,  \label{eq17}
\end{equation}
with 
\begin{equation}
\Phi \left( t\right) =\ln Tr\left\{ \exp \left[ -\sum\limits_{j}\int d^{3}r\
F_{j}\left( \mathbf{r},t\right) \hat{P}_{j}\left( \mathbf{r}\right) \right]
\right\} \ ,  \label{eq18}
\end{equation}
ensuring the normalization of $\bar{\varrho}$, and where $\left\{
F_{j}\left( \mathbf{r},t\right) \right\} $ are the set of Lagrange
multipliers that the variational method introduces, and which are determined
in terms of the basic macrovariables $Q_{j}$ by Eq. (\ref{eq16b}). Moreover,
interpreting $\Phi \left( t\right) $ as the logarithm of a nonequilibrium
ensemble partition function $\bar{Z}\left( t\right) $ it follows that 
\begin{equation}
Q_{j}\left( \mathbf{r},t\right) =-\delta \Phi \left( t\right) /\delta
F_{j}\left( \mathbf{r},t\right) =-\delta \ln \bar{Z}\left( t\right) /\delta
F_{j}\left( \mathbf{r},t\right) \ ,  \label{eq19}
\end{equation}
showing a close analogy with equilibrium, and where $\delta $ stands here
for functional differential [63].

The statistical operator of Eq. (\ref{eq17}) has the form of an
instantaneous generalized canonical distribution, and an immediate question
is if it properly provides a nonequilibrium statistical \ mechanics for
dissipative systems. The answer is on the \textit{negative.} Indeed: (1) it 
\textit{does not} satisfy Liouville equation; (2) it \textit{does not}
describe the dissipative processes that \ develop in the system; (3) it 
\textit{does not}\ provide a correct kinetic theory for the description of
the dissipative processes which are unfolding in the medium; (4) it \textit{%
does not} give the correct average values for observables, other than those
corresponding to the basic dynamical variables in Eq. (\ref{eq16b}). Taking
this statistical operator (whose meaning and relevance will be discussed
below) as the one that may properly describe the full nonequilibrium state
of the system and its dissipative evolution, has led to some unnecessary
confusion and controversy in the past.

The question is then to find the proper nonequilibrium statistical operator
that MaxEnt-NESOM should provide. The way out of the dificulties pointed out
above is contained in the idea set forward by John Kirkwood in the decade of
the forties [6]. He pointed out that the state of the system at time $t$ is
strongly dependent on all the previous evolution of the nonequilibrium
processes that have been developing in it. Kirkwood introduces this fact, in
the context of the transport theory he proposes, in the form of a so-called 
\textit{time-smoothing procedure,} which is generalized in MaxEnt-NESOM as
shown below.

Introducing in MaxEnt-NESOM [22-25] the idea that it must be incorporated
all the past history of the system (or \textit{historicity effects}), all
along the time interval going from the initial condition of preparation of
the sample in the given experiment at, say, time $t_{o}$ up to time $t$ when
a measurement is performed (i.e., when we observe the macroscopic state of
the system), we proceed to maximize Gibbs' entropy (sometimes called
fine-grained entropy) 
\begin{equation}
S_{G}\left( t\right) =-Tr\left\{ \varrho \left( t\right) \ln \varrho \left(
t\right) \right\} \qquad ,  \label{eq20}
\end{equation}
with the normalization and constraints given at any time $t^{\prime }$ in
the interval $t_{o}\leq t^{\prime }\leq t$, namely 
\begin{mathletters}
\begin{eqnarray}
Tr\left\{ \varrho \left( t^{\prime }\right) \right\}  &=&1\qquad ,
\label{eq21a} \\
&&  \nonumber \\
Q_{j}\left( \mathbf{r},t^{\prime }\right)  &=&Tr\left\{ \hat{P}_{j}\left( 
\mathbf{r}\right) \varrho \left( t^{\prime }\right) \right\} \ .
\label{eq21b}
\end{eqnarray}
Again resorting to Lagrange's procedure we find that (see second part of
Appendix \textbf{I}) 
\end{mathletters}
\begin{equation}
\varrho \left( t\right) =\exp \left\{ -\Psi \left( t\right)
-\sum\limits_{j}\int d^{3}r\int\limits_{t_{o}}^{t}dt^{\prime }\ \varphi
_{j}\left( \mathbf{r};t,t^{\prime }\right) \hat{P}_{j}\left( \mathbf{r}%
;t-t^{\prime }\right) \right\} ,  \label{eq22}
\end{equation}
where 
\begin{equation}
\Psi \left( t\right) =\ln Tr\left\{ \exp \left[ -\sum\limits_{j}\int
d^{3}r\int\limits_{t_{o}}^{t}dt^{\prime }\ \varphi _{j}\left( \mathbf{r}%
;t,t^{\prime }\right) \hat{P}_{j}\left( \mathbf{r};t-t^{\prime }\right) %
\right] \right\} ,  \label{eq23}
\end{equation}
and the $\varphi _{j}$ are the corresponding Lagrange multipliers determined
in terms of the basic macrovariables by Eq. (\ref{eq21b}), and operators $%
\hat{P}_{j}$ are given in Heisenberg representation.

An important point to be remarked is that Eqs. (\ref{eq21b}) introduce a
dynamical character in the chosen information, differently to the case of
Eqs. (\ref{eq16a}) which provide the macroscopic state at only a given time $%
t$. Both descriptions, namely the ``historical one'' via $\varrho \left(
t\right) $ and the ``instantaneous one'', via $\bar{\varrho}\left(
t,0\right) ,$ are related by the relevant fact that both describe at any
time $t$ the same macrostate [cf. Eqs. (\ref{eq16b}) and (\ref{eq21b})],
that is, it must be satisfied that 
\begin{equation}
Q_{j}\left( \mathbf{r},t\right) =Tr\left\{ \hat{P}_{j}\left( \vec{r}\right) 
\bar{\varrho}\left( t,0\right) \right\} =Tr\left\{ \hat{P}_{j}\left( \mathbf{%
r}\right) \varrho \left( t\right) \right\} \ .  \label{eq24}
\end{equation}
Another relevant observation needs be antecipated at this point to avoid
misunderstandings: The statistical operator depends on the
information-gathering interval $\left( t_{o},t\right) $, but it must be kept
in mind that this is the formal point consisting in that, as Kirkwood
pointed out, the description to be built must contain all the previous
history in the development of the macrostate of the system. This is later on
translated to the accompanying nonlinear kinetic theory (a far-reaching
generalization of those of Boltzmann and Mori), when the set of
integro-differential transport equations for the basic variables require us
to have access to \textit{only} the value of the macrovariables at time $%
t_{o},$ the initial time of preparation of the system.

Hence, following the two basic steps for the building up of the formalism
described in the first part of this section, a \textbf{third basic step} has
just been introduced, namely, the inclusion of the past history (or, as
sometimes called, \textit{retro-effects }or \textit{historicity}) of the
macrostate of the dissipative system. A \textbf{fourth basic step} needs now
be considered, being a generalization of Kirkwood's \textit{time-smoothing
procedure}. This is done introducing an extra assumption on the form of the
Lagrange multipliers $\varphi _{j}$, in such a way, we stress, that (i)
irreversible behavior in the evolution of the macroscopic state of the
system is satisfied; (ii) the instantaneous state of the system is given by
Eq. (\ref{eq24}); (iii) both $\bar{\varrho}$ and $\varrho $ are normalized
at each time $t$, and (iv) it is introduced the set \ of quantities $\left\{
F_{j}\left( \mathbf{r},t\right) \right\} $ as intensive variables
thermodynamically conjugated to basic macrovariables $\left\{ Q_{j}\left( 
\mathbf{r},t\right) \right\} $, what leads a \textit{posteriori} to generate
satisfactory Thermodynamic and Thermo-Hydrodynamic theories. This is
accomplished introducing the definition 
\begin{equation}
\varphi _{j}\left( \mathbf{r};t,t^{\prime }\right) =w\left( t,t^{\prime
}\right) F_{j}\left( \mathbf{r},t\right) \qquad ,  \label{eq26}
\end{equation}
where $w\left( t,t^{\prime }\right) $ is an auxiliary weight function,
which, to satisfy the four points just listed immediatly above, must have
well defined properties which are discussed elsewhere [23], and it is
verified that 
\begin{equation}
\Psi \left( t\right) =\int\limits_{-\infty }^{t}dt^{\prime }w\left(
t,t^{\prime }\right) \phi \left( t^{\prime }\right) \qquad .  \label{eq25}
\end{equation}
The function $w\left( t,t^{\prime }\right) $ introduces the \textit{%
time-smoothing procedure}, and, because of the properties it must have to
acomplish its purposes, it is acceptable any kernel that the mathematical
theory of convergence of trigonometrical series and transform integrals
provides. Kirkwood, Green, Mori and others have chosen what in mathematical
parlance is Fej\`{e}r (or Ces\`{a}ro-1) kernel, while Zubarev introduced the
one consisting in Abel's kernel for $w$ in Eq. (\ref{eq26}) - which
apparently appears to be the best choice, either mathematically but mainly
physically - that is, taking $w\left( t,t^{\prime }\right) =\varepsilon \exp
\left\{ \varepsilon \left( t^{\prime }-t\right) \right\} $, where $%
\varepsilon $ is a positive infinitesimal that goes to zero after the
calculation of averages has been performed, and with $t_{o}$ going to minus
infinite. Once this choice is introduced in Eq. (\ref{eq22}), in Zubarev's
approach the nonequilibrium statistical operator, designated by $\varrho
_{\varepsilon }\left( t\right) $, after integration by parts in time, can be
written in the form (see Appendix \textbf{II}) 
\begin{equation}
\varrho _{\varepsilon }\left( t\right) =\exp \left\{ -\hat{S}\left(
t,0\right) +\int\limits_{-\infty }^{t}dt^{\prime }\ e^{\varepsilon \left(
t^{\prime }-t\right) }\frac{d}{dt^{\prime }}\hat{S}\left( t^{\prime
},t^{\prime }-t\right) \right\} \ ,  \label{eq27}
\end{equation}
where 
\begin{equation}
\hat{S}\left( t,0\right) =-\ln \bar{\varrho}\left( t,0\right) =\Phi \left(
t\right) \hat{1}+\sum\limits_{j}\int d^{3}r\ F_{j}\left( \mathbf{r},t\right) 
\hat{P}\left( \mathbf{r}\right) \ ,  \label{eq28}
\end{equation}
with $\hat{1}$ being the unit operator, $\bar{\varrho}$ is\ defined in Eq. (%
\ref{eq17}), and 
\begin{equation}
\hat{S}\left( t^{\prime },t^{\prime }-t\right) =\exp \left\{ -\frac{1}{%
i\hslash }\left( t^{\prime }-t\right) \hat{H}\right\} \hat{S}\left(
t^{\prime },0\right) \exp \left\{ \frac{1}{i\hslash }\left( t^{\prime
}-t\right) \hat{H}\right\} .  \label{eq29}
\end{equation}
The operator $\hat{S}\left( t,0\right) $ is designated as the \textit{%
informational-entropy operator}, whose relevance and properties will be
evidenced later on.

Several important points can be stressed in connection with the
nonequilibrium statistical operator of Eq. (\ref{eq27}). First, \textit{the
initial condition }at time $t_{o}\rightarrow -\infty $, is 
\begin{equation}
\varrho _{\varepsilon }\left( t_{o}\right) =\bar{\varrho}\left(
t_{o},0\right) \qquad ,  \label{eq30}
\end{equation}
what implies in a kind of initial \textit{Stosszahlansatz}, in the sense
that the initial state is defined by the instantaneous generalized
canonical-like distribution $\bar{\varrho}$, thus ignoring correlations
among the basic variables prior to time $t_{o}.$ Second, $\varrho
_{\varepsilon }\left( t\right) $ can be separated into two parts, namely
(Refs. [14], [21-24], see also [8] and Appendix \textbf{II}) 
\begin{equation}
\varrho _{\varepsilon }\left( t\right) =\bar{\varrho}\left( t,0\right)
+\varrho _{\varepsilon }^{\prime }\left( t\right) \qquad ,  \label{eq31}
\end{equation}
where $\bar{\varrho}\left( t,0\right) $ is the instantaneous distribution of
Eq. (\ref{eq17}). This, and further results associated to the
MaxEnt-NESOM-based IST and Kinetic Theory described later on, clarify the
role of both $\bar{\varrho}$ and $\varrho _{\varepsilon }^{\prime }$. The
first one [cf. Eqs. (\ref{eq15}) to (\ref{eq17})] defines an instantaneous,
at time $t$, distribution, which describes a ``frozen'' equilibrium
providing at such given time the macroscopic state of the system, and for
that reason is sometimes dubbed as the \textit{quasi-equilibrium statistical
operator.} This distribution describes the macrostate of the system in a
time interval, around $t$, much smaller than the relaxation times of the
basic variables (implying in a ``frozen'' equilibrium or quasi-equilibrium
in such interval). But, of course, for larger time intervals the effect of
the dissipational processes comes into action. The dynamics that has led the
system to that state at time $t$ from \ the initial condition of preparation
at time $t_{o}$ [cf. Eq. (\ref{eq30})], as well as its continuing
dissipative evolution from that state at time $t$ to eventually a final full
equilibrium, is contained in the fundamental contribution $\varrho
_{\varepsilon }^{\prime }\left( t\right) .$ Third, there exists a
time-dependent projection operator $\mathcal{P}\left( t\right) $ with the
property that [23,24] (see Appendix \textbf{III}) 
\begin{equation}
\mathcal{P}\left( t\right) \ln \varrho _{\varepsilon }\left( t\right) =\ln 
\bar{\varrho}\left( t,0\right) \qquad .  \label{eq32}
\end{equation}

This projection procedure, a generalization of those of Zwanzig (apparently
the first to introduce projection techniques in statistical physics [10]),
Mori [9], Zubarev and Kalashnikov [17], and Robertson [15], has interesting
characteristics. We recall that the formalism involves the macroscopic
description of the system in terms of the set of macrovariables $\left\{
Q_{j}\left( \mathbf{r},t\right) \right\} $, which are the average over the
nonequilibrium ensemble of the set of dynamical quantities $\left\{ \hat{P}%
_{j}\left( \mathbf{r}\right) \right\} $. The latter, which, as already
noticed, are quasi-conserved under the dynamics generated by $\hat{H}_{o}$
[cf. Eq. (\ref{eq13})], are called the ``relevant'' variables, and we denote
the subspace they define as the \textit{informational subspace }of the space
of states of the system. The remaining quantities in the dynamical
description of the system, namely, those absent from the informational space
associated to the constraints in MaxEnt [cf; Eqs. (\ref{eq19})], are called
``irrelevant'' variables. The role of the projection operation is to
introduce what can be referred to as a \textit{coarse-graining procedure,}
in the sense that it projects the logarithm of the ``fine-grained''
statistical operator $\varrho _{\varepsilon }\left( t\right) $ onto the
subspace of the ``relevant'' (informational) variables, this projected part
being the \ logarithm of the auxiliary (or quasi-equilibrium, or
``instantaneous frozen'', or ``coarse-grained'') distribution $\bar{\varrho}%
\left( t,0\right) $, and consequently, the procedure eliminates the
``irrelevant'' variables, quite in the spirit of the Bayesian-based approach
and MaxEnt. The ``irrelevant'' variables are ``hidden'' in the contribution $%
\varrho _{\varepsilon }^{\prime }\left( t\right) $ to the full distribution $%
\varrho _{\varepsilon }\left( t\right) $ of Eq. (\ref{eq31}), since it
depends on the last term in the exponential of Eq. (\ref{eq27}), where the
differentiation in time drives $\ln \bar{\varrho}$ outside the subspace of
``relevant'' (informational) variables (see also Appendix \textbf{III)}. We
stress that the projection operation is time dependent, such dependence
corresponding to the fact that the projection $\mathcal{P}\left( t\right) $
is determined by the macroscopic state of the system at the time the
projection is performed. Further considerations of this projection procedure
will appear in the kinetic and thermodynamics theories based on this
informational approach, which are briefly discussed later on. Moreover,
geometrical-topological implications are derived and discussed in detail by
Balian et al. [64].

Two further comments are of relevance. First, for a given dynamical quantity 
$\hat{A}$, its average value in MaxEnt-NESOM, that is, the expected value to
be compared with the experimental measure, is given by 
\begin{equation}
\langle \hat{A}\mid t\rangle =\lim_{\varepsilon \rightarrow +0}\ Tr\left\{ 
\hat{A}\varrho _{\varepsilon }\left( t\right) \right\} =Tr\left\{ \hat{A}%
\bar{\varrho}\left( t,0\right) \right\} +\lim_{\varepsilon \rightarrow +0}\
Tr\left\{ \hat{A}\varrho _{\varepsilon }^{\prime }\left( t\right) \right\} ,
\label{eq33}
\end{equation}
the last equality following after the separation given by Eq. (\ref{eq31})
is introduced. This is the said generalization of Kirkwood time-smoothing
averaging [6], and we can see that \ the average value is composed of two
contributions: one is the average with the quasi-equilibrium distribution
(meaning the contribution of the state at the time $t$), plus the
contribution arising out of the dynamical behavior of the system (the one
that accounts for the past history and future dissipational evolution).
Moreover, this operation introduces in the formalism the so-called \textit{%
Bogoliubov's method of quasi-averages }[49,65]. Bogoliubov's procedure
involves a symmetry-breaking process, which is introduced in order to remove
degeneracies connected with one or several groups of transformations in the
description of the system. According to Eq. (\ref{eq33}) the regular average
with $\varrho _{\varepsilon }\left( t\right) $ is followed by the limit of
cancelling the \textit{ad hoc} symmetry-breaking introduced by the presence
of the weight function $w$ in Eq. (\ref{eq26}) (which is Abel's kernel in
Zubarev approach, cf. Eq. (\ref{eq27}), and follows for $\varepsilon $ going
to $+0$), which imposes a \textit{breaking of the time-reversal symmetry} in
the dynamical description of the system. This is mirrored in the Liouville
equation for $\varrho _{\varepsilon }\left( t\right) $: Zubarev's
nonequilibrium statistical operator \textit{does }satisfy Liouville
equation, but it must be reckoned the fact that the group of its solutions
is composed of two subsets, the one corresponding to the retarded solutions
and the one corresponding to the advanced solutions. The presence of the
weight function $w$ (Abel's kernel in Zubarev's approach) in the \textit{%
time-smoothing or quasi-average procedure} that has been introduced \textit{%
selects the subset of retarded solutions} from the total group of solutions
of Liouville equation. We call the attention (as Zubarev had; see Appendix
in the book of reference [14]) that this has a certain analogy with
Gell-Mann and Goldberger [66] procedure in scattering theory, where these
authors promote a symmetry-breaking in Bogoliubov's sense in Schroedinger
equation, in order to represent the way in which the quantum mechanical
state has been prepared during times $-\infty \leq t^{\prime }\leq t$,
adopting for the wave function a weighted time-smoothing as the one used in
Zubarev's approach to NESOM (see Appendix \textbf{IV)}. More precisely, $%
\varrho _{\varepsilon }\left( t\right) $ satisfies a Liouville equation of a
form that automatically, via Bogoliubov's procedure, selects the retarded
solutions, namely 
\begin{equation}
\frac{\partial }{\partial t}\ln \varrho _{\varepsilon }\left( t\right) +i%
\hat{\Lambda}_{\varepsilon }\left( t\right) \ln \varrho _{\varepsilon
}\left( t\right) =0\qquad ,  \label{eq34}
\end{equation}
where $\hat{\Lambda}_{\varepsilon }$ is the modified Liouville operator 
\begin{equation}
i\hat{\Lambda}_{\varepsilon }\left( t\right) =i\stackrel{\wedge }{\mathcal{L}%
}+\varepsilon \left[ 1-\mathcal{P}\left( t\right) \right] \qquad ,
\label{eq35}
\end{equation}
with $\stackrel{\wedge }{\mathcal{L}}$ being the regular Liouville operator
and $\mathcal{P}\left( t\right) $ the projection operator of Eq. (\ref{eq32}%
). Equation (\ref{eq34}) is of the form proposed by Ilya Prigogine [67],
with $\hat{\Lambda}_{\varepsilon }$ being composed of even and odd parts
under time-reversal. Therefore, the time-smoothing procedure introduces a
kind of \textit{Prigogine's dynamical condition for dissipativity} [67,68].

Using Eq. (\ref{eq32}) we can rewrite Eq. (\ref{eq34}) in the form 
\begin{equation}
\frac{\partial }{\partial t}\ln \varrho _{\varepsilon }\left( t\right) +%
\frac{1}{i\hslash }\left[ \ln \varrho _{\varepsilon }\left( t\right) ,\hat{H}%
\right] =-\varepsilon \left[ \ln \varrho _{\varepsilon }\left( t\right) -\ln 
\bar{\varrho}\left( t,0\right) \right] \ ,  \label{eq36}
\end{equation}
viz., a regular Liouville equation but with an infinitesimal source, which
introduces Bogoliubov's symmetry breaking of time reversal, and is
responsible for disregarding the advanced solutions. Equation (\ref{eq36})
is then said to have Boltzmann-Bogoliubov-Prigogine symmetry. Following
Zubarev [14], Eq. (\ref{eq36}) is interpreted as the logarithm of the
statistical operator evolving freely under Liouville operator $\stackrel{%
\wedge }{\mathcal{L}}$, from an initial condition at time $t_{o},$ and with
the system undergoing random transitions, under the influence of the
interaction with the surroundings. This is described by a Poisson
distribution ($w$ in the form of Abel's kernel), and the result at time $t$
is obtained by averaging over all $t^{\prime }$ in the interval ($t_{o},t$)
[cf. Eq. (\ref{eq22})]. This is the time-smoothing procedure in Kirkwood's
sense [cf. Eq. (\ref{eq33})], and therefore, it is introduced information
related to the past history in the thermo-hydrodynamics macrostate of the
system along its evolution from the inital $t_{o}$ (further considerations
are given in Appendix \textbf{IV).}

Two points need be considered here. One is that the initial $t_{o}$ is
usually taken in the remote past ($t_{o}\rightarrow -\infty $), and the
other that the integration in time in the interval ($t_{o},t$) is weighted
by the kernel $w\left( t,t^{\prime }\right) $ (Abel's kernel in Zubarev's
approach, Fej\'{e}r's kernel in Kirkwood, Green , Mori approaches; and
others are possible). As a consequence the procedure introduces a kind of 
\textit{evanescent history }as the system macrostate evolves toward the
future from the initial condition at time $t_{o}$ ($\rightarrow -\infty $).
Therefore, the contribution $\varrho _{\varepsilon }^{\prime }\left(
t\right) $ to the full statistical operator, that is, the one describing the
dissipative evolution of the state of the system, to be clearly evidenced -
as later described - in the resulting kinetic theory, clearly indicates that
it has been introduced a \textit{fading memory }process. This may be
considered as the statistical-mechanical equivalent of the one proposed in
phenomenological continuum-mechanical-based Rational Thermodynamics [58,69].
In Zubarev's approach this fading process occurs in an adiabatic-like form
towards the remote past: as time evolves memory decays exponentially with
lifetime $\varepsilon ^{-1}$.

We may interpret this considering that as time evolves correlations
established in the past fad away, and only the most recent ones strongly
influence the evolution of the nonequilibrium system; here again is in
action Bogoliubov's principle of correlations weakening. This establishes 
\textit{irreversible behavior} in the system introducing in a peculiar way a
kind of Eddington's \textit{time-arrow:} Colloquially speaking, we may say
that because of its fading memory, the system can only evolve irreversibly
towards the future and cannot ``remember'' how to retrive the mechanical
trajectories that would return it to the past situations (what is attained
when neglecting the advance solutions of Liouville equation). In a sense we
may say that Boltzmann original ideas are here at work in quite general
conditions [70,71], and in its evolution towards the future, once any
external perturbating source is switched off, the system tends to a final
state of equilibrium, described by the distribution of section \textbf{2},
irrespective of the nonequilibrium initial condition of preparation (this is
shown in Refs. [57] and [72] and we simply mention the fact here).

Alvarez-Romero and Garcia-Colin [25] has presented an interesting
alternative approach to the derivation of Zubarev's form of MaxEnt-NESOM,
however, which differs from ours in the interpretation of the time-smoothing
procedure, which they take as implying the connection of an adiabatic
perturbation for $t^{\prime }>t_{o}$ (we think that these authors mean
adiabatic switch on of the interactions in $H^{\prime }$ responsible for the
dissipative processes), instead of implying in a fading-memory
interpretation. We need notice that both are interpretations which we feel
are equally satisfactory and may be equivalent, but we side with the point
of view of irreversible behavior following from - in
Boltzmann-Bogoliubov-Prigogine's sense - adiabatic decorrelation of
processes in the past. This is the fading-memory phenomenon, introduced in
Zubarev's approach as a result of the postulated Poissonian random processes
(on the basis that no real system can be wholly isolated), as already
discussed. This interpretation aside, we agree with the authors in Ref.
[25], in that the method provides adequate convergence properties (ensured
by Abel's kernel in Zubarev' approach) for the equations of evolution of the
system. These properly describe the irreversible processes unfolding in the
media, with an evolution from a specific initial condition of preparation of
the system and, after remotion of all external constraints - except thermal
and particle reservoirs - tending to the final grand-canonical equilibrium
distribution of Section \textbf{2.}

Moreover, the convergence imposed by Abel's kernel in Zubarev's approach
appears as the most appropriate, not only for the practical mathematical
advantages in the calculation it provides, but mostly important, by the
attached physical meaning associated to the proposed adiabatic decoupling of
correlations which surface in the transport equations in the accompanying
MaxEnt-NESOM kinetic theory (see below). In fact, on the one hand this
kinetic theory produces, when restrictions are applied on the general
theory, the expected collision operators (as those derived in other kinetic
theories) introducing, after the time integration in the interval ($t_{o},t$%
) has been done, the expected terms of energy renormalization and energy
conservation in the collision events. Furthermore, as pointed out by Zubarev
[14], Abel's kernel ensures the convergence of the integrals in the
calculation of the transport coefficients, which in some cases show
divergences when, instead, Fej\`{e}r kernel is used (as in Green, Mori, etc.
approaches). The procedure also appears as having certain analogies with the
so-called repeated randomness assumptions [73,74] as discussed by del Rio
and Garcia-Colin [75]. In Appendix \textbf{IV} we present other alternative
derivations, using a treatment akin to the textbook one for obtaining the
ensemble algorithm in equilibrium, also following the ideas proposed by
McLennan [11], and a connection with an earlier proposal by I. Prigogine is
discussed.

A \textbf{fifth basic step }consists in the construction of a \textit{%
MaxEnt-NESOM-based Nonlinear Kinetic Theory,} that is, the transport
(evolution) equations for the basic set of macrovariables that describe the
irreversible evolution of the macrostate of the system. They are, in
principle, straightforwardly derived, consisting in Heisenberg equations of
motion for the corresponding basic dynamical variables (mechanical
observables) or Hamilton equations in the classical case, averaged over the
nonequilibrium ensemble, namely 
\begin{equation}
\frac{\partial }{\partial t}Q_{j}\left( \mathbf{r},t\right) =Tr\left\{ \frac{%
1}{i\hslash }\left[ \hat{P}_{j}\left( \mathbf{r}\right) ,\hat{H}\right] \
\varrho _{\varepsilon }\left( t\right) \right\} \qquad .  \label{eq37}
\end{equation}
Using the separation of the Hamiltonian as given by Eq. (\ref{eq13}), the
separation of the statistical operator as given by Eq. (\ref{eq31}), the
selection rule of Eq. (\ref{eq14}), and the equivalence with an
instantaneous description as given by Eq. (\ref{eq24}), it follows that Eq. (%
\ref{eq37}) can be written in the form [76,77] 
\begin{equation}
\frac{\partial }{\partial t}Q_{j}\left( \mathbf{r},t\right) =J_{j}^{\left(
0\right) }\left( \mathbf{r},t\right) +J_{j}^{\left( 1\right) }\left( \mathbf{%
r},t\right) +\mathcal{J}_{j}\left( \mathbf{r},t\right) \ ,  \label{eq38}
\end{equation}
where 
\begin{mathletters}
\begin{eqnarray}
J_{j}^{\left( 0\right) }\left( \mathbf{r},t\right)  &=&Tr\left\{ \frac{1}{%
i\hslash }\left[ \hat{P}\left( \mathbf{r}\right) ,\hat{H}_{o}\right] \ \bar{%
\varrho}\left( t,0\right) \right\} \qquad ,  \label{eq39a} \\
&&  \nonumber \\
J_{j}^{\left( 1\right) }\left( \mathbf{r},t\right)  &=&Tr\left\{ \frac{1}{%
i\hslash }\left[ \hat{P}\left( \mathbf{r}\right) ,\hat{H}^{\prime }\right] \ 
\bar{\varrho}\left( t,0\right) \right\} \qquad ,  \label{eq39b}
\end{eqnarray}
\end{mathletters}
\begin{equation}
\mathcal{J}_{j}\left( \mathbf{r},t\right) =Tr\left\{ \frac{1}{i\hslash }%
\left[ \hat{P}\left( \mathbf{r}\right) ,\hat{H}^{\prime }\right] \ \varrho
_{\varepsilon }^{\prime }\left( t\right) \right\} \qquad .  \label{eq40}
\end{equation}

As shown elsewhere [23,56,76,77] this Eq. (\ref{eq38}) can be considered as
a far-reaching generalization of Mori's equations [9,47]. It also contains a
large generalization of Boltzmann's transport theory, with the original
Boltzmann equation for the one-particle distribution retrived under
stringent asymptotic limiting conditions; details and discussions are given
in Refs. [24] and [78].

In this Eq. (\ref{eq38}), in most cases of interest the contribution $%
J^{\left( 1\right) }$ is null because of symmetry properties of the
interactions in $\hat{H}^{\prime }$, and the term $J^{\left( 0\right) }$
provides a conserving part consisting in the divergence of the flux of
quantity $Q_{j}\left( \mathbf{r},t\right) $ [55,56,79]. The last term, i.e.
the one of Eq. (\ref{eq40}), is the collision operator responsible for
relaxation processes, which, evidently, cancels if $\hat{H}^{\prime }$ or $%
\varrho _{\varepsilon }^{\prime }$ is null, what clearly indicates that
dissipative phenomena are described by these contributions to the
Hamiltonian in Eq. (\ref{eq13}), and to the statistical operator in Eq. (\ref
{eq31}), respectively. Hence, as already anticipated, dissipation is not
present in the instantaneous quasi-equilibrium operator $\bar{\varrho}\left(
t,0\right) $ of Eq. (\ref{eq17}), but in the nonequilibrium operator
containing the history and time-smoothing characteristic of $\varrho
_{\varepsilon }^{\prime }\left( t\right) $ of Eqs. (\ref{eq27}) and (\ref
{eq31}). We notice that if $\hat{H}^{\prime }$ is null, so is $\varrho
_{\varepsilon }^{\prime }\left( t\right) $, when \ $\hat{H}_{o}$ coincides
with the whole Hamiltonian corresponding to a full equilibrium condition, as
already discussed.

The collision operator of Eq. (\ref{eq40}) requires an, in general, quite
difficult, and practically unmanageable, mathematical handling. But for
practical use, it can be reformulated in the form of an infinite series of
partial collison operators in the form 
\begin{equation}
\mathcal{J}_{j}\left( \mathbf{r},t\right) =\sum\limits_{n=2}^{\infty }\Omega
_{j}^{\left( n\right) }\left( \mathbf{r},t\right) \qquad ,  \label{eq41}
\end{equation}
where quantities $\Omega ^{\left( n\right) }$ for $n=2,3,....$ can be
interpretated as corresponding to describing two-particle, three-particle,
etc., collisional processes. These partial collision operators, and then the
transport equation (\ref{eq38}), are highly nonlinear, with complete details
given in Refs. [76,77].

An interesting limiting case is the Markovian approximation to Eq. (\ref
{eq38}), consisting into retaining in the collision operator of Eq. (\ref
{eq41}) the interaction $\hat{H}^{\prime }$ strictly up to second order
(limit of weak interactions) [80,81] to obtain for a density $Q_{j}\left( 
\mathbf{r},t\right) $ the equation [12,76,77,79] 
\begin{equation}
\frac{\partial }{\partial t}Q_{j}\left( \mathbf{r},t\right) +div\vec{I}%
_{j}\left( \mathbf{r},t\right) =J_{j}^{\left( 2\right) }\left( \mathbf{r}%
,t\right) \qquad ,  \label{eq42}
\end{equation}
where $\vec{I}_{j}$ is the flux of $Q_{j}$, and 
\begin{equation}
J_{j}^{\left( 2\right) }\left( \mathbf{r},t\right) =\int\limits_{-\infty
}^{t}dt^{\prime }e^{\varepsilon \left( t^{\prime }-t\right) }Tr\left\{ \left[
\hat{H}^{\prime }\left( t^{\prime }-t\right) _{0},\left[ \hat{H}^{\prime },%
\hat{P}_{j}\left( \mathbf{r}\right) \right] \right] \ \bar{\varrho}\left(
t,0\right) \right\} ,  \label{eq43}
\end{equation}
once $J_{j}^{\left( 1\right) }$ is taken as null, and subindex nought
indicates mechanical evolution under $\hat{H}_{o}$ alone (interaction
representation). In Appendix \textbf{V} we present a derivation of a
MaxEnt-NESOM generalized Mori-like equation, showing the role of the
projection operator of Eq. (\ref{eq32}) in the kinetic theory.

Finally, the \textbf{sixth basic step} is the construction of the all
important MaxEnt-NESOM response function \ theory for systems arbitrarily
away from equilibrium, to connect theory with observation and measurement in
the experimental procedure. We do not describe it here, what is done
elsewhere [23]. We simply notice that as in the traditional response
function theory around equilibrium [44,45], the response of the system away
from equilibrium to an external probe, is expressed in terms of correlation
functions but defined over the nonequilibrium ensemble. Moreover, also in
analogy with the case of systems in equilibrium it is possible to construct
a double time nonequilibrium thermodynamic Green function formalism [28,82].

In this way, through the realization of the six basic steps we have
described, a nonequilibrium statistical ensemble formalism - the
MaxEnt-NESOM - can be built. The quite relevant case of a generalized
nonequilibrium grand-canonical ensemble is described in Appendix \textbf{VI.}
Applications of the theory are very briefly described in next section, and
further considerations presented in last section.

\section{APPLICATIONS OF THE FORMALISM}

We briefly summarize in this section some applications of the formalism.

\subsection{Theory and Experiment}

Three areas of particular interest where the formalism has full and quite
useful application are those which study ultrafast dissipative processes in
polymers, biological systems, and in highly excited semiconductor systems. A
large amount of very succesful experimental studies of these systems is
available in the scientific literature on the subject, being centered mainly
on measurement of ultrafast transport and optical properties (see for
example [53,83,84]). Before considering these cases, introducing a kind of
historical recording we describe several earlier applications of the
formalism, certainly incomplete and then we apologize to those authors whose
works have been omitted. Moreover, the description of the different
applications is in most cases summarized by writting the Abstract of the
publication, and the articles are listed in the References.

\textbf{Hydrodynamic-like equations }were derived by L. A. Prokovskii in
1968 [85], and later on by Zubarev and V. Morosov [86], and \textbf{Kinetic
equations} in a theory of relaxation phenomena by Prokovskii [87], Zubarev
and A. G. Bashkirov [88-90], and others. A \textit{double-time
nonequilibrium Green function theory} is due to V. P. Kalashnikov [91-93].
Studies of \textbf{nuclear spin diffusion} were carried by L. L. Buishvili
and Zubarev [94], Buishvili [95], G. R. Khutsishvili [96,97], and of \textbf{%
nuclear magnetic resonance} by Buishvili [98,99], Buishvili and M. D.
Zvizdadze [100-105], Buishvili and N. P. Giogadze [106], etc. \textbf{%
Spin-lattice relaxation} was considered by V. G. Grachev [107], Kalashnikov
[108-110], K. Walasek and A. L. Kuzemskii [111,112]. L. L Buishvili and M.
D. Zviadadze [113] discuss the problem of the choice of the independent
internal thermodynamic parameters for the description of nonequilibrium
systems on the basis of Bogoliubov's idea concerning the hierarchy of
relaxation times as well as successive reductions in the description of
nonequilibrium states of the system when it tends to equilibrium. The set of
parameters is shown to be specified, in principle, by the explicit form of
the Hamiltonian of the system. General conclusions were used to analyse a
number of difficulties in the phenomenological theory of magnetic relaxation
in solids. Special attention was paid to a good choice of macroscopic
parameters. The problem concerning the question why in some cases the
application of Bloch equations is incorrect was discussed.

V. P. Kalashnikov and N. V, Kozhevnikov have dealt with a two-component spin
system in [114], where the nonequilibrium statistical operator method is
used to consider the dynamics of a two-component isothermal magnetic system
in an alternating magnetic field. Exact coupled equations of motion were
obtained for the small deviations from equilibrium of the magnetizations of
the subsystems, together with dispersion relations for the spectra of the
normal modes and exact general expressions for the matrix Green's functions,
the dynamic magnetic susceptibility, and the power obsorbed in the external
field. A self-consistent approach was formulated for the calculation of
these quantities in terms of the exact characteristics of the noninteracting
subsystems. More recently M. Yu. Kovalevskii and S. V. Peletminskii [115]
have considered a hydrodynamic theory in magnetic media, where a microscopic
approach to the description of the dynamics of magnets with complete
spontaneous symmetry breaking is proposed. The structure of the source that
breaks the symmetry of the equilibrium Gibbs distribution was established,
and additional thermodynamic parameters (Cartan forms) that characterize the
equilibrium state were introduced. The quasiaverage representation was
generalized to locally equilibrium states, and the thermodynamics of such
systems is constructed. The flux densities of the additive integrals of the
motion were represented in terms of the local-equilibrium thermodynamic
potential. An expression was found for the orthogonal rotation matrix in
terms of an arbritrary statistical operator. A method of reduced description
is formulated, and ``hydrodynamic'' equations of the considered magnets were
obtained.

Now going over to the relevant area of \textbf{semiconductor physics}, first
applications seems to be due to V. P. Kalashnikov, in connection with the
so-called \textbf{`hot' electrons in semiconductors.} His work using
MaxEnt-NESOM was cited as a satisfactory approach by J. Brinkman in an
invited talk in the 21$^{st}$ International Conference on the Physics of
Semiconductors (Stuttgart, Germany, July 1994). In [116] we read in the
Abstract that the nonequilibrium density matrix, obtained from quantum
energy and momentum conservation laws, is used to describe the motion of hot
carriers in semiconductors. In an earlier publication [117] the nuclear
polarization resulting from the interaction of the nuclei with a
nonequilibrium steady-state distribution of hot conduction electrons,
generated by crossed static magnetic and strong electric fields, was
investigated theoretically. The effect of a strong electric field on the
spin-lattice relaxation times of the conduction electrons and the relaxation
time of the nuclei by hyperfine interaction with hot carriers were
considered for various types of semiconductors. Formulae for the
field-enhanced nuclear magnetization, and for the relaxation times of the
electronic and nuclear magnetic moments, were obtained as function of the
current density of the conduction electrons. Later on, in [118] the
many-electron nonequilibrium statistical operator was used to construct the
equations of momentum and energy balance. These describe the kinetics of
spatially homogeneous distributions of current carriers in conducting
crystals in strong crossed electric and magnetic fields. A study was made of
the case of high densities of the conduction electrons when their
nonequilibrium state can be described by the average values of the total
energy, total momentum, and particle number. In [119] the nonequilibrium
statistical operator method was used to derive a system of balance equations
for the momentum density and particle number in the energy space. This
system determines the kinetics of hot electrons in strong crossed electric
and magnetic fields. For quasielastic scattering the equations are
tranformed into differential equations that describe the diffusion of
electrons in the energy space; they are integrated in the lowest
approximation in the ratio of the relaxation frequency of the transverse
momentum to the frequency of the cyclotron motion.

This topic of hot carriers incorporating the idea of Jaynes' Maxent, was
further carried on by H. Sato [120], and by V. Christoph, G. Vojta, and R. R%
\"{o}pke [121], R\"{o}pke and Christoph [122], R\"{o}pke, Christoph, and
Vojta [123], where a direct calculation of the electrical resistivity of an
electron-phonon system was made by means of the force-force correlation
function method. The coupling of the system being considered to a bath is
essential for the correct physical and mathematical formulation of the
theory. It was assumed that the conduction electrons interact with the
longitudinal phonons (only normal processes were taken into account). The
interaction of the longitudinal phonons with the bath was given by a
relaxation term with a wave-vector dependent relaxation time. The
corrections to the electrical resistivity caused by the finite coupling
strength between phonons and bath were derived in lowest order and
discussed; they are connected with the phonon drag.

V. P. Semiozhenko and A. A. Yatsenko considered the kinetic of systems in
strong alternating fields [124], where an equation is obtained for the
density matrix of a system of interaction particles in a strong alternating
external field. For a system of electrons interactiong with impurities and
phonons, the limit corresponding to an arbitrarily strong constant electric
field was obtained. Examples considered were the problem of the
high-frequency conductivity of the electron gas of semiconductors and the
Cooper pairing of electrons in a strong electromagneitic field.

When dealing with transport phenomena is relevant considering the presence
of the electron-phonon interaction in nonequilibrium conditions, with
earlier studies due several authors, and we mention here Auslender and
Kalashnikov [125], who obtained an explicit expression and an integral
equation was derived for the generating functional of a nonequilibrium
electron-phonon system. A closed non-Markov equation of motion was derived
for the quasi-equilibrium generating functional, from which a system of
exact kinetic equations was derived, and also an equation for the
macroscopic displacement of the lattice. In the second order in the
interaction, collision integrals were obtained for the one-particle density
matrices. In the spatially homogeneous case and in the Markov limit
approximate expressions were constructed for the collision integrals that
take into account the nonsharp energy conservation law in collisions and
renormalization of the interaction.

Kalashnikov [126] considered the interaction of carriers with an external
electromagnetic field. In this work a study is made of the structure of the
Hamiltonian for the interaction of conduction electrons with an
electromagnetic field, and semiconducting crystals were considered whose
Hamiltonian contains terms that depend on both the spin and the kinetic
degrees of freedom of the electrons. It was shown that the previously
constructed interaction Hamiltonians for such systems break the gauge
invariance of the equations of motion of the physical quantities. By means
of a time-dependent canonical transformation of the system's Hamiltonian, a
new expression was found for the operator of the interaction with the field,
and this preserves the gauge invariance of the equations of motion.
Canonical transformations of the basis operators in the nonequilibrium
statistical operator method were considered. Expressions that are exact with
respect to the internal interactions of the subsystems were constructed for
the power absorbed by the electrons in the second order in the strengths of
the electric and the magnetic field.

This question was also considered by V. I. Emelyanov in [127] where a \
method is described for deriving kinetic equations for an electron subsystem
with a band spectrum interacting with a photon subsystem. Kinetic equations
with allowance for the polarization were obtained for the diagonal and
nondiagonal (relative to the bands) elements of the density matrix. The
polarization leads to renormalization of the photon frequencies and
interaction of the electrons with one another through the exchange of
virtual photons. The relaxation of the band populations and the polarization
due to an external field were studied in the two-band approximation by means
of the collision integrals obtained in the method.

This system of so-called `hot' carriers in semiconductors belongs to the
area of the \textbf{photoinjected highly excited two-component plasma in
semiconductors (HEPS),} a \ physical phenomenon of large relevance reviewed
in some of its aspects by V. S. Vorobev [128] (see also Proceedings of the
International Conferences on Hot Carriers in Semiconductors [129]). In
Vorobev's work are described the physical phenomena which result in the
appearance of plasma near the surface of a solid heated by laser radiation.
They include: the dynamics of heating and vaporization of the solid surface
as a whole; separate defects of the surface, and aerosol particles; the
hydrodynamics of the expansion of the vaporized material into the gas
surrounding the target; and, the kinetics of ionization in the vapors or in
their mixture with the surrounding gas. Plasma formation is linked, first,
to the necessity of having sufficiently high vapor density in the
interaction zone or to heating of the target up to a definite temperature,
which imposes definite requirements on the energy parameters of the laser
pulse. It is also linked to the possibility of development of ionization in
the vapors or their mixtures with the surrounding gas; this dictates a
definite laser-pulse intensity. Starting from these requirements, the
boundary of the plasma \ region is found in the energy-intensity plane. The
moment at which plasma appears is determined as the point of intersection of
this boundary by the laser pulse in the same plane. Different cases of
plasma formation in diffusion and hydrodynamic regimes of vapor efflux,
associated with heating and vaporization of the target as a whole and its
microdefects or aerosol particles, are described on the basis of this
approach for different materials, pressures, and composition of the gas
surrounding the target, size of the focusing spot, and durations, shapes,
and wavelengths of the laser pulses.

We notice that the statistical characteristics of a regular plasma was
considered by M. V. Tokarchuk [130] in a paper in an issue of Teor. Mat.
Fiz. dedicated in memory of D. N. Zubarev, who\ describes how Zubarev's
nonequilibrium statistical operator method is used to give a statistical
description of a nonequilibrium plasma in its electromagnetic self-field.
Generalized transport equations were obtained for the charged particles and
the oscillators of the electromagnetic field with allowance made for the
local conservation laws. The case of a nonequilibrium plasma was considered.

On the question of optical properties in solid state matter A. Zehe and G. R%
\"{o}pke considered the radiative recombination of hot carriers in
cathodo-excited semiconductors [131], where application of the theory of
nonequilibrium processes was made to study the relaxation and recombination
of high-excited electrons in semiconductors. Following the scheme of a
hot-electron theory based on the method of nonequilibrium statistical
operator developed by Zubarev and, independently, by McLennan, expressions
for relaxation and recombination times of hot carriers has been established,
resulting directly from a microscopical description of the system. As an
example a free-to-bound transition is discussed (transitions between donor
levels and the valence band or between acceptor levels and the conduction
band) in (Ga,Al)As. With this, on the one hand, the concept of hot electrons
is shown to be related to the general theoretical description and, on the
other hand, intensity curves of time-resoved spectra, i.e. the
time-variation of spectra after an excitation pulse, are given. Moreover, R%
\"{o}pke [132] uses Zubarev's nonequilibrium statistical operator method to
obtain balance equations for a system of interacting electrons in the field
of a scattered potential in which both localized and delocalized states are
possible. An explicit expression for the conductivity is obtained by
calculating the correlation function by the method of thermal Green's
functions. In [133] R\"{o}pke and Christoph derive a general linear response
formula for the isothermal electrical resistivity by means of the
nonequilibrium statistical operator formalism of Zubarev-McLennan. The
coupling between system and heat bath is taken into account. R\"{o}pke and
F. E. H\"{o}hne in [134] consider that the quantum transport theory of
electrical conductivity of a model system of electrons interacting with
fixed ions should unify the usual theories of electrical conductivity of
doped semiconductors, liquid metals, and dense ion plasmas. Starting from
the method of nonequilibrium statistical operator, the conductivity is
expressed by correlation functions which can be evaluated using the Green's
function technique. Especially, the influence of electron-electron
scattering on the conductivity is considered. Results are given for
arbitrary degree of degeneration and are compared with results obtained from
the solution of the Boltzmann equation by standard methods.

We may mention at this point, on the question of\textbf{\ quantum transport}
related to hot carriers, a couple of brief reviews \ and comments, one by J.
R. Barker [135] and the other by L. Reggiani [136]. Furthermore, a \textbf{%
higher-order Einstein relation} for nonlinear charge transport in an
approach of the type we are considering, is due to S. A. Hope, G. Feat, and
P. T. Landsberg [137], who consider nonlinear terms in relations for current
densities which are treated macroscopically, semi-microscopically and
microscopically. In the macroscopic treatment, terms in $\phi ^{2}$, $E^{2}$%
, $\left( \nabla n\right) ^{2},$ $\nabla ^{2}n$ and $\vec{E}\mathbf{\cdot }%
\nabla n$ are included, where $\phi $ is the electrostatic potential, $n$ is
the carrier concentration and $\vec{E}$ is the electric field. The power
series expansion of the current density is valid for equilibrium and yields
conductivity-diffusion type Einstein relations. In the semi-microscopic
approach a perturbation theory for the density matrix is used, and Einstein
relations were then derived by equating the average of the current density
operator to zero. In the microscopic approach a Kubo formalism is developed,
based on a local nonequilibrium distribution function due to Mori. This
leads to Einstein relations via correlation functions and Liouville's
equation. A set of such relations which emerge consistently from such a
treatment was given.

We notice the quite interesting phenomenon which is present in the
photoinjected plasma in semiconductors - at very low temperatures and mainly
for indirect gap semiconductors -, consisting in the so-called \textbf{%
Keldysh's condensation,} or formation of \textbf{metallic electron-hole
droplets} (e.g. [138]). The question was treated by some authors within a
scheme akin to MaxEnt-NESOM. We may mention the work of G. Mahler and J. L.
Birman [139], where the gas-liquid transition of an electron-hole plasma is
studied under the influence of a donor electron system of density $n_{D}^{-}.
$ It is found that the density of holes within the drop decreases with
increasing $n_{D}^{-}.$ The width of the electron-hole recombination line
calculated from the joint density of states, is found to go through a
minimum in agreement with experiment. It is further shown that the behavior
of the linewidth reflects the nature of the impurity-induced
semiconductor-metal transition. It is therefore possible to construct the
underlying change of the free-carrier density with doping which was
demonstrated for Si:P. Condensation seems to occur up to high doping levels.
It might also be expected in a number of heavily doped (metallic)
semiconductors, for which the metallic phase is not stable with respect to
excitation formation under normal conditions. This was followed by a theory
of the electron-hole plasma in highly excited Si and Ge [140], where in a
macroscopic-model calculation the electron-hole system was treated as an
interacting free-carrier system in thermal equilibrium with a nonideal
exciton gas. Renormalization of the excitons was approximately taken into
account. The chemical potential as a function of the total electron-hole
density was discussed with respect to possible unstable regions also in the
low-density regime. Making simplifying assumptions, the phase diagram for
the matallic condensation was derived for Ge and Si and compared to
experimental data.

Some additional work is also due to R. N. Silver in a series of papers: In
[141] a theory of electron-hole condensation in germanium and silicon is
developed in which recombination, evaporation, and exciton condensation are
treated as random processes. An expression was derived for the probability
distribution of the number of electron-hole pairs bound to a nucleation
center. Consideration of the effects of lifetime, surface tension, and
nucleation centers leads to: (i) the relation of multiexciton complexes to
electron-hole droplets; (ii) stable droplet sizes which are strong functions
of temperature but only weak functions of pair-generation rate; (iii) the
possibility of determining the surface tension from size measurements; and
(iv) deviations from the usual phase diagram for a liquid-gas transition at
very low tempeatures. Detailed numerical calculations were carried out for
uniform excitation of germanium and compared to data available from laser
excitation experiments. In [142] the dominant relaxation rates in
electron-hole condensation are calculated from the stochastic rate equations
proposed in a previous paper. These govern the time scale for the nucleation
of electron-hole droplets and for fluctuations in the number of
electron-hole pairs bound to a nucleation center. The calculational
procedure makes use of exact recursion relations for the temporal moments of
the probability distribution. It was found that metastable states of
electrons and holes can exist at high temperatures ($T\gtrsim 2K$ in Ge),
for a limited range of exciton densities. At low temperatures the finite
carrier lifetimes lead to measurably short relaxation times. The importance
of the time dependence of the exciton density in experiments at fixed
generation rates was stressed.

Moreover, in [143] is shown that expressions for electron-hole droplet
nucleation and decay currents, derived in earlier papers by Silver and
others, can be applied to problems of time-dependent exciton densities only
in restricted conditions. The induction time, which characterizes the
response of nucleation and decay currents to changes in the exciton density,
must be shorter than the inverse logarithmic time derivative of the
nucleation and decay currents rising high-intensity pulses, and for most
decay phenomena except for extremely slow changes in the laser intensity.
Finally in [144] a hydrodynamic model for the nonequilibrium thermodynamics
of electron-hole droplets in semiconductors was presented. It predicts
droplet properties at densities and temperatures where the assumptions of
nucleation kinetics fail. It is an extension to finite lifetimes of the
Cahn-Hilliard theories of critical droplets and spinodal decomposition. As
in finite-lifetime nucleation kinetics both critical and stable droplets are
found to exist above a minimum supersaturation which must become large at
low temperatures. However, stable droplets differ both quantitatively and
qualitatively from the capillarity approximation commonly assumed in
nucleation kinetics. For example, they may be characterixed by a velocity
profile which peaks in the surface region. Among other novel predictions are
that: (i) a maximum supersaturation before phase separation is given by the
spinodal line; (ii) stable droplets continue to exist at temperatures
approaching $T_{c}$ but their properties are strongly affected by impurity
and phonon scattering; and (iii) at very low temperatures critical droplets
are too small for hysteresis, but stable droplet properties are calculated.
Quantitative predictions are made by the principle of corresponding states.

Let us now proceed with a recollection of more recent applications of
MaxEnt-NESOM in the case of the \textbf{photoinjected two-component plasma
in semiconductors.}

Consider first the case of \textbf{quantum transport.} D. K. Ferry and
collaborators have considered in Zubarev's approach the case of transient
transport in bulk semiconductors and submicron devices. Ferry et al. [145]
call the attention to the fact that essentially all investigations of
hot-electron transport in semiconductors are based on a one-electron
transport equation, usually the Boltzmann transport equation (BTE). Indeed,
the overriding theoretical concern in such high-field transport is primarily
the solution of the transport equation to ascertain the form that the
nonequilibrium distribution function takes in the presence of the electric
field. However, for transport purposes, this is not an end product, since
integrals must be carried out over the distribution function in order to
evaluate the transport coefficients. In applications to semiconductor
devices, however, the full solution of the BTE is usually too complicated to
be determined at each spatial point within the device, and transport
equations for relevant observables, such as energy and momentum, are
preferred. Such transport equations are obtained by taking moments of the
kinetic equation, and these often relate directly to the normal hydrodynamic
semiconductor equations. In general, the complicated nature of the kinetic
equation precludes solving it analytically, and the existence of the various
moment equations is based upon a number of assumptions, the most common of
which is that the distribution function can be represened as a displaced
Maxwellian.

In small semiconductor devices, the time scales are such that the use of the
most common kinetic equation, the BTE, must be questioned. Traditional
semiclassical approaches, such as that of the BTE, assume that the response
of the carriers to any applied force occurs simultaneously with the applied
force, even though the system may undergo subsequent relaxation to a
nonequilibrium steady state. On the short time scale of interest though, a
truly causal theory introduces memory effects that lead to convolution
integrals in the transport coefficients, so that the resultanting kinetic
equation is not of the Markovian type. For the steady state, this results in
a collision operator that depends upon the frequency of the driving field.

The concerns over the detailed form of the moment equations can be removed
by deriving these equations directly from the quantum transport equations.
The exact solutions of the Liouville equations describe the time evolution
of a statistical ensemble at any time interval after an external field is
applied. If the rate of intercarrier scattering is high, then after a short
time interval $\tau _{1}$, smaller than any appropriate time scale of
interest, the evolution of the nonequilibrium density matrix must be
independent of the initial distribution, and there should be a reduction in
the number of parameters necessary to describe the nonequilibrium response
of the system [52]. It is therefore possible to assume a nonequilibrium
statistical operator that is smoothed in its microfluctuations and from the
very beginning describes the slow evolution of the system for time intervals
that are larger than $\tau _{1}$ [21-24,76-78]. By utilizing such an
approach, both the relevant moment equations and the form of the
distribution function itself are obtained directly prior to the extension to
the semiclassical transport properties. When one introduces a retarded
kinetic equation to describe the transport on a short time scale, this
retarded equation is a significant deviation from the usual assumption of a
simultaneous response to driving forces, and is a consequence of extending
the concept of a causal response to the short time scale. Causal behavior is
usually associated with ignoring a large percentage of the individual
dynamic variables. This is covered by the MaxEnt-NESOM nonlinear quantum
kinetic theory described in the previous section.

J.\ L. Birman and collaborators have used MaxEnt-NESOM for the study of 
\textbf{nonlinear (non-Ohmic) transport} by the carrier system in the
photoinjected plasma. In [146] X. L. Ling et al. consider the
balance-equation approach for hot-electron transport systems composed of two
groups of carriers, each of different effective mass. This is the simplest
model for a real band structure of a multivalley semiconductor. The
separation of the center-of-mass (c.m.) motion from the relative motion of
the electrons is incomplete due to the possibility of exchange of particle
number between the two systems and this is taken into account in the
Liouville equation for the density matrix. General expressions for the rates
of change of the center-of-mass momenta, electron system energies, and
particle numbers are obtained. \ These equations in their classical forms
are used for a model calculation for the high-field steady-state transport
in GaAs and the calculated results show reasonably good agreement with
experiments.

This paper was followed by additional publications: Xing et al. in [147]
employ the method of the nonequilibrium statistical operator developed by
Zubarev to investigate steady-state hot-electron transport in a strong
electric field. The momentum and energy balance equations which are
nonlinear in the drift velocity $v_{e}$ are derived to the lowest order in
the electron-impurity and electron-phonon interactions. It is shown that
these balance equations are exactly identical to those obtained by Ling et
al. [146]. However, their equations cannot be reduced to those of
Kalashnikov and Ferry, who previously applied a similar approach to study
warm-electron transport. The origin of this difference was demonstrated in
detail. These results yield better agreement with the experimental data in $p
$-type Ge for hot heavy-hole transport. D. Y. Xing and C. S. Ting [148]
present a Green's-function approach for steady-state hot-electron transport
from the transient region: coupled differential and integral equations are
constructed for the evolution of the drift velocity $v\left( t\right) $ and
the electron temperature $T_{e}\left( t\right) $. The effect of current
fluctuations on these equations was included and discussed qualitatively.
When the evolution equation for $v\left( t\right) $ is linearized, a
Langevin equation is obtained for the drift velocity. For electric fields
with moderate strengths, it is shown that the calculated values for $v\left(
t\right) $ as a function of $t$ with memory and without memory differ very
little. Thus the memory effect is neglegible and the original evolution
equations are reduced to a pair of differential nonlinear equations for $%
v\left( t\right) $ anf $T_{e}\left( t\right) $. Numerical computation has
been carried out for hole transport in $p$-type Ge. Also, scatterings due to
both acoustic and nonpolar-optical phonons were considered. For a step field
at $t=0,$ $v\left( t\right) $ shows the well known ballistic and overshoot
behavior. When a rectangular pulse field is applied, both overshoot and
undershoot phenomena may appear in $v\left( t\right) .$

In [149] M. Liu et al. carry further the study of high-field carrier
transport in many-valley semiconductors. They derive for steady states the
balance equations for momenta, energies, and populations of the hot
electrons in various valleys. Taking $n$-type silicon as an example, it is
calculated the drift velocity, electron temperatures and repopulations of
both cold and hot valleys as functions of electric field ($1-10^{5}$ V/cm) \
at several temperatures between $T=8\ K$ and $T=300\ K$. By applying the
electric field parallel to the $\langle 111\rangle $, $\langle 100\rangle $,
and $\langle 110\rangle $ crystallographic directions, the anisotropic
effect for the drift velocity was investigated. Furthermore, when
considering the transient repopulation effect for a given sample with a
certain length, a negative-differential-mobility region was obtained. It was
shown that these results are not only in excellent agreement with the
results of Monte Carlo method but also quantitatively comparable with
experimental data in all temperature ranges.

Also in this area of nonlinear transport, has also been considered
time-dependent conductivity in HEPS, when under the action of a
high-frequency oscillating field and in the presence of a constant electric
field [150].

\textbf{Ultrafast transient transport} (in the picosecond time scale) has
also been considered within the scope of MaxEnt-NESOM. In [151] a study of
ultrafast transient transport in nonequilibrium direct-gap polar
semiconductors under high levels of excitation is presented. The dynamic
equation for the drift velocity was derived. A numerical application,
appropriate for the case of photoexcited carriers distributed in the
zone-center valleys of GaAs, was done. The time evolution of the momentum
relaxation time and drift velocity is discussed, and it is shown that,
depending on experimental conditions, a velocity overshoot may result. V. N.
Freire et al. [152] present an analytical study of the ultrafast-mobility
transient of central-valley nonequilibrium carriers in a highly photoexcited
plasma in semiconductors. General expressions for the mobility of the
photoinjected carriers were derived. Numerical results were obtained in the
case of low to moderately high fields in GaAs. It was shown that the
mobility transient has a structure (maxima and minima) depending on the
degree of photoexcitation and electric field intensity. Three different
regimes are present, corresponding to (i) structure without overshoot and an
Ohmic steady state, (ii) structure with overshoot and a non-Ohmic steady
state, and (iii) normal evolution and an Ohmic steady state. A brief
discussion of the diffusion coefficient was given.

Transport in HEPS also requires to look for a \textbf{generalization of
Einstein relation,} linking conductivity and diffusivity, for
far-from-equilibrium systems. The diffusion coefficient in a nonlinear
transport theory has been considered by Hope et al. as already noticed
[137]. Vasconcellos et al. 153] have presented a quasi-thermohydrodynamic
theory built in the framework of MaxEnt-NESOM for the study of effects of
diffusion in the photogenerated carrier system in highly photoexcited polar
semiconductors. It was developed a quantum quasi-hydrodynamic description of
the system based on the nonequilibrium statistical operator formalism. It
was presented a generalized Fick's diffusion equation for the charge density
of the carriers, with the ambipolar diffusion coefficient obtained at the
microscopic level and depending on the evolving macroscopic (nonequilbrium
thermodynamic) state of the sample. A detailed numerical calculation for the
case of GaAs was done, obtaining good agreement with experimental data.

These conductivity and diffusion coefficient in HEPS, calculated in
MaxEnt-NESOM, were used to derive a generalized Einstein relation for the
nonequilibrium carriers (electrons and holes) for weak and intermediate
electric field strengths. It was discussed the effect of the irreversible
evolution of the system and of the non-Ohmic behavior on such generalized
Einstein relation [154]. This is further analized in [155] where a
MaxEnt-NESOM-based response theory to thermal perturbations was applied to a
simple model. It was obtained the evolution equation for the particle
density, which becomes of the form of a propagating wave with a damping
dependent on the diffusion coefficient. The latter was calculated at the
microscopic level. For a charged system it is obtained the mobility
coefficient for arbitrarily intense electric fields, resulting in a
generalized Ohm's law for nonlinear charge transport. Using the expressions
for both transport coefficients an Einstein relation in the nonlinear
nonequilibrium thermodynamic state of the system was derived.

We turn now our attention to the quite relevant area of \textbf{optical
properties in HEPS. }The optical properties of semiconductors have been
extensively studied, because of both the interest in the comprehension of
their microscopic properties and their applications in devices. A
considerable amount of information is available on the subject, and
presently the mounting interest in the areas of development of quantum
generators and other devices based on semiconductors at high density of
excitations has made this one of the dominant fields in the area of
semiconductor optics. The hot excitation model has been applied to the study
of these problems [the hot electron model has been amply used in the area of
high field transport in semiconductors since the pioneering work of Fr\"{o}%
hlich(see for example [156])], beginning with the concept of hot excitons at
weak to moderate excitation densities, as reviewed by Permogorov [157], and
going to some present work on solid state plasma in doped or photoinjected
semiconductors at high excitation densities.

Plasmas in semiconductors are quite interesting physical systems, among
other reasons, because of the flexibility in the choice of a number of
parameters such as Fermi energy, plasma frequency, cyclotron frequency,
energy dispersion relation, effective masses, different types of carriers,
etc. [158,159]. Furthermore, the carrier concentration $n$ can be controlled
either by doping or by the excitation intensity of a pumping laser,
producing a variation of $n$ \ which permits the parameter $r_{s}$, the
intercarrier spacing measured in units of the excitonic Bohr radius, run
through the metallic region ($1\lesssim r_{s}\lesssim 5$), which is of
particular interest to the study of the electron gas many-body problem
[160]. A nonequilibrium distribution of carriers and of optical phonons,
with which they interact, can also be obtained in these semiconductor
plasmas. The participation of nonequilibrium distributions of carriers and
optical phonons and many-body effects in this ``hot'' high density gas are
manifested in the behaviour and shape of optical spectra.

It ought to be noticed that such nonequilibrium distributions can be
characterized by well defined Lagrange multipliers that the variational
MaxEnt-NESOM introduces, and whose evolution while the experiment is being
performed, evolution which is a consequence\ of the dissipative effects that
are developing in the system, are completely determined within the scope of
the nonlinear quantum kinetic theory based on MaxEnt-NESOM. It is worth
recalling that the Lagrange multipliers associated with the energy operator
can be interpreted as the reciprocal a nonequilibrium thermodynamic
intensive variable playing the role of a temperature-like quantity, usually
refered to as a \textit{quasitemperature. }Hence, within MaxEnt-NESOM is
given a more rigorous meaning to such concept which was introduced on
phenomenological basis, and used in different contexts, by several authors:
H.\ B. G. Casimir and F. K. du Pre [161] for nuclear spins; C. S.
Wamg-Chang, G. E. Uhlenbeck, and J. de Boer [162] for molecules; L. D.
Landau [163] for plasmas; H. Fr\"{o}hlich [164] for electrons excited in
strong electric fields; V. A. Shklovskii [165] for electrons in
superconductors; J. Shah and R. C. C. Leite [166] for photoexcited carriers;
J. Shah, R. C. C. Leite, and J. F. Scott [167] for photoexcited phonons.

We also call the attention to the fact that HEPS are, at the microscopic
quantum-mechanical level, dealt with in the Random Phase Approximation (RPA,
sometimes referred to as the generalized time-dependent Hartree-Fock
approach); on this we can highlight the books by D. Pines [160], Pines and
Nozieres [168], Hedin and Lindquist [169], among others.

On the question of the optical properties of HEPS, we begin recalling that
in [52] is provided an illustrative discussion of the role of \textbf{%
Bogoliubov hierarchy of relaxation times in HEPS}, that is, it is considered
the question of the contraction of the macroscipic nonequilibrium
thermodynamic description of dissipative dynamic systems. As argumented by
Bogoliubov, this process is associated to the determination of a spectrum
(hierarchy) of relaxation times. Studies of the irreversible evolution of
highly photoexcited plasmas in polar semiconductors to provide a way to
exemplify and test the existence of such a spectrum of characteristic times
were presented. It was shown that, in fact, several kinetic stages can be
characterized, each accompanied by a successive contraction in the
description of the macroscopic state of the system. A proper choice,
depending on the experimental conditions, leads to good agreement with
observational data.

Moreover, in [170] is presented a description of the time-dependent
thermodynamic properties of the carrier system in HEPS, but resorting to a
quite simplified model for better visualization of the illustration: In that
paper is described the nonequilibrium thermodynamics of a model for
semiconductors under high levels of excitation using the nonequilibrim
statistical operator method. The thermodynamic functions in terms of
thermodynamic variables that are accesible to experimental measurements via
ultrafast laser spectroscopy were obtained. Calculations of entropy
production and the rate of entropy production were performed, kinetic
equations for the relaxation processes derived and Onsager-like coefficients
defined. It was shown that for the system what is considered Prigogine's
theorem of minimum entropy production holds even in the non-linear regime,
and the that Glansdorff-Prigogine thermodynamic (sometimes referred to as
universal) evolution criterion is verified.

In \textbf{optical spectroscopy} we need to consider either time-integrated
measurements, with collection of data in the, say, nanosecond scale, and
time-resolved measurements, \ with collection of data in the pico- to
femto-second scale. We begin with a description of application of
MaxEnt-NESOM to the study of data in \textbf{time-integrated experiments.}
We summarize some earlier work which was based on a simplified application
of MaxEnt-NESOM consisting in assuming an averaged - over the
time-resolution of the detection apparatus - of the nonequilibrium
statistical operator. That is, the details of the evolution of the system in
such interval of time are not evaluated, paying the price of leaving as open
parameters the averaged, over such time interval, nonequilibrium
thermodynamic variables (the Lagrange multipliers in MaxEnt-NESOM).

In [171] is investigated the effect of plasmon-phonon coupling on the
spectrum of laser light scattered by doped semiconductors under
high-excitation conditions. A model consisting of the plasmon and LO-phonon
subsystems in conditions of quasiequilibrium, characterized by uniform
internal temperatures, was used. The scattering cross section was derived
through an appropriate generalization of the fluctuation-dissipation theorem
and the use of Bogoliubov's Green's-function formalism. The Coulomb
interaction between electrons was treated within the random-phase
approximation and the Fr\"{o}hlich Hamiltonian was used to describe the
electron-phonon interaction. The phonon lifetime and plasma lifetime, other
than Landau damping, were introduced in a phenomenological way. Numerical
calculations for the case of a nondegenerate conduction-electron gas in GaAs
were presented. The analysis was completed with a discussion of resonant
Raman scattering when ``hot'' polarons are considered for the intermediate
states of the scattering amplitude.

The dependence of photoluminescence on hot excitations (in carrier and
phonon systems) is considered in [172], where the high energy tail of the
photoluminescence spectra of CdS in conditions of high photoexcitation
intensities is studied. It was shown that the increasing excitation is a
result of the simultaneous effect of nonequilibrium distribution of
electrons and LO-phonons. This has also been further considered by E. A.
Meneses et al. in [173], where experimental results concerning the
photoluminescence spectrum of CdS under high excitation intensities were
considered. With increasing excitation intensity the single emission band
observed at $77K$ shows a peak displacement towards lower energies, an
enhancement of its low-energy side as well as of its high-energy side. These
effects increase with excitation intensity. It was described a complete
model that can account for the above behavior and the emission band shape.
The broadening is mainly due to nonequilibrium distributions of LO phonons,
though contributions to the high-energy side from hot electrons is quite
significant. Self-energy corrections account for the energy-gap reduction
and consequently for the shift of the luminescence band peak towards lower
energies.

The above and other resuts are described in [174], where progress (up to the
end of the decade of the seventies) in the investigation of optical
responses from highly photo-excited semiconductor plasma is reviewed. The
theoretical interpretation of the spectroscopy data is outlined on the basis
of the coupling of the usual scattering theory with Zubarev's nonequilibrium
statistical operator method. Connection with appropriate extension to the
nonequilibrium state of the thermodynamic double-time Green functions was
also presented. \textbf{Magneto-optical effects }in HEPS have been also
considered in the above mentioned simplified form of MaxEnt-NESOM and we can
mention inelastic scattering of light from a hot magnetoplasma in direct-gap
polar semiconductors [175], analized for an arbitrary experimental geometry;
the geometry-dependent mixing between the hybrid-plasma first cyclotron mode
and the first Berstein mode was studied. Also, field-dependent alternative
effects are discussed. Moreover, in [176] is considered the inelastic
scattering of light by single-particle spin-flip electron excitations and
paramagnetic spin-wave in semiconductor magneto-plasmas. The generalized
Landau quasiparticle picture was used to calculate the Raman cross section.
Particular attention is given to $n$-doped GaAs for carrier concentrations
in the metallic region ($r_{s}=1$ to $6$). Scattering of the carriers with
lattice vibrations was considered to discuss the spin-diffusion coefficient,
in uniform quasi-equilibrium \ conditions, as a function of the magnetic
field and the electron effective temperature. Spin-flip magneto-Raman
investigations in doped semiconductors can produce valuable information on
the ``hot metallic electron liquid'' and its interactions.

S. Frota-Pessoa and R. Luzzi [177] have considered resonant Raman scattering
of longitudinal optical phonons in HEPS in magnetic fields. The influence of
the electron-phonon interaction on the Raman line was studied. It was shown
that significant effects appear when, for certain experimental geometries, a
harmonic of the cyclotron frequency equals the LO-phonon frequency. The
frequency shift and line shape were discussed and application to GaAs made.

Another kind of resonant magneto-Raman scattering was considered by A. J.
Sampaio and R. Luzzi [178], that is, direct gap, small effective mass, $n$%
-doped semiconductors were considered in conditions in which the laser and
the scattered frequencies are of the order and larger than the harmonics of
the cyclotron frequency. In these resonant and near resonant conditions the
lifetime of the intermediate states gives important contribution to the
scattered intensity. A calculation was presented of the scattering
cross-section for the magneto-Raman scattering by LO phonons which includes
the electron-phonon interaction contribution to the electron lifetimes.
Numerical study in the case of InSb was done.

Let us now consider the case of \textbf{ultrafast laser spectroscopy }%
[53,83,84,179]$.$ Inelastic scattering of light by `hot' carriers is
considered in [180], where a theoretical study of time-resolved Raman
scattering by a highly photoexcited semiconductor plasma was described. The
scattering cross-section was calculated with numerical results appropriate
for GaAs presented as an illustration. These results were derived resorting
to a MaxEnt-NESOM-based formalism to determine the response function of a
sample in conditions far from thermal equilibrium [181]. The scattering
cross section was expressed in terms of double-time correlation functions,
which were related to appropriate nonequilibrium thermodynamic Green's
functions.

These results were extended to include the phenomenon of hybridization of
plasma waves and longitudinal optical phonon [182]. In this work the
ultra-fast time-resoved Raman spectra of a highly photoexcited semiconductor
plasma is calculated. Information is provided on the time-evolution of
irreversible relaxation processes that develop in the system. It was
described how plasmon-phonon interference effects in the Raman spectra are
affected in the far-from-equilibrium state of the system. A numerical
analysis appropriate for the case of GaAs samples was presented.

As already noticed ultrafast time-resolved Raman scattering, luminescence,
reflectivity, etc., provide a notable amount of information on the
dissipative relaxation processes in matter [53,83,84,179]. We do not go in
length on a number of results presently available, simply mentioning the
review article [183] to be followed by an up-to-date addendum [184]. Also
general considerations are given in [185], where is presented a summarized
discussion of the question of the very rapid relaxation processes that
follow in the photoinjected plasma in semiconductors, phenomena that can be
experimantally studied via ultrafast laser spectroscopy.

Moreover, in [186] is discussed the ultrafast kinetics of evolution of
optical phonons in a photoinjected highly excited plasma in semiconductors.
The state of the nonequilibrium (``hot'') phonon system is described in
terms of the concept of a nonequilibrium temperature, referred to as \textit{%
quasitemperature}, per mode, which can be experimentally characterized and
measured. The phonon emission time shows that optical phonons are
preferentially produced, well in excess of equilibrium, in a reduced
off-center region of the Brillouin zone. The phonons in this region are
responsible for the phenomenon referred to as ``hot-phonon temperature
overshoot''. Most of the phonons, namely, those outside such region, are
only weakly to moderately excited, and mutual thermalization of the
nonequilibrium carriers and optical phonons follows, typically, in the
tenfold picosecond scale. All these results are influenced by the
experimental conditions, which were discussed on the basis of calculations
specialized for GaAs. Comparison with experimental data was presented,
showing good \ agreement.

Recently an interesting phenomenon was observed in experiments measuring
time-resolved changes in the reflectivity of GaAs and others materials, in
which a distinct oscillation of the signal in real time was observed [187].
Such phenomenon was attributed to the generation of coherent lattice
vibrations, and some theoretical approaches were presented. The question is
reconsidered in [188] on the basis of a description in the framework of
MaxEnt-NESOM, where it is shown that the phenomenon originates in nonlinear
coupling of plasma waves and coherent lattice vibrations, in the
nonequilibrium conditions created in the pump-probe experiment. The plasma
waves are strongly excited through coupling with the coherent photons in the
laser-electromagnetic field, and become the source for the creation of the
coherent phonons.

Besides the above mentioned pump-probe experiments, which allow to follow
the ultrafast development of dissipative processes in the medium, we can
also mention the case of continuous excitation leading to \textbf{%
nonequilibrum dissipative-steady states.} In the same vein used to describe
the topics above, we summarize several results in this area which have been
treated in MaxEnt-NESOM.

S. A. Hassan et al. [189] have developed an analysis of the nonequilibrium
thermodynamics of a dissipative system, dealt with at the
mechano-statistical level provided by informational-statistical
thermodynamics. In particular the case of a highly excited photoinjected
plasma in a polar semiconductor was considered. Under continuous constant
illumination, a steady state sets in, whose macroscopic state was
characterized, and the main mechanisms for pumping and dissipation that are
involved were discussed. It is shown that such evolution displays a
phenomenon of quasitemperature overshoot.

The influence of nonlinear contributions in the kinetic equations that
govern the evolution of the macrostate in HEPS under continuous laser
illumination, which appear as one considers higher order collision processes
in the nonlinear kinetic theory in MaxEnt-NESOM, are considered in [190]. It
was analized how dissipative nonlinear processes influence the
photoproduction of optical phonons in $n$-doped polar semiconductors. It was
demonstrated that such nonlinear effects may lead to interesting and
remarkable results. First it was shown that, through free-carrier absorption
processes, LO phonons in excess of equilibrium are produced in a small
off-center region of reciprocal space, second, that nonlinear relaxation
effects are unimportant at weak to intermediate laser intensities but come
to play a role in the range of strong excitations. On the one hand, they
reduce the amplification of the preferentially pumped modes, and at very
strong levels of photoexcitation the nonlinear contributions are responsible
for a channeling of energy from the pumped modes to the modes with lowest
frequency which grow enormously: it follows a phenomenon akin to a
nonequilibrium Bose-Einstein condensation. Even though this effect is hardly
accessible experimentally in real semiconductors under laser-light
excitation, it may follow in other system, like biological polymers.

The possible emergence of complex behavior in HEPS under constant excitation
is considered in [191] where the nonequilibrium steady state of a direct gap
semiconductor is studied under high levels of photoexcitation by continuous
laser light. The stability of the uniform steady state of itinerant carriers
is probed resorting to linear normal mode analysis of the nonlinear
equations of evolution for the carrier charge density. Such analysis leads
to the determination of the wavevector dependent electronic contribution to
the dielectric function. Examination of its behavior allows us to show that
in the extremely degenerate regime the carrier system becomes nonmetallic,
and displays a coexistence of metallic and nonmetallic phases on leaving
that regime: itinerant carriers move in the background of an extended state
of bounded electron and hole charge densities. This introduces a new view of
Mott transition in photoinjected semiconductors. This complex behavior is a
result of collective together with dissipative effects in the far from
equilibrium carrier system governed by nonlinear dynamic laws.

The elementary excitations in the carrier's system in HEPS has been
determined [192] resorting to MaxEnt-NESOM. It is calculated of the Raman
spectrum of the double (electrons and holes) photoinjected plasma in
direct-gap polar semiconductors. It allows to identify four types of
elementary excitations in the photoinjected carrier system. Besides the
expected two Raman bands corresponding to single quasi-particle excitations
and a higher frequency plasma wave (labelled optical), there appears a
low-frequency band. It shows linear energy dispersion relation and is
labelled an acoustic plasma oscillation. While the optical plasma branch is
ascribed to the out of phase collective movement of both types of carriers
interacting through the bare Coulomb interaction, the acoustic branch is
ascribed to the in-phase oscillation of each type of carriers interacting
through the screened part of Coulomb interaction. This type of excitation
has been experimentally evidenced.

The interesting possibility of spatial ordering - formation of a
steady-state charge density wave - in the carrier's system in HEPS is
considered in [193]. The case of $n$-doped direct-gap polar semiconductors
under continuous laser illumination was studied. It is shown that the
solution consisting of the homogeneous population of electrons and holes,
that develops from the equilibrium state with increasing values of the laser
power flux, presents a bifurcation point. A new solution emerges from it
corresponding to the formation of a steady-state spatially organized
structure in the carrier charge density in the form of a polarization wave.
This dissipative structure occurs beyond a critical distance from
equilibrium, when a fluctuation drives the carrier system to such
self-organized macroscopically ordered state. It is stabilized by the joint
action of collective and dissipative effects.

The physics of low-dimensional semiconductors is acquiring continuous
relevance (see for example [194]). On the basis of MaxEnt-NESOM, Hassan et
al. [195] describe the production and the optical properties of the
nonequilibrium photoinjected plasma in semiconductor quantum wires under
continuous UV-light illumination. The wavenumber-dependent dynamic
dielectric function of this system was derived and the Raman scattering
cross-section calculated. From the latter it is identified the contributions
from different types of elementary excitations. They are composed of,
besides the single-particle excitations, two types of collective
oscillations: An upper one, consisting of an intrasubband-like plasmon and a
lower one, identified as an acoustic-like plasmon. The dependence of both on
the nonequilibrium (dissipative) macroscopic state of the system is
evidenced and discussed.

The case of steady states far from equilibrium also encompasses an area of
particular interest, namely systems that may display the so-called \textbf{%
complex behavior}, which can arise in systems whose evolution is governed by
nonlinear kinetic laws (see for example [196]). A possible role of
Predictive Statistical Mechanics in the realm of Complexity Theory is
considered in [196,197], where some aspects of the physics of dissipative
phenomena are analized. Attention is called to its connection with the
emerging theory of complexity, and its place in the realm of nonlinear
physics. This nonlinearity has a fundamental role in determining complex
behavior in open systems far away from equilibrium. In many cases it shall
lead to the formation of \textit{self-organized synergetic behavior} at the
macroscopic level, in the form of the so called Prigogine's dissipative
structures. Dissipation is then not a source of decay but has a constructive
role, maybe including the emergence of life, natural evolution, and the
astounding functioning of living systems. This is the question of order out
of chaos in complex system, i.e. synergetic processes leading to
self-organization in open systems far-from-equilibrium, when, as noticed,
are governed by nonlinear kinetic laws.

Such kind of complex behavior can arise in solid state matter, and we
briefly summarized a few examples in semiconductors, polymers, and
biological system. In [198] is given an overview of the role of the
irreversible thermodynamics based on MaxEnt-NESOM in dealing with
biosystems. On the question of the relation of Biology and Physics we may
say that the sometimes mentioned gap between both disciplines has been
slowly shrinking along recent decades. The old difficulty with this sought
after connection, which resided into looking exclusively at the microscopic
mechanistic level of Physics, becomes to disappears when one begins to
consider the macroscopic level of Physics. An important role is played by
the thermodynamics of nonlinear nonequilibrium open dissipative systems. In
that paper it was succinctly described a statistical approach to
irreversible thermodynamics of nonlinear nonequilibrium dissipative systems.
MaxEnt-NESOM is applied in the determination of the complex behavior that
may result in biopolymers, namely, Fr\"{o}hlich-Bose-Einstein-like
condensation and propagation of Davidov's solitary-wave-like excitations,
which are of relevance in Bioenergetics. The question of the lifetime of the
latter at physiological conditions was discussed, and the case was
illustrated in a comparison with experiments performed in the case of an
organic molecular polymer.

The interesting phenomenon of the so-called Fr\"{o}hlich-Bose-Einstein-like
condensation - as viewed from the MaxEnt-NESOM angle - has been considered
in several articles. Mesquita et al. [199] have considered the
nonequilibrium and dissipative evolution, and the steady state of the
population of vibrational polar modes in a chain of biomolecules. These
polar modes are excited through a coupling with a pumping source metabolic
energy and are in anharmonic interaction with an elastic continuum. Groups
of polar modes are coupled in this way through nonlinear terms in the
kinetic equations. This nonlinearity is shown to be the source of an
unexpected phenomenon characterizing complex behavior in this kind of
system: after a threshold of intensity of the pumping source is achieved,
polar modes with the lowest frequencies increase enormously their population
in a way reminiscent of a \textit{Bose-Einstein condensation (Fr\"{o}hlich
effect)}. The transient time for the steady-state condensate to follow is
very short (picosecond time scale) and the condensation appears even for
weak values of the anharmonic coupling strength responsible for its
occurrence. Further, it seemingly requires accessible levels of pumping
power of metabolic energy in order to be produced and sustained.

The question was pursued in [200-202], where the above mentioned \textit{Fr%
\"{o}hlich effect} - consisting, as noticed, of a coherent behavior of
boson-like excitations in biological and molecular polymers - is fully
derived and analyzed in terms of MaxEnt-NESOM. It was shown that when double
(or multiple) processes of excitation of the boson system are possible there
follows a \textit{positive -feedback phenomenon} that greatly favors and
enhances the effect.

The transient regime in Fr\"{o}hlich condensation phenomenon was considered
by M. A. Tenan et al. [203]. It was analyzed the dynamics of (i) the
transient stage before the establishment of the steady state and (ii) the
relaxation to the final thermodynamic equilibrium state after the pumping
source is turned off.

An alternative possibility for Fr\"{o}hlich condensation to follow due to
higher order relaxation processes mediated by carriers - as for example in
proteins -, was considered by L. Lauck et al. [204], who take into account a
model of a biological system, for example, a long chain of proteins
possessing polar modes of vibration and where energy is pumped through
metabolic processes. It was considered the effect produced by free electrons
that are usually present as hole carriers in proteins with electron-donor
molecules. MaxEnt-NESOM was used for the derivation of the kinetic equations
and introducing non-linearities due to interactions of the polar vibrations
with the carriers and with the thermal bath. These non-linearities arising
out of high order relaxation processes lead to the emergence of the Fr\"{o}%
hlich effect in the polar modes, i.e., as noticed, the occurence of a
peculiar (nonequilibrium) Bose-Einstein-like condensation. It also points to
an instability of the system that seems to be followed by a morphological
transformation in the form of a spatially ordered dissipative structure.

Another phenomenos is closely related to this phenomenon of Fr\"{o}%
hlich-Bose-Einstein condensation in a nonequilibrium phase, in the sense
that it is a result of the same many-body system microdynamics being
governed by the same Hamiltonian operator, and nonlinear kinetic effects,
which consists in the propagation of solitary waves in several systems.
Solitary waves are ubiquitous and of relevance in a number of situations in
applied science [205]. They play a fundamental role in the case of
conducting polymers (for use in electric car batteries, microcircuits, etc.)
[206-209]. In biological systems, the so-called Schroedinger Davydov's
solitary waves [210] may have a relevant role; they have been considered
within MaxEnt-NESOM in [211], where it was analized Davydov's biolophysical
model in the context of a nonequilibrium statistical thermodynamics. It is
shown that excitations of the Davydov-soliton type that can propagate in the
system, which are strongly damped in near-equilibrium conditions, become
near dissipation-free in the Fr\"{o}hlich-Bose-Einstein-like condensate and
that this occurs after a certain threshold of pumped metabolic energy is
reached. This implies in the propagation of excitations at long distances in
such biosystems. The question have been further analyzed and extended in
[212], and in [213] is considered the case of the organic polymer
acetanilide, which mimics some aspects of certain biopolymers and has been
subjected to extensive experimental analysis. In that letter it is presented
an analysis of \ the behavior of the macroscopic thermodynamic state of
polymers, however, as noticed, centering on acetanilide. The nonlinear
equations of evolution for the populations and the statistically averaged
field amplitudes of polar modes of vibration are derived. The existence of
excitations of the solitary wave type was evidenced. The infrared spectrum
was calculated and compared with the experimental data of Careri et al.
[214], resulting in a good agreement. It was also considered the situation
of a nonthermally highly excited sample, predicting the occurence of a large
increase in the lifetime of the solitary wave excitation, as we have already
noticed above.

The case of acetanilide and biopolymers involve the study of the polar
vibrations resulting of the so-called CO-stretching, corresponding to
frequencies in the infrared. But it is also of relevance to consider the
case of acoustical vibrations (propagation of sound): Mesquita et al. [215]
have shown that accoustic vibronic excitations propagating in highly excited
matter present, as a result of nonlinear interactions, a large increase in
the populations of the normal modes lowest in frequency, the above mentioned
Fr\"{o}hlich effect. This phenomenon makes the solitary waves which can be
produced in the medium to have associated three relevant phenomena when in
sufficiently away-from-equilibrium conditions:(1) A large increase in the
populations of the normal modes lowest in frequency, (2) a large increase of
the solitary wave lifetime, and (3) emergence of a Cherenkov-like effect,
consisting in a large emission of phonons in privileged directions, when the
velocity of propagation of the soliton is larger than the group velocity of
the normal vibrations. Comparison with experiments was presented, which
points out to the corroboration of the theory. The eventual relevance of the
phenomenon in medical imaging is considered in [216], in conection with some
recent research in ultrasonography evidencing the propagation of a peculiar
kind of excitation in water [217]. Such excitation, dubbed a X-wave, has
characteristics resembling that of a solitary-wave type. Considering its
possible relevance for improving ultrasound medical imaging, the probem in a
medium consisting of a biological material of the like of $\alpha $-helix
proteins was reconsidered. It can be shown that in this case is expected an
excitation of the Davidov's solitary wave type, however strongly damped in
normal conditions. The case of acoustic (sound) vibrations was considered,
where, as in the case of polar vibrations, a damped Davidov-like solitary
wave may be excited, which when traveling in conditions sufficiently away
from equilibrium, has its lifetime largely enhanced. Moreover, a soliton
moving in bulk with a velocity larger than that of the group velocity of the
normal vibrational wave would produce a Cherenkov-like emission of phonons
giving rise to the observed X-wave-like pattern.

We close this subsection mentioning some applications of MaxEnt-NESOM to the
study of complex behavior - of the type of formation of Prigogine's
dissipative structures - in far-from-equilibrium many-body systems.
Morphological ordering of the reaction-diffusion Turing type is considered
in [218], where is analysed the dissipative thermodynamic regime of a
electron system in bulk matter under the action of an external source of
energy, which generate electron-hole pairs with a nonequilibrium
distribution in energy space. It was shown that with increasing values of
the source power (furthering the distance from equilibrium), and strictly in
the case of a $p$-doped material, the carrier system displays complex
behavior characterized by undergoing a succession of transitions between
synergetically self-organized dissipative structures. The sequence goes from
the homogeneous steady state (or stochastic thermal chaos), to sinusoidal
spatial deviations (morphological ordering), to an intricate ordered states
(subharmonic bifurcations), and finally to deterministic
turbulent-like-chaos (large amount of nonlinear periodic spatial
organization of the Landau-Prigogine's type). The phenomenon may arise, for
example, in semiconductor systems, molecular polymers, and protein molecular
chains in biosystems.

Moreover, in [219] is considered the nonlinear kinetics of evolution of the
carrier system in the photoinjected plasma in semiconductors under the
action of constant illumination with ultraviolet light. It was shown that
the spatially homogeneous steady-state becomes unstable, and a charge
density wave emerges after a critical intensity of the incident radiation is
achieved. For intensities beyond this critical threshold an increasing
number of modes provide further contributions (subharmonics) to the space
inhomogeneity, leading the system to display chaotic-like behavior, as it is
in [211]. It was shown that this phenomenon can only follow in doped $p$%
-type materials as semiconductors and some molecular and biological
polymers, the latter when under dark biochemical excitation.

\subsection{Informational Statistical Thermodynamics}

Meixner has stated that Thermodynamics of Irreversible Processes has many
faces [69]. This has been reinforced by Tisza [220], who catalogued its main
four approaches, one of them being the mechanical-statistical-based Gibbs'
Thermodynamics. In Tisza's words the latter is considered to be by itself
richer, and to constitute a point of departure of a whole away of
generalizations. Evidently, as Gibbs'ensemble algorithm in equilibrium gives
microscopic (mechanical-statistical) bases for Thermostatics, the
nonequilibrium statistical ensemble formalism MaxEnt-NESOM so does for the
Thermodynamics of Irreversible Processes. This is the so-called \textit{%
Informational Statistical Thermodynamics }(IST), sometimes also referred-to
as Information-theoretic Thermodynamics. IST originated with the pioneering
work of Hobson [221], after publication of Jaynes' seminal papers on
Statistical Mechanics founded on Information Theory [39] (a brief and
partial reviews are given in Refs. [60,61]; see also Ref. [222]; a more
detailed account in Refs. [60-62] and in Refs. [223,224]).

IST is the statistical thermodynamics in which the space of states is
defined by the set of macrovariables $\left\{ Q_{j}\left( \mathbf{r}%
,t\right) \right\} $ of Eq. (\ref{eq19}) [also (\ref{eq21b}) and (\ref{eq24}%
)]. In Thermostatics a fundamental state function is the entropy, defined
over the Gibbs space consisting of the energy, particle numbers (or molar
fractions) of the chemical species that are present, and the volume, which
is\ given at the mechanical-statistical level by the Gibbs entropy of Eq. (%
\ref{eq3}) (in units of Boltzmann constant). But Gibbs entropy, as given by
Eq. (\ref{eq20}), the straight forward generalization of the case in
equilibrium, cannot represent an appropriate entropy-like function for
dissipative systems since it is a constant of motion, that is $dS_{G}\left(
t\right) /dt=0.$ This is a consequence that $\varrho _{\varepsilon }\left(
t\right) $ satisfies Liouville equation (we recall that the inhomogenous
term on the right of Eq. (\ref{eq36}) does not introduce any spurious
external mechanical perturbation, but simply selects the retarded solutions,
and goes to zero - in the spirit of Boguliobov's quasi-averages - after the
calculation of averages has been performed), what can be considered as a
manifestation of the fact that $S_{G}$ is a fine-grained entropy in the
sense that it preserves information. This is the \textit{information
provided at the initial time of preparation of the system }[as given by the
initial condition of Eq. (\ref{eq30})], which defines, via Eq. (\ref{eq24})
at time $t_{o}$, the initial point for trajectories in Gibbs (nonequilibrium
thermodynamic) space of states $\left\{ Q_{j}\left( \mathbf{r},t_{o}\right)
\right\} $. The IST-entropy that is introduced needs be a coarse-grained one
which is associated, at each time $t\geq t_{o},$ to the information provided
by the constraints of Eq. (\ref{eq25}). This is accomplished by the use of
the projection operator of Eq. (\ref{eq32}), that is, we introduce 
\begin{eqnarray}
\bar{S}\left( t\right)  &=&-Tr\left\{ \varrho _{\varepsilon }\left( t\right) 
\mathcal{P}\left( t\right) \ln \varrho _{\varepsilon }\left( t\right)
\right\} =-Tr\left\{ \varrho _{\varepsilon }\left( t\right) \ln \bar{\varrho}%
\left( t,0\right) \right\} =  \nonumber \\
&&  \nonumber \\
&=&-Tr\left\{ \varrho _{\varepsilon }\left( t\right) \hat{S}\left(
t,0\right) \right\} =\Phi \left( t\right) +\sum\limits_{j}\int
d^{3}rF_{j}\left( \mathbf{r},t\right) Q_{j}\left( \mathbf{r},t\right) \ ,
\label{eq44}
\end{eqnarray}
where we have used $\bar{\varrho}$ as given by Eq. (\ref{eq17}), and $\hat{S}
$ as given by Eq. (\ref{eq29}). This last quantity, namely $\hat{S}$, is
called the informational-entropy operator, whose properties are discussed
in\ Ref. [225], and its connection with a Rosenfeld-Prigogine's
complementary principle between micro- and macro-descriptions [67,226], in
Bohr's sense [227], is considered elsewhere [228].

Consequently, the difference between IST-entropy $\bar{S}\left( t\right) $
and Gibbs entropy $S_{G}\left( t\right) $ is a kind of measurement of the
information lost when the macroscopic state of the system is described at
any time $t$ by the reduced set of basic macroscopic variables (the
``relevant'' or informational ones), that is, in terms of what we have
called the informational subspace of the full space of states. The point is
illustrated in Figure \textbf{1}: Starting from the inital condition of Eq. (%
\ref{eq30}) (which defines the point $\left\{ Q_{j}\left( \mathbf{r}%
,t_{o}\right) \right\} $ in the nonequilibrium-thermodynamic space) the
fine-grained informational-statistical Gibbs entropy remains constant in the
full space of states (the union of the informational and supplementary
subspaces), while the IST-entropy is its projection over the informational
subspace; see also the topological-geometrical approach of Balian et al.
[64].

Without going into further details (given in the Ref. [62]), we summarize
the main properties of the IST-entropy. First, defining a local entropy
density $\bar{s}\left( \vec{r},t\right) $, such that 
\begin{equation}
\bar{S}\left( t\right) =\int d^{3}r\ \bar{s}\left( \mathbf{r},t\right) =\int
d^{3}r\left[ \varphi \left( \mathbf{r},t\right) +\sum\limits_{j}F_{j}\left( 
\mathbf{r},t\right) Q_{j}\left( \mathbf{r},t\right) \right] \ ,  \label{eq45}
\end{equation}
with [54] 
\begin{equation}
\Phi \left( t\right) =\int d^{3}r\ \varphi \left( \mathbf{r},t\right) \qquad
,  \label{eq46}
\end{equation}
there follows a generalized Gibbs' relation [57,60-62] 
\begin{equation}
d\bar{s}\left( \mathbf{r},t\right) =\sum\limits_{j}F_{j}\left( \mathbf{r}%
,t\right) \ dQ_{j}\left( \mathbf{r},t\right) \qquad ,  \label{eq47}
\end{equation}
introducing the important fact that the Lagrange multipliers are the
functional differentials of the IST-entropy, that is 
\begin{equation}
F_{j}\left( \mathbf{r},t\right) =\delta \bar{S}\left( t\right) /\delta
Q_{j}\left( \mathbf{r},t\right) \qquad ,  \label{eq48}
\end{equation}
where, we recall, $\delta $ stands for functional differential [63].

These Eqs. (\ref{eq48}) constitute the set of equations of state for
nonequilibrium situations, and go over those in equilibrium when the latter
is attained. We stress the fact that $\bar{S}\left( t\right) $, given in
terms of $\bar{\varrho}$ of Eq. (\ref{eq17}), is the quasi-entropy
associated at time $t$ with the instantaneous or ``frozen'' equilibrium
described by this quasi-equilibrium distribution, as already discussed in
the previous section.

The IST-entropy density satisfies a continuity equation, which, once Eqs. (%
\ref{eq38}) and (\ref{eq42}) are taken into account, takes the form
[57,60-62] 
\begin{equation}
\frac{\partial }{\partial t}\bar{s}\left( \mathbf{r},t\right) +div\vec{I}_{%
\bar{s}}\left( \mathbf{r},t\right) =\sigma _{\bar{s}}\left( \mathbf{r}%
,t\right) \qquad ,  \label{eq49}
\end{equation}
where $\vec{I}_{\bar{s}}$ is the flux of IST-entropy, 
\begin{equation}
\vec{I}_{\bar{s}}\left( \mathbf{r},t\right) =\sum\limits_{j}F_{j}\left( 
\mathbf{r},t\right) \vec{I}_{j}\left( \mathbf{r},t\right) \qquad ,
\label{eq50}
\end{equation}
and $\sigma _{\bar{s}}$ is the IST-entropy production density 
\begin{equation}
\sigma _{\bar{s}}\left( \mathbf{r},t\right) =\sum\limits_{j}\left[ \vec{I}%
_{j}\left( \mathbf{r},t\right) \cdot \nabla F_{j}\left( \mathbf{r},t\right)
+F_{j}\left( \mathbf{r},t\right) \mathcal{J}_{j}\left( \mathbf{r},t\right) %
\right] \ ,  \label{eq51}
\end{equation}
obtained after some calculus and the use of Eqs. (\ref{eq38}).

We call the attention to the fact that for the sake of brevity, we have
oversimplified the presentation including only the densities $Q_{j}$ and
their first (vectorial) fluxes, but, as shown elsewhere [55,56,79] the
selection law of Eq. (\ref{eq14}) requires to introduce the full set or
fluxes of all order $r=2,3,...,$ of these densities, besides the first
vectorial one.

In this Eq. (\ref{eq51}) the first term on the right takes the traditional
form of a product of fluxes and thermodynamics forces, the latter given by
the gradient of the Lagrange multipliers, which are fields of nonequilibrium
intensive variables in IST as defined by the nonequilibrium equations of
state, Eqs. (\ref{eq48}). The other term is the one associated to the
interactions that are contained in $\hat{H}^{\prime }$ of Eq. (\ref{eq13}),
present in the collision operator of Eq. (\ref{eq40}), and, as noticed,
accounted for the presence of the contribution $\varrho _{\varepsilon
}^{\prime }\left( t\right) $ in Eq. (\ref{eq31}). When in Eq. (\ref{eq51})
we perform integration in space over the volume of the system, the first
contribution (the one involving fluxes and forces), which arises from the
term $J^{\left( 0\right) }$ in Eq. (\ref{eq38}), cancels out, only remaining
the second one which constitutes the IST-entropy production 
\begin{equation}
\bar{\sigma}\left( t\right) =d\bar{S}\left( t\right) /dt=\sum\limits_{j}\int
d^{3}r\ F_{j}\left( \mathbf{r},t\right) \mathcal{J}_{j}\left( \mathbf{r}%
,t\right) \qquad .  \label{eq52}
\end{equation}
This reinforces our previous observation that no dissipation is present in
the description given by the quasi-equilibrium distribution $\bar{\varrho}%
\left( t,0\right) $, the dissipative evolution being described by $\varrho
_{\varepsilon }^{\prime }\left( t\right) $ in Eq. (\ref{eq31}). We close the
description of IST by simply pointing some additional results, consisting in
that the IST-entropy satisfies generalizations of Glansdorff-Prigogine's
criteria for thermodynamic evolution and (in)stability, and, in the strictly
linear regime, can be derived a Prigogine-like theorem of minimum production
of IST-entropy [62]. Moreover, it is derived a generalized Clausius relation
(introducing a generalized heat-density function), and a Boltzmann-like
relation stating that, in the thermodynamic limit, is given by [57] 
\begin{equation}
\bar{S}\left( t\right) =-\ln W\left( t\right) \qquad .  \label{eq53}
\end{equation}
In this Eq. (\ref{eq53}) $W\left( t\right) $ is the number of complexions
(number of quantum mechanical states, or volume in phase space in the
classical level) corresponding, at time $t,$ to the microscopic states
compatible with the constraints imposed by the information introduced by Eq.
(\ref{eq21b}), quite in analogy with the case in equilibrium described in
section \textbf{2} (se also Appendix \textbf{VI}).

Finally, at present we cannot make any definitive statement on the sign of
the local in space IST-entropy production density of Eq. (\ref{eq51}).
However. it is possible to derive what we call a \textit{weak principle of
non-negative informational-entropy production,} namely [57,62] 
\begin{eqnarray}
\Delta S\left( t\right)  &=&\bar{S}\left( t\right) -\bar{S}\left(
t_{o}\right) =\bar{S}\left( t\right) -S_{G}\left( t\right) =  \nonumber \\
&&  \nonumber \\
&=&\int\limits_{t_{o}}^{t}dt^{\prime }\int d^{3}r\ \sigma _{\bar{s}}\left( 
\mathbf{r},t\right) \geq 0\qquad ,  \label{eq54}
\end{eqnarray}
which is an expression indicating that the global IST-entropy of the
dissipative system always increases in time, from its initial value, along
the trajectory of evolution of the nonequilibrium macrostate of the system.
This result is equivalent to the one obtained by del Rio and Garcia-Colin
[229], who have interpreted the resulting inequality as the fact that a
sequence of observations performed, within a specific time interval, results
in a loss of information in Shannon-Brillouin's sense [37,38] This is
illustrated in Figure \textbf{1}: The average of the difference of both
kinds of entropy, namely the fine-grained and the coarse-grained, ones,
represents the increase in the statistical-informational entropy along the
irreversible evolution of the system namely the average of $\left[ 1-%
\mathcal{P}\left( t\right) \right] \ln \varrho _{\varepsilon }\left(
t\right) $ [we recall that $S_{G}$ is conserved and \ equal to $\bar{S}%
\left( t_{o}\right) $, what justifies the second equal sign in Eq. (\ref
{eq53})]. Of course, it is equal to zero in Eq. (\ref{eq54}) when $\varrho
_{\varepsilon }^{\prime }\left( t\right) =0,$ for example in equilibrium
conditions, but care should be taken to noticed that $\varrho _{\varepsilon
}^{\prime }\left( t\right) $ is non-null in any nonequilibrium steady-state:
This is a condition different from equilibrium, and, precisely, responsible
for the possible complex behavior consisting into the emergence of
Prigogine-like dissipative structures, as commented in subsection \textbf{4.1%
}. It is relevant to call the attention to the fact that we do not make any
attempt to relate the law of Eq. (\ref{eq54}) to the second law of
Thermodynamics; in its own right it can be only considered as a $\mathcal{H}-
$theorem in Jancel's sense [230]. As an ending observation in this section,
we notice that IST is accompanied by a MaxEnt-NESOM nonclassical
hydrodynamics, which may be dubbed as Informational Thermo-Hydrodynamics
[54-56,79]. A quite simplified form of it applied to the analysis of a
techno-industrial problem is presented elsewhere [231].

\section{SOME OPEN QUESTIONS AND CRITICISMS}

Let us briefly consider some general questions and criticisms associated to
the formalism. Further considerations are presented in Ref. [25], and
extensive ones can be found in Jaynes' ouvre (see for example Refs.
[4,28-32] and [232]).

Rolf Landauer [233] has argued that ``advocacy of MaxEnt is perpertuated by
selective decision making in the generation of papers [...] MaxEnt is likely
to be sound, but often it is dreadfully difficult to understand what the
constraints are'' (see Apendix \textbf{VI}). Mario Bunge stated [234] that
``when confronted with a random or seemingly random process, one attempts to
build a probabilistic model that could be tested against empirical data, no
randomness, no probability. Moreover, as Poincar\'{e} pointed out long ago,
talk of probability involves some knowledge; it is not a substitute for
ignorance [and Bunge adds, not correctly, in what refers to predictive
statistical mechanics as we are discussing here, that] this is not how the
Bayesian or personalists view the matter: when confronted with ignorance or
uncertainty, they use probability -- or rather their own version of it. This
allows them to assign prior probabilities to facts and propositions in an
arbitrary manner [again, this is not the case in MaxEnt-NESOM] -- which is a
way of passing off mere intuition, hunch, or guess for scientific hypothesis
[...]; it is all a game of belief rather than knowledge''.

Sometimes arguments against MaxEnt in terms of playing dices have been
advanced. To this, it must be recalled that the question we are addressing
here does not deal with gambling, but with many-body theory. That is, we
deal with systems with very many degrees of freedom, and then is necessary
to have in mind the distinction between interpretations in terms of
microscopic and macroscopic variables.

Concerning the arguments that knowledge arises out of ignorance, this is
simply unnecessary confusion coming from a wrong interpretation of, maybe, a
sometimes not correct phrasing used by some practitioners of MaxEnt in areas
other than Many-Body Physics. Quite to the contrary, the spirit of the
formalism is to make use of the restricted knowledge available, but without
introducing any spurious one. Quoting Laplace [235], ``the curve described
by a molecule of air or of vapor is following a rule as certainly as the
orbits of the planets:\ the only difference between the two is our
ignorance. Probability is related, in part to this ignorance, in part to our
knowledge''. Also, as pointed out by Bricmont [236], ``the part `due to our
ignorance' is simply that we\textit{\ use} probabilistic reasoning. If we
were omniscient, it would not be needed (but the averages would remain what
they are, of course). The part `due to our knowledge' is what makes
reasoning work [...]. But this is the way things are: our knowledge \textit{%
is }incomplete, and we have to live with that. Nevertheless, probabilistic
reasoning is extraordinarly sucessful in practice, but, when it works, this
is due to our (partial) knowledge. It would be wrong to attribute any
constructive role to our ignorance.\ And it is also erroneous to assume that
the system must be somehow indeterminate when we apply probabilistic
reasoning to it.'' (See also Ref. [35]).

Moreover, we stress a point illustrated in section \textbf{2}: To derive the
behavior of the macroscopic state of the system from partial knowledge has
been already present in the original genial work of Gibbs. This is at the
roots of the well established, fully accepted, and exceedingly successful
statistical mechanics in equilibrium; as noted, the statistical distribution
which should depend on all constants of motion is built, in any of the
canonical ensembles, in terms of the available information we do have,
namely, the preparation of the sample in the given experimental conditions
in equilibrium with a given (and quite reduced) set of reservoirs. Werner
Heisenberg wrote [237], ``Gibbs was the first to introduce a physical
concept which can only be applied to an object when our knowledge of the
object is incomplete''.

Finally on this point, the dismissal of a theoretical approach in Physics
cannot (and should not) be done on the basis of general verbal arguments,
which may or may not be sensible, but which need be strongly fundamented on
the scientific method. In other words, the merits, or rather dismerits, of a
theory reside in establishing its domain of validity (see for example Refs.
[1,238,239]), when tested against the experimental results it predicts. This
point has recently been forcefully stressed by Stephen Hawkings as we stress
in next section. Moreover, as we have already described in section \textbf{4}%
, MaxEnt-NESOM appears to provide a successful statistical mechanics and
thermodynamics of a very large scope (that is, of a large domain of validity
accompained with good intuitive physical interpretations) for the Physics of
many-body system arbitrarily away from equilibrium. In this way, as
antecipated in the Introduction, it confirms Zwanzig's prediction about the
merits of the formalism.

Returning to the question of the Bayesian approach in statistical mechanics,
Sklar [241 has summarized that Jaynes firstly suggested that equilibrium
statistical mechanics can be viewed as a special case of the general program
of systematic inductive reasoning, and that, from this point of view, the
probability distributions introduced into statistical mechanics have their
bases not so much in an empirical investigation of occurences in the world,
but, instead in a general procedure for determining appropriate \textit{a
priori }\ subjective probabilities in a systematic way. Also, Jaynes'
prescription was to choose the probability distribution which maximizes the
statistical entropy (now thought in the information-theoretic vein) relative
to the known macroscopic constraints, using the standard measure over the
phase space to characterize the space of possibilities. This probability
assignment is a generalization of the probabilities determined by the 
\textit{Principle of Indifference} (or Bernoulli's ``principle of
insufficient reason'', Maynard Keynes' ``principle of ignorance'', Tolman's
``principle of equal \textit{a priori} probabilities''), specifying one's
rational choice \textit{of a priori} probabilities. In equilibrium this is
connected with ergodic theory, as known from classical textbooks. Of course
it is implied to accept the justification of identifying averages with
measured quantities using the time in the interval of duration of the
experiment. This cannot be extended to nonequilibrium conditions involving
ultrafast relaxation processes. Therefore, there remains the explanatory
question: Why do our probabilistic assumptions work so well in giving us
equilibrium values? [241].

The Bayesian \ approach attempts an answer which, apparently, works quite
well in equilibrium, and then it is tempting to extend it to nonequilibrium
conditions. Jaynes rationale for it is, again, that the choice of
probabilities, being determined by a Principle of Indifference, should
represent maximum uncertainty relative to our knowledge as exhausted by our
knowledege of the macroscopic constraints with which we start [31]. This has
been described in previous sections.

At this point, it can be rised the question of the possibility of situations
when the Principle of Indifference may be seen as not applying. This seems
to be the case of some phenomena said to follow L\'{e}vy's distributions
[242]. In condensed matter, Fick (or Fourier) diffusion equations are, for
certain systems, nonvalid and a so-called anomalous L\'{e}vy-like diffusion
is present. A generalized concept of entropy has been proposed to deal with
these cases [243]. According to this approach the Principle of Indifference
is violated in a nontrivial way. The motion is in some sense fractal, what
is thought to explain the appearence of the many fractal structures in
nature. It is considered that many, if not all, fractal structures could
emerge through a self-organized criticality [242]. An accompanying, say,
statistics for fractal classical systems, sometimes called Tsallis'
statistics, is being proposed and devised along some kind of an
informational MaxEnt approach (see for example Refs. [244,245]).

We call the attention fo the fact that the principle of maximization of
informational entropy in Shanon-Jaynes sense, can be alternatively covered
by the principle of Minimization of cross-entropy distances [246] ( the \
``distance'' appropriately defined of two types of distributions adjudicated
to a given statistical problem), \textit{MinxEnt }for short. MaxEnt-NESOM
follows from minimization of the so-called Kullback-Leibler measure
(distance from the sought after distribution to the uniform distribution).
But some problems can require non-exponential distributions, such that
additivity - or better to say Euler homogeneity of order one - in the
statistical functions is not satisfied. We have already mentioned Levy
distributions, requiring generalized measures of cross-entropy. The case of
the above cited Tsallis statistics is based on a particular case of
Csiszer's measure [247], namely Havrada-Charvat measure [248].

We also call the attention to an attempt to derive, mainly on
topological-geometrical bases, a fundamental theory for the realm of the
physics of macroscopic systems by unifying, and properly characterizing, the
levels of description involving micro- and macro-physics. The approach
called Generics aspires to be all-embracing and multifarious. Its implicit
programme is in development in the general aspects and in what regards its
practicality [249].

Another point of contention is the long standing question about \textit{%
macroscopic irreversibility in nature}. As discussed in section \textbf{3},
it is introduced in the formalism via the generalization of Kirkwood's
time-smoothing procedure, after a specific initial condition [cf. Eq. (\ref
{eq30})] -- implying in a kind of generalized \textit{Stosszahlansatz} --
has been defined. This is a working proposal that goes in the direction
which was essentially suggested by Boltzmann, as quoted in Ref. [236]:
``Since in the differential equations of mechanics themselves there is
absolutely nothing analogous to the second law of thermodynamics, the latter
can be mechanically represented only by means of assumptions regarding
initial conditions''. Or, in other words [236], that the laws of physics are
always of the form: given some initial conditions, here is the result after
some time. But they never tell us how the world \textit{is or evolves.} In
order to account for that, one always needs to assume something, first on
the initial conditions and, second, on the distinction of the description
being macroscopic and the system never isolated (damping of correlations).
In this vein Hawkings [240] has manifested that ``It is normally assumed
that a system in a pure quantum state evolves in a unitary way through a
succession of [such] states. But if there is loss of information through the
appearence and disappearance of black holes, there can't be a unitary
evolution. Instead, the [...] final state [...] will be what is called a 
\textit{mixed quantum state.} This can be regarded as an ensemble of
different pure quantum states, each with its own probability'' (see Appendix 
\textbf{IV}).

Needless to say that this question of Eddington's time-arrow problem has
produced a very extensive literature, and lively controversies. We do not
attempt here to add any considerations to this difficult and, as said,
controversial subject. We simply list in the references
[67,70,71,240,250-264] some works on the matter which we have selected. As
commented by Sklar [241], Nicolai S. Krylov (the Russian scientist
unfortunately prematurely deceased) was developing an extremely insightful
and careful foundational study of nonequilibrium statistical mechanics [19].
Krylov believed that he could show that in a certain sense, neither
classical nor quantum mechanics provide an adequate foundation for
statistical mechanics. Krylov's most important critical contribution is
precisely his emphasis on the importance of initial ensembles. Also, that we
may be utterly unable to demonstrate that the correct statistical
description of the evolution of the system will have an appropriate finite
relaxation time, much less the appropriate exact evolution of our
statistical correlates of macroscopic parameters, unless our statistical \
approach includes an appropriate constraint on the initial ensemble with
which we choose to represent the initial nonequilibrium condition of the
system in question. Moreover, it is thought that the interaction with the
system from the outside at the single moment of preparation, rather than the
interventionists ongoing interaction, is what grounds the asymmetric
evolution of many-body systems. It is the ineluctable interfering
perturbation of the system by the mechanism that sets it up in the first
place that guarantees that the appropriate statistical description of the
system will be a collection of initial states sufficiently large,
sufficiently simple in shape, and with a uniform probability distribution
over it. Clearly, a question immediatly arises, namely: Exactly how does
this initial interference lead to an initial ensemble of just the kind we
need? [241] We have seen in section \textbf{3} how MaxEnt-NESOM, mainly in
Zubarev's approach, tries to heuristically address the question. Also
something akin to these ideas seems to be in the earlier work of the
Russian-Belgian Nobel Prize Ilya Prigogine [67,250,265], and also in the
considerations in Refs. [241,259-261,266,267]. Certain kind of equivalence -
at least partial - seems to exists between Prigogine's approach and
MaxEnt-NESOM, as pointed out by Dougherty [268,269], and we side with
Dougherty's view [270] (see also Appendix \textbf{IV}). More recently,
Prigogine and his School have extended those ideas incorporating concepts,
at the quantum level, related to dynamical instability and chaos (see for
example Refs. [271,272]). In this direction, some attempts try to
incorporate time-symmetry breaking, extending quantum mechanics to a general
space state, a ``rigged'' (or ``structured'') Hilbert space or Gelfand
space, with characteristics (superstructure) mirroring the internal
structure of collective and cooperative macroscopic systems [273-275]. This
formulation of dynamics constitutes an effort towards including the second
law of thermodynamics, as displayed explicitly by a $\mathcal{H}$-function
of the Boltzmann type, which decreases monotonically and takes its minimum
value when unstable systems have decayed or when the system reaches
equilibrium [276,277].

Finally, and in connection with the considerations presented so far, we
stress that in the formalism as by us described in previous sections, no
attempt is made to establish any direct relation with thermodynamic entropy
in, say, the classical Clausius-Carnot style, with its increase between
initial and final equilibrium states defining irreversibility. Rather, it
has been introduced a statistical-informational entropy with an evolution as
given by the laws of motion of the macrovariables, as provided by the
MaxEnt-NESOM-based kinetic theory. Irreversible transport phenomena are
described by the fluxes of energy, mass, etc..., which can be observed, but
we do not see entropy flowing. We have already stressed in subsection 
\textbf{5.2} that the increase of IST-entropy amounts to a $\mathcal{H}$%
-like theorem, that is, a manifestation on the irreversible character of the
transport equations, in close analogy with Boltzmann's $\mathcal{H}$-theorem
which does so for Boltzmann equation. Moreover, as stated in other papers
(see for example Ref. [57]), we side with Meixner's point of view [69,278]
that, differently to equilibrium, does not exists a unique and precisely
defined concept of thermodynamic entropy out of equilibrium. The one of
subsection \textbf{5.2}, is the one peculiar to IST, and depending on the
nonequilibrium thermodynamic state defined by Zubarev-Peletminskii selection
law altogether \ with the use of Bogoliubov's principle of correlation
weakening. The IST entropy, we recall, has several properties listed in
section \textbf{5.2}, and one is that it takes (in the thermodynamic limit)
a typical Boltzmann-like expression [cf. Eq. (\ref{eq53})], implying that
the macroscopic constraints imposed on the system (the informational bases)
determine the vast majority of microscopic configurations that are
compatible with them and the initial conditions. It is worth noticing that
then, according to the weak principle of increase of the IST-entropy, as the
dissipative system evolves, such number of microscopic configurations keeps
increasing up to a maximum when final full equilibrium is achieved. Further,
MaxEnt-NESOM recovers in the appropriate limit the distribution in
equilibrium, and in IST one recovers the traditional Clausius-Carnot results
for increase of thermodynamics entropy between an initial and a final
equilibrium states, as shown in Ref. [72,279].

\section{CONCLUDING REMARKS}

In the preceding sections we have described, in terms of a general overview,
a theory that attempts a particular answer to the long-standing sought-after
question about the existence of a Gibbs-style statistical ensemble formalism
for nonequilibrium systems. Such formalism, providing microscopic
(mechanical-statistical) bases for the study of dissipative processes,
heavily rests on the fundamental ideas and concepts devised by Gibbs and
Boltzmann. It consists into the so-called MaxEnt-NESOM formalism, which, as
noticed in the Introduction, appears to be covered under the theoretical
umbrella provided by Jaynes' Predictive Statistical Mechanics. We have
already called the attention to the fact that it is grounded on a kind of
scientific inference approach, Jeffrey's style, based on Bayesian
probability and information theory in Shannon-Brillouin's sense [33,34]. It
has been improved and systematized mainly by the Russian School of
Statistical Physics, and the different approaches have been brought under a
unified description based on a variational procedure. It consists in the use
of the principle of maximization of the informational entropy, meaning to
rely exclusively on the available information and avoiding to introduce any
spurious one. The aim is to make predictions on the behavior of the dynamics
of the many-body system on the basis of only that information. On this,
Jeffreys, at the begining of Chapter I in the book of reference [279],
states that: ``The fundamental problem of scientific progress, and the
fundamental of everyday life, is that of learning from experience. Knowledge
obtained in this way is partly merely discription of what we have already
observed, but part consists of making inferences from past experience to
predict future experience. This may be called generalization of induction.
It is the most important part.'' Jeffreys also quotes J. C. Maxwell who
stated that the true logic for this world is the Calculus of Probability
which takes account of the magnitude of the probability that is, or ought to
be, in a reasonable man's mind.

Some authors conjecture that this may be the revolutionary thought in modern
science (see for example Refs. [33,34,250,280]): It replaces the concept of
inevitable effects (trajectories in a mechanicist point of view of many-body
(large) systems) by that of the probable trend (in a generalized theory of
dynamical systems). Thus, the different branches of science that seem to be
far apart, may, within such new paradigm, grow and be hold together
organically [281]. These points of view are the subject of controversy,
mainly on the part of the adepts of the mechanicist-reduccionist school. We
call the attention to the subject but we do not take any particular
position, simply adhering to the topic here presented from a pragmatical
point of view. In that sense, we take a position coincident with the one
clearly stated by Stephen Hawkings [282]: ``I do not demand that a theory
corresponds to reality. But that does not bother me. I do not demand that a
theory correspond to reality because I do not know what reality is. Reality
is not a quality you can test with litmus paper. All I am concerned with is
that the \textit{Theory should predict the results of measurement}''
[emphasis is ours].

MaxEnt-NESOM is the construtive criterion for deriving the probability
assignment for the problem of dissipative processes in many-body systems, on
the bases of the available information (provided, as Zwanzig pointed out
[5], on the knowledge of measured properties and of sound theoretical
considerations). The fact that a certain probability distribution maximizes
the informational entropy, subject to certain constraints representing our
incomplete information, is the fundamental property which justifies the use
of that distribution for inference; it agrees with everything that is known,
but carefully avoids assuming anything that is not known. In that way it
enforces - or gives a logico-mathematical viewpoint - to the principle of
economy in logic, known as Occam's Razor, namely ``Entities are not to be
multiplied except of necessity''. Particularly, in what concerns Statistical
Thermodynamics (see subsection \textbf{5.2}), MaxEnt-NESOM, in the context
of Jaynes' Predictive Statistical Mechanics, allows to derive laws of
thermodynamics, not on the usual viewpoint of mechanical trajectories and
ergodicity of classical deductive reasoning, but by the goal of using
inference from incomplete information rather than deduction: the
MaxEnt-NESOM distribution represents the best prediction we are able to make
from the information we have [4,28-32].

In Section \textbf{3,} we have briefly described the use of MaxEnt-NESOM to
rederive the old-vintage equilibrium statistical mechanics. That is, to
retrive the original Gibbs-ensemble canonical distributions, and where -
what is considered to be already contained in the work of Gibbs [30] - it is
clearly visualized the use of the specific information used, and only this,
consisting into the knowledge of the existence of equilibrium with well
specified reservoirs.

In section \textbf{4, }we have, along general lines, shown how to derive a
nonequilibrium ensemble formalism, in the MaxEnt-NESOM framework, which
unifies and gives basic structure to the different approaches attempted on
either heuristic arguments or projection operator techniques. Without
entering into details, given in the references in each case indicated, we
have described the main six basic steps that the formalism requires.
Summarizing them:

$\bullet $ i) On the basis of Bogoliubov's principle of correlation
weakening one chooses the basic set of dynamical variables, introducing the
separation of the total Hamiltonian in the form of Eq. (\ref{eq13}),
eliminating fast variables (associated to correlations that have died down),
and retaining only the slow variables (with relaxation times larger than the
characteristic time for the description of the evolution of the
nonequilibrium macroscopic state of the system).

$\bullet $ ii) Introduction of the rule of Eq, (\ref{eq14}), which selects
the relevant set of slow variables.

$\bullet $ iii) Introduction of retro-effects, making the state at time $t$
be dependent on the past history of evolution of the system from a well
established initial condition of preparation.

$\bullet $ iv) Introduction of an \textit{ad hoc} time-smoothing procedure,
with properties which allow to complete the construction of a satisfactory
nonequilibrium statistical operator. In particular, the fading-memory
characteristic it produces leads to the irreversible evolution of the
macrostate of the system, from the initial condition of preparation (the
information available to build the formalism) and to obtain the description
of the system on the basis of such information and its evolution to a final
state of equilibrium in the future.

And two final steps, those indispensable to obtain the macroscopic
properties of the system, namely,

$\bullet $ v) A nonlinear quantum kinetic theory; and

$\bullet $ vi) A response function theory for system arbitrarily away from
equilibrium.

In section \textbf{5}, we listed (given the corresponding references) some
applications of MaxEnt-NESOM to the analysis and interpretation of
experimental studies in the photoinjected plasma in highly excited
semiconductors. Also, some considerations concerning complex behavior in
biopolymers and organic polymers were briefly mentioned. In subsection 
\textbf{5.2}, we have described the construction, on the basis of the
MaxEnt-NESOM, of a Statistical Thermodynamics, the so-called Informational
Statistical Thermodynamics, and its main aspects have been described.
Moreover, one can develop a nonclassical Thermo-Hydrodynamics, and, it may
be noticed, the scope of the formalism makes it suitable for applications in
many areas other than that of the physics and chemistry of semiconductors
and polymers, as in rheology, chemical engineering, food engineering,
hydraulics etc...., and we have already mentioned the particular cases of
the techno-industrial processes of laser-thermal estereo-lithography and
medical imaging.

As ending considerations, we stress that, in this paper we have given a
brief descriptional presentation of the MaxEnt-NESOM, which is an approach
to a nonequilibrium statistical ensemble algorithm in Gibbs' style,
seemingly as a very powerful, concise, soundly based, and elegant formalism
of a broad scope apt to deal with systems arbitrarily away from equilibrium.
We may say that it constitutes a promising tentative to fulfill a Programme
for nonequilibrium statistical mechanics and thermodynamics, consisting of
the itens:

$\bullet $ I. Construction of a relevant statistical operator corresponding
to a nonequilibrium ensemble formalism, which can provide a satisfactory
description of the \textit{irreversible evolution} of macroscopic many-body
systems.

$\bullet $ II. To build a \textit{Statistical Irreversible Thermodynamics }%
derived from this nonequilibrium ensemble formalism.

$\bullet $ III. To derive a confiable nonlinear quantum \textit{Kinetic
Theory}, providing for a description of the evolution of the nonequilibrium
system (the trajectory in the nonequilibrium thermodynamic state space).

$\bullet $ IV. To derive a \textit{Response Function Theory}, for systems
arbitrarily away from equilibrium, the all important step to establish the
connection between theory and experiment.

$\bullet $ V. The final closing step of \textit{applying the theory to real
situations} in laboratory conditions for comparison with experimental data.

\bigskip 

{\LARGE ACKNOWLEDGEMENTS}

Our Research Group gratefully acknowledges finantial support provided in
different opportunities by the S\~{a}o Paulo State Research Foundation
(FAPESP), the Brazilian National Research Council (CNPq), the Ministry of
Planning (Finep), the Ministry of Education (CAPES), Unicamp Foundation
(FAEP), IBM Brasil, the USA-National Science Foundation (USA-Latinamerica
Collaboration Project; NSF, Washington); and the John Simon Guggenheim
Memorial Foundation (New York, USA). We also avow enlightening and useful
discussions with Leopoldo Garc\'{i}a-Col\'{i}n (UAM, Mexico), Jos\'{e}
Casas-V\'{a}zquez and David Jou (UAB, Spain), and Mario E. Foglio (Unicamp,
Brazil).

\newpage

\appendix

\renewcommand{\theequation}{\Roman{section}.\arabic{equation}}

\renewcommand{\thesection}{Appendix \Roman{section}:}

\section{The variational procedure}

\setcounter{equation}{0}

Consider a many-body system in contact and in equilibrium with a set of $n$
reservoirs with which exchange the physical quantities $\hat{P}_{1}$, ..., $%
\hat{P}_{n}$ (e.g. energy, particles, etc.). The equilibrium is
characterized by the equalization of the corresponding intensive
thermodynamic variables, say, $F_{1}$, ...., $F_{n}$ (e.g. temperature,
chemical potential, etc.), which determine the average value of the former, $%
Q_{1}$, ..., $Q_{n}$. In the variational method to Gibbs' ensemble formalism
the statistical operator $\varrho $ for the system in equilibrium follows
from the maximization of Gibbs' entropy 
\begin{equation}
S_{G}=-Tr\left\{ \varrho \ln \varrho \right\}  \label{eqI1}
\end{equation}
under the constraints of (1) normalization 
\begin{equation}
Tr\left\{ \varrho \right\} =1\qquad ,  \label{eqI2}
\end{equation}
and (2) the average values fixed by the condition of equilibrium with the
reservoirs, namely 
\begin{equation}
Q_{j}=Tr\left\{ \hat{P}_{j}\varrho \right\}  \label{eqI3}
\end{equation}
for $j=1,2,....,n,$ which of course depend on the intensive variables $F_{j}$%
.

As known, the method of Lagrange multipliers requires to make extremal the
functional 
\begin{equation}
I\left\{ \varrho \right\} =Tr\left\{ \varrho \left[ \ln \varrho +\left( \phi
-1\right) +\sum\limits_{j=1}^{n}F_{j}\hat{P}_{j}\right] \right\} \quad ,
\label{eqI4}
\end{equation}
what produces Gibbs' generalized canonical distribution 
\begin{equation}
\varrho =\exp \left\{ -\phi -\sum\limits_{j=1}^{n}F_{j}\hat{P}_{j}\right\}
\qquad ,  \label{eqI5}
\end{equation}
with 
\begin{equation}
\phi =\ln Z=\ln Tr\left\{ \exp \left[ -\sum\limits_{j=1}^{n}F_{j}\hat{P}_{j}%
\right] \right\}  \label{eqI6}
\end{equation}
ensuring the normalization of $\varrho $ and where $Z$ is the corresponding
partition function. The usual canonical distribution follows for the case of
a thermal reservoir when $\hat{P}_{1}=\hat{H}$ and $F_{1}=\beta =1/k_{B}T$;
the grand-canonical distribution for the case of a thermal and a particle
reservoir when $\hat{P}_{1}=\hat{H}$, $\hat{P}_{2}=\hat{N},$ and $%
F_{1}=1/k_{B}T$ , $F_{2}=-\mu /k_{B}T$ .

For nonequilibrium conditions, as noticed in the main text, once the choice
of the set of basic variables $\left\{ \hat{P}_{j}\left( \mathbf{r}\right)
\right\} $ has been done according to the procedure we have described, and
the initial condition at time $t_{o}$ has been set, and following Kirkwood
time-smoothing \ procedure, we maximize the time-dependent Gibbs' entropy 
\begin{equation}
S_{G}\left( t\right) =-Tr\left\{ \varrho \left( t\right) \ln \varrho \left(
t\right) \right\} \qquad ,  \label{eqI7}
\end{equation}
under the constraints 
\begin{equation}
Tr\left\{ \varrho \left( t^{\prime }\right) \right\} =1\qquad ,  \label{eqI8}
\end{equation}
\begin{equation}
Q_{j}\left( \mathbf{r},t^{\prime }\right) =Tr\left\{ \hat{P}_{j}\left( 
\mathbf{r}\right) \varrho \left( t^{\prime }\right) \right\} \quad ,
\label{eqI9}
\end{equation}
for $t_{o}\leq t^{\prime }\leq t$, that is, keeping the information on the
history of evolution of the system from the inital condition of preparation
at time\ $t_{o}$ time up to the time $t$ when the measurement is performed.
Hence, following the spirit of the method of Langrange multipliers, we need
to make extremal the functional 
\[
I\left\{ \varrho \left( t\right) \right\} =Tr\left\{ \varrho \left( t\right)
\ln \varrho \left( t\right) +\int\limits_{t_{o}}^{t}dt^{\prime }\left[ \psi
\left( t^{\prime }\right) -\delta \left( t-t^{\prime }\right) \right]
\varrho \left( t^{\prime }\right) +\right. 
\]
\begin{equation}
\left. +\sum\limits_{j=1}^{n}\int\limits_{t_{o}}^{t}dt^{\prime }\int d^{3}r\
\varphi _{j}\left( \mathbf{r},t,t^{\prime }\right) \hat{P}_{j}\left( \mathbf{%
r}\right) \right\} \qquad ,  \label{eqI10}
\end{equation}
leading to the result that 
\begin{equation}
\varrho \left( t\right) =\exp \left\{ -\Psi \left( t\right)
-\sum\limits_{j=1}^{n}\int\limits_{t_{o}}^{t}dt^{\prime }\int d^{3}r\
\varphi _{j}\left( \mathbf{r},t,t^{\prime }\right) \hat{P}_{j}\left( \mathbf{%
r}\right) \right\} \quad ,  \label{eqI11}
\end{equation}
where 
\begin{equation}
\Psi \left( t\right) =\int\limits_{t_{o}}^{t}dt^{\prime }\psi \left(
t^{\prime }\right) =\ln Tr\left\{ \exp \left[ -\sum\limits_{j=1}^{n}\int%
\limits_{t_{o}}^{t}dt^{\prime }\int d^{3}r\ \varphi _{j}\left( \mathbf{r}%
,t,t^{\prime }\right) \hat{P}_{j}\left( \mathbf{r}\right) \right] \right\} 
\label{eqI12}
\end{equation}
ensures the normalization of $\varrho \left( t\right) $.

The Lagrange multipliers $\varphi _{j}\left( \mathbf{r},t,t^{\prime }\right) 
$ are determined by the conditions of Eq. (\ref{eqI9}), which, as indicated
in the main text, are redefined in the form 
\begin{equation}
\varphi _{j}\left( \mathbf{r},t,t^{\prime }\right) =w\left( t,t^{\prime
}\right) F_{j}\left( \mathbf{r},t^{\prime }\right) \qquad ,  \label{eqI13}
\end{equation}
introducing the weight function $w$ in the integration in time, with the
properties discussed in Ref. [23].

\setcounter{equation}{0}

\section{Zubarev's NESOM}

Zubarev's approach to MaxEnt-NESOM follows from the choice of the weight
function $w\left( t,t^{\prime }\right) $ of Eq. (\ref{eqI13}) in the form of
Abel's kernel $\varepsilon \exp \left\{ \varepsilon \left( t^{\prime
}-t\right) \right\} $ and $t_{o}$ taken in the remote past, Hence, 
\begin{equation}
\varrho _{\varepsilon }\left( t\right) =\exp \left\{ -\varepsilon
\int\limits_{-\infty }^{t}dt^{\prime }\ e^{\varepsilon \left( t^{\prime
}-t\right) }\ \hat{S}\left( t^{\prime },t^{\prime }-t\right) \right\} \quad ,
\label{eqII1}
\end{equation}
which after part integration takes the form 
\begin{equation}
\varrho _{\varepsilon }\left( t\right) =\exp \left\{ -\hat{S}\left(
t,0\right) +\hat{\zeta}_{\varepsilon }\left( t\right) \right\} \qquad ,
\label{eqII2}
\end{equation}
where 
\begin{equation}
\hat{\zeta}_{\varepsilon }\left( t\right) =\int\limits_{-\infty
}^{t}dt^{\prime }\ e^{\varepsilon \left( t^{\prime }-t\right) }\frac{d}{%
dt^{\prime }}\ \hat{S}\left( t^{\prime },t^{\prime }-t\right) \quad .
\label{eqII3}
\end{equation}

Using the operator identity 
\begin{equation}
e^{-\hat{A}+\hat{B}}=Y\left( \hat{B}\mid 1\right) e^{-\hat{A}}\qquad ,
\label{eqII4}
\end{equation}
where 
\begin{equation}
Y\left( \hat{B}\mid x\right) =1+\int\limits_{0}^{x}du\ Y\left( \hat{B}\mid
u\right) e^{-u\hat{A}}\ B\ e^{u\hat{A}}\quad ,  \label{eqII5}
\end{equation}
the statistical operator of Eq. (\ref{eqII2}) can be written as 
\begin{equation}
\varrho _{\varepsilon }\left( t\right) =\bar{\varrho}\left( t,0\right)
+\varrho _{\varepsilon }^{\prime }\left( t\right) \qquad ,
\end{equation}
where 
\begin{equation}
\bar{\varrho}\left( t,0\right) =\exp \left\{ -\hat{S}\left( t,0\right)
\right\} \qquad ,  \label{eqII6}
\end{equation}
\begin{equation}
\varrho _{\varepsilon }^{\prime }\left( t\right) =\hat{D}_{\varepsilon
}\left( t\right) \ \bar{\varrho}\left( t,0\right) \qquad ,  \label{eqII7}
\end{equation}
\begin{equation}
\hat{D}_{\varepsilon }\left( t\right) =\int\limits_{0}^{1}du\ Y\left( \hat{%
\zeta}_{\varepsilon }\left( t\right) \mid u\right) \left[ \bar{\varrho}%
\left( t,0\right) \right] ^{u}\ \hat{\zeta}_{\varepsilon }\left( t\right) %
\left[ \bar{\varrho}\left( t,0\right) \right] ^{-u}\quad ,  \label{eqII8}
\end{equation}
\begin{equation}
Y\left( \hat{\zeta}_{\varepsilon }\left( t\right) \mid u\right)
=1+\int\limits_{0}^{x}du\ Y\left( \hat{\zeta}_{\varepsilon }\left( t\right)
\mid u\right) \left[ \bar{\varrho}\left( t,0\right) \right] ^{u}\ \hat{\zeta}%
_{\varepsilon }\left( t\right) \left[ \bar{\varrho}\left( t,0\right) \right]
^{-u}\ ,  \label{eqII9}
\end{equation}
and we recall that 
\begin{equation}
\hat{S}\left( t,0\right) =\phi \left( t\right) \hat{1}+\sum\limits_{j=1}^{n}%
\int d^{3}r\ F_{j}\left( \vec{r},t\right) \hat{P}_{j}\left( \vec{r}\right)
\qquad .  \label{eqII10}
\end{equation}

We call the attention to the fact that because of Eq. (\ref{eq24}) 
\begin{equation}
Tr\left\{ \varrho _{\varepsilon }^{\prime }\left( t\right) \right\}
=Tr\left\{ \hat{D}_{\varepsilon }\left( t\right) \bar{\varrho}\left(
t,0\right) \right\} =0\qquad ,  \label{eqII11}
\end{equation}
\begin{equation}
Tr\left\{ \hat{P}_{j}\varrho _{\varepsilon }^{\prime }\left( t\right)
\right\} =Tr\left\{ \hat{P}_{j}\hat{D}_{\varepsilon }\left( t\right) \bar{%
\varrho}\left( t,0\right) \right\} =0\qquad .  \label{eqII12}
\end{equation}

Moreover, we notice that the statistical operator in Green-Mori's approach
follows from the choice 
\begin{equation}
\varrho _{\tau }\left( t\right) =\exp \left\{ -\frac{1}{\tau }%
\int\limits_{t-\tau }^{t}dt^{\prime }\hat{S}\left( t^{\prime },t^{\prime
}-t\right) \right\} \qquad ,  \label{eqII13}
\end{equation}
with $\tau $ going to infinity after the trace operation in the calculation
of averages has been performed. We may see that Eq. (\ref{eqII13}) implies
in a kind of time average over the time interval of extension of $\tau $,
while the one in Eq. (\ref{eqII1}) is the so-called causal average. We
stress that \textit{a posteriori} the latter introduces in the kinetic
equations a fading memory and produces finite transport coefficients, while
the use of the statistical operator of Eq. (\ref{eqII13}) may produce
divergent integrals in the calculation of transport coefficients.

\setcounter{equation}{0}

\section{Time-Dependent Projection Operator}

Let us consider the set of dynamical variables $\hat{P}_{j}$, to which we
further add $\hat{P}_{o}=\hat{1}$ whose associated Lagrange multiplier is $%
F_{o}\left( t\right) =\phi \left( t\right) $, in terms of which we define
the supercorrelation functions 
\begin{equation}
\mathcal{C}_{ij}\left( t\right) =\left\{ \hat{P}_{i},\hat{P}_{j}\mid
t\right\} \qquad ,  \label{eqIII1}
\end{equation}
with $i,j=0,1,2,...,n$, and where for any pair of operators $A$ and $B$ 
\begin{equation}
\left\{ \hat{A},\hat{B}\mid t\right\} =Tr\left\{ \int\limits_{0}^{1}du\ \hat{%
A}\ Y\left( \hat{\zeta}_{\varepsilon }\mid u\right) \left[ \bar{\varrho}%
\left( t,0\right) \right] ^{u}\hat{B}\ \left[ \bar{\varrho}\left( t,0\right) %
\right] ^{-u}\ \bar{\varrho}\left( t,0\right) \right\} \quad ,
\label{eqIII2}
\end{equation}
where $Y$ is given in Eq. (\ref{eqII9}).

In terms of this particular metric - in the informational subspace of
quantities $\hat{P}_{j}$ - we define the time-dependent projection operator 
\begin{equation}
\mathcal{P}_{\varepsilon }\left( t\right) \hat{A}=\sum\limits_{i,j=0}^{n}%
\hat{P}_{i}\mathcal{C}_{ij}^{\left( -1\right) }\left( t\right) \left\{ \hat{P%
}_{j},\hat{A}\mid t\right\} \quad ,  \label{eqIII3}
\end{equation}
where $\mathcal{C}^{\left( -1\right) }$ is the inverse of the matrix with
elements given by Eq. (\ref{eqIII1}). This projection operator has the
property that 
\begin{equation}
\mathcal{P}_{\varepsilon }\left( t\right) \ln \varrho _{\varepsilon }\left(
t\right) =\ln \bar{\varrho}\left( t,0\right) \qquad .  \label{eqIII4}
\end{equation}
In fact, 
\begin{equation}
\mathcal{P}_{\varepsilon }\left( t\right) \ln \varrho _{\varepsilon }\left(
t\right) =\mathcal{P}_{\varepsilon }\left( t\right) \left[ \ln \bar{\varrho}%
\left( t,0\right) +\hat{\zeta}_{\varepsilon }\left( t\right) \right] \quad ,
\label{eqIII5}
\end{equation}
because of Eq. (\ref{eqII2}), and 
\begin{eqnarray}
\mathcal{P}_{\varepsilon }\left( t\right) \ln \bar{\varrho}\left( t,0\right)
&=&\sum\limits_{i,j,k=0}^{n}\hat{P}_{i}\mathcal{C}_{ij}^{\left( -1\right)
}\left( t\right) \left\{ \hat{P}_{j},F_{k}\left( t\right) \hat{P}_{k}\mid
t\right\} =  \nonumber \\
&=&\sum\limits_{i,j,k=0}^{n}F_{k}\left( t\right) \hat{P}_{k}\mathcal{C}%
_{ij}^{\left( -1\right) }\left( t\right) \mathcal{C}_{jk}\left( t\right) = 
\nonumber \\
&=&\sum\limits_{k=0}^{n}F_{k}\left( t\right) \hat{P}_{k}=\ln \bar{\varrho}%
\left( t,0\right) \qquad ,  \label{eqIII6}
\end{eqnarray}
\begin{equation}
\mathcal{P}_{\varepsilon }\left( t\right) \hat{\zeta}_{\varepsilon }\left(
t\right) =\sum\limits_{i,j,k=0}^{n}\hat{P}_{i}\mathcal{C}_{ij}^{-1}\left(
t\right) \left\{ \hat{P}_{j},\hat{\zeta}_{\varepsilon }\left( t\right) \mid
t\right\} \qquad .  \label{eqIII7}
\end{equation}
But 
\begin{eqnarray}
\left\{ \hat{P}_{j},\hat{\zeta}_{\varepsilon }\left( t\right) \mid t\right\}
&=&Tr\left\{ \int\limits_{0}^{1}du\ \hat{P}_{j}\ Y\left( \hat{\zeta}%
_{\varepsilon }\mid u\right) \left[ \bar{\varrho}\left( t,0\right) \right]
^{u}\ \hat{\zeta}_{\varepsilon }\left( t\right) \left[ \bar{\varrho}\left(
t,0\right) \right] ^{-u}\ \bar{\varrho}\left( t,0\right) \right\} = 
\nonumber \\
&=&Tr\left\{ \hat{P}_{j}\ \varrho _{\varepsilon }^{\prime }\left( t\right)
\right\} =0
\end{eqnarray}
because of Eq. (\ref{eq24}).

\setcounter{equation}{0}

\section{Alternative Derivations in NESOM}

We present some alternative derivations of the statistical operator,
beginning with one following a path completely similar to that used to
obtaining the usual Liouville equation, but first introducing only the
retarded solutions of Schroedinger equation as proposed by Gell-Mann and
Goldberger in scattering theory [66]

We recall that in Gell-Mann and Goldberger approach the wavefunction
satisfies a modified Schroedinger equation for the state function $\mid \psi
\left( t\right) \rangle $, namely 
\begin{equation}
i\hslash \frac{\partial }{\partial t}\mid \psi \left( t\right) \rangle -\hat{%
H}\mid \psi \left( t\right) \rangle =f_{\eta }\left( t\right) \qquad ,
\label{eqIV1}
\end{equation}
where $f_{\eta }$ is an infinitesimal source which selects the retarded
solutions of the equation, which is 
\begin{equation}
f_{\eta }\left( t\right) =\eta \left( \mid \psi \left( t\right) \rangle
-\mid \phi \left( t\right) \rangle \right) \qquad ,  \label{eqIV2}
\end{equation}
and $\eta $ is an infinitesimal positive real number which goes to zero
after the calculation of averages over the quantum state of observables of
the system, and $\mid \phi \left( t\right) \rangle $ is the unperturbed
state function. Given the initial condition at $t_{o}=-\infty $, $\mid \psi
\left( t_{o}\right) \rangle =\mid \phi \left( t_{o}\right) \rangle $, and
the evolution operator $U\left( t\right) $, i.e. $\mid \psi \left( t\right)
\rangle =U\left( t\right) \mid \phi \left( t_{o}\right) \rangle $, which in
the interaction representation is $U\left( t\right) =U_{o}\left( t\right)
U^{\prime }\left( t\right) $ with $U_{o}\left( t\right) $ being the
unperturbed evolution operator, i.e. $U_{o}\left( t\right) \mid \phi \left(
t_{o}\right) \rangle =$\linebreak $\mid \phi \left( t\right) \rangle $,
after some algebra we find that 
\begin{equation}
f_{\eta }\left( t\right) =\eta \left[ \hat{1}-\hat{K}\left( t\right) \right]
\mid \psi \left( t\right) \rangle \qquad ,  \label{eqIV3}
\end{equation}
where 
\begin{equation}
\hat{K}\left( t\right) =U_{o}\left( t\right) U^{\prime \dagger }\left(
t\right) U_{o}^{\dagger }\left( t\right) \qquad .  \label{Eq.IV4}
\end{equation}
On the basis of these results we can write for the retarded solutions that
the modified Schroedinger Eq. (\ref{eqIV1}) can be written as a normal
Schroedinger equation, but for the modified wavefunction 
\begin{equation}
\mid \psi _{\eta }\left( t\right) \rangle =\mathcal{U}_{\eta }\left(
t\right) \mid \psi \left( t\right) \rangle \qquad ,  \label{eqIV5}
\end{equation}
where we have defined the operator 
\begin{equation}
\mathcal{U}_{\eta }\left( t\right) =\exp \left\{ -\eta \int\limits_{-\infty
}^{t}dt^{\prime }\hat{M}(t^{\prime })\right\}  \label{eqIV6}
\end{equation}
and 
\begin{equation}
\hat{M}(t^{\prime })=\hat{1}-\hat{K}\left( t^{\prime }\right)  \label{eqIV6a}
\end{equation}

Let us consider next the statistical ensemble which can be built on the
basis of these wavefunctions. We designate by $\mid \psi _{j}\left( t\right)
\rangle $ the wavefunction of the $j$-th replica, and by $\mathbf{P}%
_{j}\left( t\right) =\ \mid \psi _{j}\left( t\right) \rangle \langle \psi
_{j}\left( t\right) \mid $ the corresponding statistical operator in pure
space. The normal (reversible) statistical operator is given by 
\begin{equation}
\varrho \left( t\right) =\sum\limits_{j}p_{j}\mathbf{P}_{j}\left( t\right)
\qquad ,  \label{eqIV7}
\end{equation}
where $p_{j}$ is the statistical weight of the $j$-th replica compatible
with the macroscopic constraints imposed on the system. Normalization needs
be verified, implying in that $\sum\limits_{j}p_{j}=1$.

Let us now introduce for each replica a caracterization in terms of the
wavefunction of Eq. (\ref{eqIV5}), that is, the retarded wavefunctions of
Gell-Mann and Goldberger proposal. Then 
\begin{equation}
\varrho _{\eta }\left( t\right) =\sum\limits_{j}p_{j}^{\prime }\mathcal{U}%
_{\eta }\left( t\right) \mathbf{P}_{j}\left( t\right) \mathcal{U}_{\eta
}^{\dagger }\left( t\right)  \label{eqIV8}
\end{equation}
with the statistical weight $p_{j}^{\prime }$ ensuring its normalization.
This statistical operator satisfies the modified Liouville equation 
\begin{equation}
\frac{\partial }{\partial t}\varrho _{\eta }\left( t\right) +\frac{1}{%
i\hslash }\left[ \varrho _{\eta }\left( t\right) ,\hat{H}\right] =-\eta %
\left[ \hat{1}-\Bbb{P}\left( t\right) \right] \varrho _{\eta }\left(
t\right) \quad ,  \label{eqIV9}
\end{equation}
where we have introduced the definition 
\begin{equation}
\left[ \hat{1}-\Bbb{P}\left( t\right) \right] \varrho _{\eta }\left(
t\right) =\left[ \hat{1}-\hat{K}\left( t\right) \right] \varrho \left(
t\right) +\varrho \left( t\right) \left[ \hat{1}-\hat{K}^{\dagger }\left(
t\right) \right] \quad .  \label{eqIV10}
\end{equation}
Hence, we can rewrite Eq. (\ref{eqIV9}) in the form 
\begin{equation}
\frac{\partial }{\partial t}\varrho _{\eta }\left( t\right) +\frac{1}{%
i\hslash }\left[ \varrho _{\eta }\left( t\right) ,\hat{H}\right] =-\eta %
\left[ \varrho _{\eta }\left( t\right) -\tilde{\varrho}\left( t\right) %
\right] \quad ,  \label{eqIV11}
\end{equation}
where 
\begin{equation}
\tilde{\varrho}\left( t\right) =\Bbb{P}\left( t\right) \varrho _{\eta
}\left( t\right) \qquad ,  \label{eqIV12}
\end{equation}
and whose solution is 
\begin{equation}
\varrho _{\eta }\left( t\right) =\tilde{\varrho}\left( t\right)
-\int\limits_{-\infty }^{t}dt^{\prime }e^{\eta \left( t^{\prime }-t\right) }%
\frac{d}{dt^{\prime }}\tilde{\varrho}\left( t^{^{\prime }}-t\right) \quad .
\label{eqIV13}
\end{equation}

This statistical operator differs from the one of Eq. (\ref{eq36}) in two
points. One is the difference in the form of the time-dependent projection
operator, and the other that it is the logarithm of $\varrho _{\varepsilon
}\left( t\right) $ which satisfied a modified Liouville equation.

However, we notice first that if in Eq. (\ref{eqIV9}) we replace $\Bbb{P}%
_{\eta }\left( t\right) $ for $\Bbb{P}\left( t\right) $, such that 
\begin{equation}
\Bbb{P}_{\eta }\left( t\right) \varrho _{\eta }\left( t\right) =\bar{\varrho}%
\left( t,0\right)  \label{eqIV14}
\end{equation}
we obtain Zubarev's modified Liouville equation 
\begin{equation}
\frac{\partial }{\partial t}\varrho _{\eta }\left( t\right) +\frac{1}{%
i\hslash }\left[ \varrho _{\eta }\left( t\right) ,\hat{H}\right] =-\eta %
\left[ \varrho _{\eta }\left( t\right) -\bar{\varrho}\left( t,0\right) %
\right] \quad .  \label{eqIV15}
\end{equation}

Second, as shown by Zubarev and Kalashnikov [283] and by Tishenko [284] (see
also the books of Refs. [14] and [21]), the description provided by the
statistical operator of Eqs. (\ref{eqIV15}) and the one of Eq. (\ref{eq36})
are identical in the sense of providing the same average values of
observables.

Consequently, the nonequilibrium statistical operator of Section \textbf{4}
can be considered as resulting also from an equivalent alternative
construction via a nonequilibrium ensemble in the traditional style, but
postulating an \textit{ad hoc} hypothesis consisting in requiring evolution
of the system towards the future from a given initial condition of
preparation. Moreover, besides introducing irreversibility in the evolution
of the macroscopic state of the system \ -- built resorting to Boltzmann's
proposal by the use of kind of generalized Stosszahlansatz in the initial
conditions -- the form of the projection operator ensures a description in
terms of only the states contained in the informational subspace, as
described in Fig. \textbf{1}. The infinitesimal contribution on the right of
the modified Liouville equation, we stress, discards the advanced solutions
responsible for the reversibility: Its particular form - analogous to the
one in Gell-Mann and Goldberger's Schroedinger equation - introduces a
fading memory effect accounted for the use of Abel's kernel (in Zubarev's
approach; see Appendix \textbf{II}) in the definition of the Lagrange
multipliers incorporated by the variational formalism, which, we recall,
express the intensive nonequilibrium thermodynamic variables. Thus, the
particular construction of the statistical operator is a result of deriving
it in terms of wavefunctions, or better to say statistical operators for the
pure state, for the different replicas in the statistical ensemble
satisfying a modified equation of evolution which selects the retarded
solutions and, for these, the part contained in the informational subspace,
that is, giving a description based only on the basic set of informational
(or ``relevant'') variables.

Let us consider Eq. (\ref{eq36}). Introducing Gibbs-entropy operator in
MaxEnt-NESOM and the informational-entropy operator, respectively 
\[
\hat{S}_{\varepsilon }\left( t\right) =-\ln \varrho _{\varepsilon }\left(
t\right) \quad \quad and\quad \quad \hat{S}\left( t,0\right) =-\ln \bar{%
\varrho}\left( t,0\right) \qquad , 
\]
then Eq. (\ref{eq36}) reads as 
\begin{equation}
\frac{\partial }{\partial t}\hat{S}_{\varepsilon }\left( t\right) +\frac{1}{%
i\hslash }\left[ \hat{S}_{\varepsilon }\left( t\right) ,\hat{H}\right]
=-\varepsilon \left[ \hat{1}-\mathcal{P}_{\varepsilon }\left( t\right) %
\right] \hat{S}_{\varepsilon }\left( t\right) \qquad ,  \label{eqIV16}
\end{equation}
where, we recall, $\mathcal{P}_{\varepsilon }\left( t\right) \hat{S}%
_{\varepsilon }\left( t\right) =\hat{S}\left( t,0\right) $. This Eq. (\ref
{eqIV16}) also follows from Liouville equation for the regular Gibbs entropy
operator, $S_{G}\left( t\right) =-\ln \varrho \left( t\right) $, after
noticing that 
\begin{equation}
\hat{S}_{\varepsilon }\left( t\right) =\mathcal{U}_{\varepsilon }\left(
t\right) \hat{S}_{G}\left( t\right) \qquad ,  \label{eqIV17}
\end{equation}
with 
\begin{equation}
\mathcal{U}_{\varepsilon }\left( t\right) =\exp \left\{ \varepsilon
\int\limits_{-\infty }^{t}dt^{\prime }\hat{M}_{\varepsilon }(t^{\prime
})\right\}  \label{eqIV18}
\end{equation}
and 
\begin{equation}
\hat{M}_{\varepsilon }(t^{\prime })=\hat{1}-\mathcal{P}_{\varepsilon }\left(
t^{\prime }\right)  \label{eqIV18a}
\end{equation}

It is worth mentioning that there exists a certain analogy with earlier
proposals by I. Prigogine [67], namely, that irreversible behavior at the
macroscopic level may follow modifying the theory of representation in
microscopic mechanic, by introducing the so-called star-unitary
transformations $\mathcal{U}_{p}^{\star }\left( t\right) $ instead of the
regular unitary ones, in that way the statistical operator to be used takes
the form 
\begin{equation}
\varrho _{p}\left( t\right) =\mathcal{U}_{p}^{\star }\left( t\right) \varrho
\left( t\right) \qquad ,  \label{eqIV19}
\end{equation}
where $\varrho \left( t\right) $ is the regular statistical operator, and 
\begin{equation}
\mathcal{U}_{p}^{\star }\left( t\right) =\exp \left\{ \frac{1}{i\hslash }%
\hat{V}\left\{ \mathcal{L}\right\} \hat{\tau}\right\} \qquad ,
\label{eqIV20}
\end{equation}
where $\hat{V}$ is a superoperator depending on the Liouvillian operator $%
\mathcal{L}$ and $\hat{\tau}$ is an operator with the property that $\left[ 
\hat{\tau},\mathcal{L}\right] =\hslash $. The analogy with, for example, Eq.
(\ref{eqIV17}), or with 
\begin{equation}
\varrho _{\eta }\left( t\right) =\exp \left\{ \eta \int\limits_{-\infty
}^{t}dt^{\prime }\left[ \hat{1}-\Bbb{P}_{\eta }\left( t\right) \right]
\right\} \varrho \left( t\right) \quad ,  \label{eqIV21}
\end{equation}
is evident. However, it can be noticed that while in Eq. (\ref{eqIV21}) the
projection operator depends on the state of the system at any time $t$,
Prigogine's star-unitary transformation introducing irreversibility is a
universal one.

Moreover, we introduce another derivation of MaxEnt-NESOM-Zubarev's
statistical operator following a line akin to the one proposed by McLennan
[11]. Consider the nonequilibrium system with the Hamiltonian $H$ of Eq. (%
\ref{eq13}), weakly interacting with all the surrounding media via a
potential energy operator $V$, and let $H_{R}$ be the Hamiltonian of these
surroundings (``the rest of the universe''). We write $R\left( t\right) $
for the whole statistical operator, depending on the degrees of freedom of
the system and surroundings. It satisfies - for the closed system -
Liouville equation 
\begin{equation}
\frac{\partial }{\partial t}R\left( t\right) +\frac{1}{i\hslash }\left[
R\left( t\right) ,\hat{H}+\hat{H}_{R}+\hat{V}\right] =0\quad \quad .
\label{eqIV22}
\end{equation}

Definig the reduced statistical operator 
\begin{equation}
\varrho \left( t\right) =Tr_{R}\left\{ R\left( t\right) \right\} \qquad ,
\label{eqIV23}
\end{equation}
where the trace is taken over the space of states of the surroundings, and
introducing the definition 
\begin{equation}
R\left( t\right) =\frac{1}{2}\left[ \varrho \left( t\right) \Xi \left(
t\right) +\Xi \left( t\right) \varrho \left( t\right) \right] \qquad ,
\label{eqIV24}
\end{equation}
we find for $\varrho \left( t\right) $ the Liouville equation with sources 
\begin{equation}
\frac{\partial }{\partial t}\varrho \left( t\right) +\frac{1}{i\hslash }%
\left[ \varrho \left( t\right) ,\hat{H}\right] =\hat{\Phi}\left( t\right)
\varrho \left( t\right) \qquad ,  \label{eqIV25}
\end{equation}
where 
\begin{equation}
\hat{\Phi}\left( t\right) \varrho \left( t\right) =Tr_{R}\left\{ \frac{1}{2}%
\left( \frac{1}{i\hslash }\left[ V,\varrho \left( t\right) \Xi \left(
t\right) \right] +\frac{1}{i\hslash }\left[ V,\Xi \left( t\right) \varrho
\left( t\right) \right] \right) \right\} \quad .  \label{eqIV26}
\end{equation}

In the classical limit Eq. (\ref{eqIV25}) goes over the equation derived by
McLennan [11]. Furthermore, we may observe that the procedure we have used
parallels the one proposed to build the BBGKY chain of equations for reduced
density matrices [48]. Equation (\ref{eqIV25}) involves on the left the
evolution of the system under its internal interactions, while the right
side accounts for the action of the surroundings. As pointed out by
Bogoliubov there are two ways to deal with this equation: in the weak
coupling case one can use perturbation theory or one can introduce
propositions for the form of this term. A detailed study is left for a
future publication, here suffice it to say that in the weak coupling limit
and using some assumptions it can be obtained an expression of the type of a
relaxation-time approximation, namely 
\begin{equation}
\hat{\Phi}\left( t\right) \varrho \left( t\right) \simeq -\frac{1}{\tau }%
\left[ \varrho \left( t\right) -\bar{\varrho}\left( t,0\right) \right]
\qquad ,  \label{eqIV27}
\end{equation}
where the reciprocal of the relaxation time $\tau $ is proprtional to the
square modulus of the interaction potential $V$. Clearly, writing $%
\varepsilon =\tau ^{-1}$ and taking the limit of $V$ going to zero we obtain
a modified Liouville equation for $\varrho \left( t\right) $ of the type of
Eq. (\ref{eqIV15}).

In the spirit of this point of view we can then say that in Zubarev's
approach to MaxEnt-NESOM the infinitesimal source in Eq. (\ref{eq36}), which
selects the retarded solutions of the regular Liouville equation and
introduces irreversible evolution from an initial condition of preparation
of the nonequilibrium system, is a result of the inevitable interaction of
the system with the surroundings in the weak coupling limit. Irreversible
dissipative behavior is, in this interpretation, a result of such
interaction, that is, that a \textit{system in nature is never isolated}.

\setcounter{equation}{0}

\section{Generalized Nonlinear \\Mori-Heisenberg-Langevin Equations for
\\the Lagrange Multipliers}

Taking into account that because of the nonequilibrium equations of state 
\begin{equation}
Q_{j}\left( t\right) =-\delta \phi \left( t\right) /\delta F_{j}\left(
t\right) \qquad ,  \label{eqV1}
\end{equation}
and that 
\[
\frac{d}{dt}Q_{j}\left( t\right) =\sum\limits_{k=1}^{n}\frac{\delta
Q_{j}\left( t\right) }{\delta F_{j}\left( t\right) }\frac{d}{dt}F_{j}\left(
t\right) =-\sum\limits_{j,k=1}^{n}\left[ \frac{\delta ^{2}\phi \left(
t\right) }{\delta F_{j}\left( t\right) \delta F_{k}\left( t\right) }\right] 
\frac{d}{dt}F_{j}\left( t\right) =
\]
\begin{equation}
=-\sum\limits_{j,k=1}^{n}C_{jk}\left( t\right) \frac{d}{dt}F_{j}\left(
t\right) \ \qquad ,  \label{eqV2}
\end{equation}
where 
\begin{equation}
C_{jk}\left( t\right) =Tr\left\{ \int\limits_{0}^{1}du\ \hat{P}_{j}\ \left[ 
\bar{\varrho}\left( t,0\right) \right] ^{u}\ \Delta \hat{P}_{k}\ \left[ \bar{%
\varrho}\left( t,0\right) \right] ^{-u+1}\right\}   \label{eqV3}
\end{equation}
with 
\begin{equation}
\Delta \hat{P}_{k}=\hat{P}_{k}-Tr\left\{ \hat{P}_{k}\ \bar{\varrho}\left(
t,0\right) \right\} \qquad ,  \label{eqV4}
\end{equation}
being the elements of a correlation matrix which we designate by $\mathbf{%
\hat{C}}\left( t\right) $, we find that Eq. (\ref{eq37}), after neglecting
the dependence on the space variable for simplicity, can be written as 
\begin{equation}
\frac{d}{dt}\mathbf{Q}\left( t\right) =\mathbf{\hat{C}}\left( t\right) \ 
\frac{d}{dt}\mathbf{F}\left( t\right) =Tr\left\{ \mathbf{\dot{P}}\ \bar{%
\varrho}\left( t,0\right) \right\} +\left\{ \mathbf{\dot{P}},\zeta
_{\varepsilon }\left( t\right) \mid t\right\} \quad ,  \label{eqV5}
\end{equation}
where we have introduced 
\begin{equation}
\mathbf{\dot{P}=}\frac{1}{i\hslash }\left[ \mathbf{P},\hat{H}\right] \qquad ,
\label{eqV6}
\end{equation}
Eq. (\ref{eq31}) has been used and the fact that 
\begin{equation}
Tr\left\{ \mathbf{\dot{P}}\ \varrho _{\varepsilon }^{\prime }\left( t\right)
\right\} =\left\{ \mathbf{\dot{P}},\zeta _{\varepsilon }\left( t\right) \mid
t\right\} \qquad ,  \label{eqV7}
\end{equation}
using the supercorrelation function defined in Eq. (\ref{eqIII2}), and
finally $\mathbf{Q}$, $\mathbf{F}$, $\mathbf{P}$, and $\mathbf{\dot{P}}$ are
vectors with components $Q_{j}$, $F_{j}$, $P_{j}$, and $\dot{P}_{j}$, $%
j=1,2,...,n$. From this Eq. (\ref{eqV5}) we obtain that 
\begin{equation}
\frac{d}{dt}\mathbf{F}\left( t\right) =\mathbf{\hat{C}}^{-1}\left( t\right)
\ Tr\left\{ \frac{1}{i\hslash }\left[ \mathbf{P},\hat{H}\right] \ \bar{%
\varrho}\left( t,0\right) \right\} +\mathbf{\hat{C}}^{-1}\left( t\right) \
\left\{ \mathbf{\dot{P}},\zeta _{\varepsilon }\left( t\right) \mid t\right\}
\quad .  \label{eqV8}
\end{equation}
But 
\begin{equation}
Tr\left\{ \frac{1}{i\hslash }\left[ \mathbf{P},\hat{H}\right] \ \bar{\varrho}%
\left( t,0\right) \right\} =Tr\left\{ \mathbf{P\ }\frac{1}{i\hslash }\left[ 
\hat{H},\bar{\varrho}\left( t,0\right) \right] \right\} =\left( \mathbf{P,%
\dot{P}}\mid t\right) \quad ,  \label{eqV9}
\end{equation}
where 
\begin{equation}
\left( \mathbf{P,\dot{P}}\mid t\right) =Tr\left\{ \int\limits_{0}^{1}du\ 
\mathbf{P}\ \left[ \bar{\varrho}\left( t,0\right) \right] ^{u}\ \Delta 
\mathbf{\dot{P}}\ \left[ \bar{\varrho}\left( t,0\right) \right]
^{-u+1}\right\}   \label{eqV10}
\end{equation}
is a tensor acting on the $n$-component vector on the right of it, and 
\begin{equation}
\Delta \mathbf{\dot{P}}=\mathbf{\dot{P}}-Tr\left\{ \mathbf{\dot{P}}\ \bar{%
\varrho}\left( t,0\right) \right\} \qquad .  \label{eqV11}
\end{equation}
Using Eq. (\ref{eqV9}) and the expression for $\zeta _{\varepsilon }\left(
t\right) $ as given by Eq. (\ref{eqII2}), we can write Eq. (\ref{eqV8}) in
the form 
\begin{eqnarray}
\frac{d}{dt}\mathbf{F}\left( t\right)  &=&-\mathbf{\hat{C}}^{-1}\left(
t\right) \ \left( \mathbf{P,\dot{P}}\mid t\right) \mathbf{F}\left( t\right) -
\nonumber \\
&&\!-\int\limits_{-\infty }^{t}dt^{\prime }e^{\varepsilon \left( t^{\prime
}-t\right) }\mathbf{\hat{C}}^{-1}\left( t\right) \left[ \left\{ \mathbf{\dot{%
P},P}\left( t^{\prime }-t\right) \mid t\right\} \frac{d}{dt^{\prime }}%
\mathbf{F}\left( t^{\prime }\right) \right. +  \nonumber \\
&&+\left. \left\{ \mathbf{\dot{P},\dot{P}}\left( t^{\prime }-t\right) \mid
t\right\} \mathbf{F}\left( t^{\prime }\right) \right] \qquad .  \label{eqV12}
\end{eqnarray}
This Eq. (\ref{eqV12}) implies in a set of coupled highly nonlinear
integrodifferential equations for the Lagrange multipliers (intensive
nonequilibrium thermodynamic variables) $F_{j}\left( t\right) $, the
nonlinearity being present in the correlation functions which are defined in
terms of the statistical operator \noindent $\bar{\varrho}\left( t,0\right) $%
.

At this point we introduce an approximation consisting into retaining in the
supercorrelation functions the lowest order in the dissipative relaxation
processes, that is, we take in them [cf. eq. (\ref{eqII5})] $Y\left( \zeta
_{\varepsilon }\mid u\right) =1$, and then 
\begin{equation}
\left\{ \hat{A},\hat{B}\mid t\right\} \simeq \left( \hat{A},\hat{B}\mid
t\right) \equiv Tr\left\{ \int\limits_{0}^{1}du\ \hat{A}\ \left[ \bar{\varrho%
}\left( t,0\right) \right] ^{u}\hat{B}\ \left[ \bar{\varrho}\left(
t,0\right) \right] ^{-u+1}\right\} \quad .  \label{eqV13}
\end{equation}
In this approximation the projection operator of Eq. (\ref{eqIII3}) becomes 
\begin{equation}
\mathcal{P}_{\varepsilon }\left( t\right) \hat{A}=\mathbf{P}\Bbb{\ }\mathbf{C%
}^{-1}\left( t\right) \left( \mathbf{P},\hat{A}\mid t\right) \qquad .
\label{eqV14}
\end{equation}

Moreover, in Eq. (\ref{eqV12}) we keep the time derivations of the basic
variables $P$ only up to second order in the interactions, introducing in
the second term on the right of Eq. (\ref{eqV12}) 
\begin{equation}
\frac{d}{dt^{\prime }}\mathbf{F}\left( t^{\prime }\right) \simeq -\mathbf{%
\hat{C}}^{-1}\left( t^{\prime }\right) \left( \mathbf{P},\mathbf{\dot{P}}%
\mid t\right) \mathbf{F}\left( t^{\prime }\right) \qquad ,  \label{eqV15}
\end{equation}
and further taking into account the form of the projection operator of Eq. (%
\ref{eqIII3}) and its property that 
\begin{equation}
\mathcal{P}_{\varepsilon }\left( t\right) \hat{\zeta}_{\varepsilon }\left(
t\right) =0\qquad ,  \label{eqV16}
\end{equation}
we can write Eq. (\ref{eqV12}) in the compact form 
\begin{equation}
\frac{d}{dt}\mathbf{F}\left( t\right) =i\ \hat{\Omega}\left( t\right) 
\mathbf{F}\left( t\right) +\int\limits_{-\infty }^{t}dt^{\prime
}e^{\varepsilon \left( t^{^{\prime }}-t\right) }\hat{\Gamma}\left(
t^{^{\prime }}-t\mid t\right) \mathbf{F}\left( t^{\prime }\right) \quad ,
\label{eqV17}
\end{equation}
where 
\begin{equation}
\hat{\Omega}\left( t\right) =i\ \mathbf{\hat{C}}^{-1}\left( t\right) \left( 
\mathbf{P},\mathbf{P}\mid t\right) \qquad ,  \label{eqV18}
\end{equation}
\begin{equation}
\hat{\Gamma}\left( t^{^{\prime }}-t\mid t\right) =\mathbf{\hat{C}}%
^{-1}\left( t\right) \left( \mathbf{\dot{P}}\left[ \hat{1}-\mathcal{P}%
_{\varepsilon }\left( t\right) \right] e^{i\left( t^{^{\prime }}-t\right) 
\mathcal{L}}\left[ \hat{1}-\mathcal{P}_{\varepsilon }\left( t\right) \right] 
\mathbf{\dot{P}}\Bbb{\mid }t\right) \quad ,  \label{eqV19}
\end{equation}
where 
\begin{equation}
\hat{A}\left( t^{^{\prime }}-t\right) =\exp \left\{ i\left( t^{^{\prime
}}-t\right) \mathcal{L}\right\} \hat{A}\qquad ,  \label{eqV20}
\end{equation}
with $\mathcal{L}$ being the Liouvillian operator.

We can see that Eq. (\ref{eqV17}) for the Lagrange multipliers takes the
form of a generalized nonlinear Mori-like equation [9], where, following
Mori's terminology, the first term on the right is a precession term and the
second contains the memory function $\hat{\Gamma}$ (a fading memory in this
case because the presence of the exponential in the time integration). The
nonlinearity is hidden in the fact that both $\Omega $ and $\Gamma $ are
highly nonlinear expressions in the intensive nonequilibrium thermodynamic
variables $F_{j}\left( t\right) $, which are contained in the operator $\bar{%
\varrho}$.

The presence in Eq. (\ref{eqIV19}) of the projection operators $\left[ 1-%
\mathcal{P}_{\varepsilon }\right] $ at left and right of the correlation
function in the ``memory function'' $\Gamma $ cancels the contribution $%
\left[ \mathbf{P},H_{o}\right] $ to the change in time of variables $\mathbf{%
P}$, leaving only at both ends of the correlation function only the
contribution $\left[ \mathbf{P},H^{\prime }\right] $. Thus, the operator $%
\left[ 1-\mathcal{P}_{\varepsilon }\left( t\right) \right] $ projects $%
\mathbf{P}$ on the subspace which is complementary to the informational
subspace containing the basic set of variables $\left\{ P_{j}\right\} $: In
Mori's words, $\Gamma $ is a nonequilibrium correlation matrix of the
rapidly fluctuating forces $\mathbf{\dot{P}}=\left( 1/i\hslash \right) \left[
\mathbf{P},H^{\prime }\right] $. This is achieved through the difference
between the nonequilibrium correlation matrix of the total generalized
forces [last contribution on the right of Eq. (\ref{eqV12})] minus the
projection of these generalized forces over the informational subspace [a
contribution that arises from the second contribution on the right of Eq. (%
\ref{eqV12})]. This point has been stressed by V. P. Kalashnikov who showed
that the use of the total generalixed forces $\mathbf{\dot{P}}$, i.e. not
corrected by the substraction of the contributions of their secular parts
(the change under $H_{o}$), leads to wrong results like the so-called
paradox of vanishing damping at zero frequency in the Markovian limit of
this equation [80].

Equation (\ref{eqV17}) is similar to the one obtained by B. Robertson [15],
and also to those of Kawasaki and Gunton [285] and of Grabert [16], because
of the equivalence of these with that of Robertson. However, it must be
stressed a difference consisting in the presence in the correlation matrix
(or memory matrix) of the weight function $\exp \left\{ \varepsilon \left(
t^{\prime }-t\right) \right\} $, which, we recall, fixes the inital state of
preparation of the system (characterized by $\bar{\varrho}\left( t,0\right) $%
) and provides irreversible behavior for the evolution of the macroscopic
state of the system.

As final words in this Appendix, we call the attention to the fact that a
more practical and manageable way to calculate the equations of evolution of
the basic variables is the one provided by the MaxEnt-NESOM kinetic theory
[76,77], and then using Eqs. (\ref{eq41}) and (\ref{eqV6}) we can write 
\begin{equation}
\frac{d}{dt}\mathbf{F}\left( t\right) =\mathbf{\hat{C}}^{-1}\left( t\right) %
\left[ \mathbf{J}^{\left( o\right) }\left( t\right) +\mathbf{J}^{\left(
1\right) }\left( t\right) \right] +\sum\limits_{n=2}^{\infty }\mathbf{\hat{C}%
}^{-1}\left( t\right) \ \mathbf{\Omega }^{\left( n\right) }\left( t\right)
\quad ,  \label{eqV21}
\end{equation}
where $\mathbf{J}^{\left( o\right) }$, $\mathbf{J}^{\left( 1\right) }$, and $%
\mathbf{\Omega }^{\left( n\right) }$ are vectors with components $%
J_{j}^{\left( o\right) }$, $J_{j}^{\left( 1\right) }$, and $\Omega
_{j}^{\left( n\right) }$ respectively, with $j=1,2,....,n$, all of these
quantities defined in [23].

\setcounter{equation}{0}

\section{Nonequilibrium Grand-Canonical Ensemble}

For the study of many-body systems out of equilibrium, particularly the case
of solid state matter, and in its realm the area of semiconductor physics,
the relevant ensemble to consider is the grand-canonical one generalized to
nonequilibrium conditions. It is described in Ref. [55], with applications
in Ref. [56] where, on its foundations is derived a kinetic theory
appropriate for giving bases to a nonequilibrium thermo-hydrodynamics. We
briefly describe it here.

On the basis of the relevance for almost all problems in many-body physics
of considering energy and particle numbers, we begin selecting the densities
of particles $\hat{n}\left( \mathbf{r}\right) $ and of energy $\hat{h}\left( 
\mathbf{r}\right) $ for the set of basic variables. But Zubarev-Peletminskii
selection rule of Eq. (\ref{eq14}) requires that the fluxes of all order of
these densities needs be added to the set of basic variables, which is then
composed of 
\begin{equation}
\left\{ \hat{n}\left( \mathbf{r}\right) ,\ \hat{h}\left( \mathbf{r}\right)
,\left\{ \hat{I}_{n}^{\left[ r\right] }\left( \mathbf{r}\right) \right\} ,\
\left\{ \hat{I}_{h}^{\left[ r\right] }\left( \mathbf{r}\right) \right\}
\right\} \qquad ,  \label{eqVI1}
\end{equation}
where $\hat{I}_{n\left( h\right) }^{\left[ r\right] }\left( \mathbf{r}%
\right) $ is the flux of order $r$ (which is also its tensorial rank) of
particle number (energy). As discussed in Ref. [55] these fluxes are
Hermitian operators expressed in terms of single-particle dynamical
operators, and are a \ functional of a generating velocity of the wavepacket
of the particles involved.

We stress that in the case of solid state physics, and particularly of
semiconductor physics, the description of the system can be done in terms of
only single-particle operators, i.e. in terms of independent quasiparticles
like phonons, Bloch-band electrons, plasmons, magnons, polarons, polaritons,
plasmaritons, etc. Because of this fact there follows an important
consequence, namely, that since the fluxes -- which are imposed as basic
variables by Zubarev-Peletminskii selection rule of Eq. (\ref{eq14}) -- are
linear combination of the single-particle Wigner-Landau-Dirac density
matrices (that is, the average over the nonequilibrium ensemble of the
single-particle dynamical operators), the latter can, in principle, be
obtained once the equations of evolution for the fluxes are solved. In this
way, and we reinforce that only for systems amenable to be described by a
single-particle representation, we may say that Zubarev-Peletminskii rule
consists into a \textit{closure condition }for the nonequilibrium
thermodynamic description of the system: Once the single-particle density
matrices have been obtained we can calculate any macroscopic observable of
the system. This is not the case for liquids, when two-particle density
matrices are required to be added to the description (see for example [24]).
But in this case a closure condition cannot be obtained since the
two-particle potential, which differently to Coulomb interaction in solids,
cannot be dealt with in a molecular field approximation, couples the reduced
density matrices of all order leading to a kind of BBGKY hierarchy in
nonequilibrium conditions requiring the introduction at a certain point of a
decoupling approximation like in the theory of Green functions.

The basic set of dynamical quantities (mechanical observables) of Eq. (\ref
{eqIV1}), in this case the basic $P_{j}$ of the general theory of Section 
\textbf{3,} has associated the set of macrovariables (or nonequilibrium
thermodynamic variables in IST) 
\begin{equation}
\left\{ n\left( \mathbf{r},t\right) ,\ h\left( \mathbf{r},t\right) ,\left\{
I_{n}^{\left[ r\right] }\left( \mathbf{r},t\right) \right\} ,\ \left\{
I_{h}^{\left[ r\right] }\left( \mathbf{r},t\right) \right\} \right\}
\label{eqVI2}
\end{equation}
and the Lagrange multipliers 
\begin{equation}
\left\{ F_{n}\left( \mathbf{r},t\right) ,\ F_{h}\left( \mathbf{r},t\right)
,\ \left\{ F_{n}^{\left[ r\right] }\left( \mathbf{r},t\right) \right\} ,\
\left\{ F_{h}^{\left[ r\right] }\left( \mathbf{r},t\right) \right\} \right\}
\label{eqVI3}
\end{equation}
and the auxiliary (``frozen'' or instantaneous quasi-equilibrium)
statistical operator in the nonequilibrium grand-canonical ensemble is 
\begin{eqnarray}
\bar{\varrho}\left( t,0\right) &=&\exp \left\{ -\phi \left( t\right) -\int
dr^{3}\left[ F_{h}\left( \mathbf{r},t\right) \hat{h}\left( \vec{r}\right)
+F_{n}\left( \mathbf{r},t\right) \hat{n}\left( \mathbf{r}\right) +\right.
\right.  \nonumber \\
&&\left. +\mathbf{F}_{h}\left( \mathbf{r},t\right) \cdot \stackrel{\wedge }{%
\mathbf{I}}_{h}\left( \mathbf{r}\right) +\mathbf{F}_{n}\left( \mathbf{r}%
,t\right) \cdot \stackrel{\wedge }{\mathbf{I}}_{n}\left( \mathbf{r}\right) %
\right] +  \nonumber \\
&&\left. +\sum\limits_{r\geq 2}\int dr^{3}\left[ F_{h}^{\left[ r\right]
}\left( \mathbf{r},t\right) \times \hat{I}_{h}^{\left[ r\right] }\left( 
\mathbf{r}\right) +F_{n}^{\left[ r\right] }\left( \mathbf{r},t\right) \times 
\hat{I}_{n}^{\left[ r\right] }\left( \mathbf{r}\right) \right] \right\}
\label{eqVI4}
\end{eqnarray}
where $\times $ stands for fully contracted product of tensors.

The nonequilibrium grand-canonical statistical operator $\varrho
_{\varepsilon }\left( t\right) $ is the one given by Eq. (\ref{eq27}) once
we use in this expression the grand-canonical instantaneous
quasi-equilibrium operator of Eq. (\ref{eqVI4}). As shown in Ref. [72] such
nonequilibrium grand-canonical operator goes over Gibbs' grand-canonical
distribution in equilibrium when the latter is achieved after the perturbing
external sources are switched off while the contact with the thermal and
particle reservoirs is maintained. This suggests to rewrite the Lagrange
multipliers in a particular form, namely, 
\begin{equation}
F_{h}\left( \mathbf{r},t\right) =\beta \left( \mathbf{r},t\right)
=1/k_{B}T^{\star }\left( \mathbf{r},t\right) \qquad ,  \label{eqVI5}
\end{equation}
\begin{equation}
F_{n}\left( \mathbf{r},t\right) =-\beta \left( \mathbf{r},t\right) \mu
\left( \mathbf{r},t\right) \qquad ,  \label{eqVI6}
\end{equation}
\begin{equation}
\mathbf{F}_{h}\left( \mathbf{r},t\right) =-\beta \left( \mathbf{r},t\right) 
\mathbf{\alpha }_{h}\left( \mathbf{r},t\right) \qquad ,  \label{eqVI7}
\end{equation}
\begin{equation}
\mathbf{F}_{n}\left( \mathbf{r},t\right) =-\beta \left( \mathbf{r},t\right) 
\mathbf{\alpha }_{n}\left( \mathbf{r},t\right) \qquad ,  \label{eqVI8}
\end{equation}
\begin{equation}
F_{h}^{\left[ r\right] }\left( \mathbf{r},t\right) =-\beta \left( \mathbf{r}%
,t\right) \alpha _{h}^{\left[ r\right] }\left( \mathbf{r},t\right) \qquad ,
\label{eqVI9}
\end{equation}
\begin{equation}
F_{n}^{\left[ r\right] }\left( \mathbf{r},t\right) =-\beta \left( \mathbf{r}%
,t\right) \alpha _{n}^{\left[ r\right] }\left( \mathbf{r},t\right) \qquad ,
\label{eqVI10}
\end{equation}
introducing a field of nonequilibrium temperature (usually referred to as 
\textit{quasitemperature}) $T^{\star }\left( \mathbf{r},t\right) $, a field
of nonequilibrium chemical potential (dubbed quasichemical potential) $\mu
\left( \mathbf{r},t\right) $, the vectorial fields of drift velocities of
particles and of energy, $\mathbf{\alpha }_{n}\left( \mathbf{r},t\right) $
and $\mathbf{\alpha }_{h}\left( \mathbf{r},t\right) $ respectively, and some
kind of fields of tensorial drift velocities also of particles and energy $%
\alpha _{n}^{\left[ r\right] }\left( \mathbf{r},t\right) $ and $\alpha _{h}^{%
\left[ r\right] }\left( \mathbf{r},t\right) $, with $r=2,3,......,$
associated to the fluxes with an order higher that one.

We call the attention to the fact that in nonequilibrium conditions we can
find situations requiring to define different quasitemperatures for
different sets of quasiparticles, a point apparently originally stressed by
Lev Landau [163]. This is for example the case in highly excited
semiconductors as described in Section \textbf{5}, where we have also
mentioned some studies of a partial thermo-hydrodynamics for the fluid of
carriers and phonons in the photoinjected plasma in semiconductors.

\end{document}